\documentclass[a4paper,fleqn,usenatbib,useAMS]{mnras} 

\usepackage{graphicx,multicol} 
\usepackage{times}
\def\etal{\it et~al.}

\title[Classification of Subpulse Drifting in Pulsars]{Classification of Subpulse Drifting in Pulsars}
\author[Basu, Mitra, Melikidze \& Skrzypczak]{Rahul Basu$^{1}$, Dipanjan Mitra$^{2,3}$, George I. Melikidze$^{3,4}$, Anna Skrzypczak$^{3}$\\
$^{1}$ Inter-University Centre for Astronomy and Astrophysics, Pune, 411007, India; rahulbasu.astro@gmail.com \\
$^{2}$ National Centre for Radio Astrophysics, Tata Institute of Fundamental Research, Pune 411007, India \\
$^{3}$ Janusz Gil Institute of Astronomy, University of Zielona G\'ora, ul. Szafrana 2, 65-516 Zielona G\'ora, Poland \\
$^{4}$ Abastumani Astrophysical Observatory, Ilia State University, 3-5 Cholokashvili Ave., Tbilisi, 0160, Georgia \\
}
\begin{document}

%\date{Accepted\ldots Received\ldots ; in original form\ldots}

%\pagerange{\pageref{firstpage}--\pageref{lastpage}} \pubyear{2017}

\maketitle

\label{firstpage}

\begin{abstract}
In this study we propose a classification scheme for the phenomenon of subpulse
drifting in pulsars. We have assembled an exhaustive list of pulsars which 
exhibit subpulse drifting from previously published results as well as recent 
observations using the Giant Meterwave Radio Telescope. We have estimated 
detailed phase variations corresponding to the drifting features. Based on 
phase behaviour the drifting population was classified into four groups : 
coherent phase-modulated drifting, switching phase-modulated drifting, diffuse 
phase-modulated drifting and low-mixed phase-modulated drifting. We have 
re-established the previous assertion that the subpulse drifting is primarily 
associated with the conal components in pulsar profile. The core components 
generally do not show the drifting phenomenon. However, in core emission of 
certain pulsars longer periodic fluctuations are seen, which are similar to 
periodic nulling, and likely arise due to a different physical phenomenon. In 
general the nature of the phase variations of the drifting features across the 
pulsar profile appears to be associated with specific pulsar profile classes, 
but we also find several examples that show departures from this trend. It has 
also been claimed in previous works that the spin-down energy loss is 
anti-correlated with the drifting periodicity. We have verified this dependence
using a larger sample of drifting measurements.
\end{abstract}


\section{Introduction}
\noindent
The single pulse emission from pulsars consist of one or more components which
are known as subpulses. Shortly after the discovery of pulsars it was observed
that the subpulses in some cases carry out systematic drift motion within the 
pulse window and the phenomenon was termed subpulse drifting \citep{dra68}. A 
systematic approach to measuring drifting features was developed by 
\cite{bac73,bac75}, which included the fluctuation spectral studies. The 
drifting is usually characterised by two periodicities, $P_2$, the longitudinal
separation between two adjacent drift bands, and $P_3$, the interval over which
the subpulses repeat at any specific location within the pulse window. 
Fluctuation spectral studies estimate the Fourier transform of the subpulse 
intensities along specific longitude ranges within the pulse window. The 
frequency peak of the amplitude corresponds to $P_3$, while the phase 
variations across the pulse window, corresponding to the peak amplitude, give a
more detailed indication of the subpulse motion across the window compared to 
$P_2$. The periodic subpulse variations can be broadly categorised into two 
classes : phase modulated drifting which is associated with subpulse motion and
periodic amplitude modulation, similar to periodic nulling, which is expected 
to be a different phenomenon distinct from subpulse drifting \citep{bas16,
mit17,bas17,bas18b}. Phase modulated drifting also shows a wide variety of 
features which can be characterised by their drifting phase variations. 
For instance, positive drifting corresponds to the case where the phases show a
decreasing trend from the leading to the trailing edge of the profile, and vice
versa for negative drifting.

Subpulse drifting has been a topic of intensive research with the phenomenon 
expected to be present in 40-50\% of the pulsar population. There are around 
120 pulsars known at present to exhibit some form of periodic modulations in 
their single pulse sequence \citep{wel06,wel07,bas16}. The drifting effects are
very diverse and associations are seen with other phenomena like mode changing 
and nulling in some cases \citep{wri81,dei86,han87,viv97,red05,bas18b}. 
However, to gain a deeper understanding of the phenomenon at large, more 
comprehensive studies involving classification of the drifting population and 
identification of underlying traits are essential. The first such attempt was 
undertaken by \citet{ran86} who estimated the drifting behaviour within the 
empirical core-cone model of the emission beam. The pulsar profile, which is 
formed after averaging several thousand single pulses, has a highly stable 
structure and is made up of one or more (typically 5) components. The 
components can be classified into two distinct categories, the central core 
component and the adjacent conal components. A number of detailed studies 
suggest that the radio emission beam consists of a central core emission 
surrounded by conal emission arranged in nested rings \citep{ran90,ran93,
mit99}. The differences in the observed profiles are a result of the different 
line of sight (LOS) traverses across the emission beam. \citet{ran86} suggested
subpulse drifting to be primarily a conal phenomenon, and the drifting to arise
due to the circulation of the conal emission around the magnetic axis. This 
would result in an association between the drifting properties (particularly 
the phase variation of the drifting feature across the pulse) and the profile 
classification. The most prominent drifting with clear drift bands and large 
phase variations was associated with the conal single (S$_d$) and barely 
resolved conal double (D) profile classes. This corresponded to the LOS 
traversing the emission beam towards the outer edge. Progressively more 
interior LOS traverses of the emission beam result in well resolved conal 
double (D), conal Triple ($_c$T) and conal Quadruple ($_c$Q) profile shapes. In
the above classification scheme such profile classes were expected to show 
primarily longitude stationary drift with very little phase variations. The 
core dominated profiles were associated with central LOS traverses of the 
emission beam. The principal profile classes were categorised as core single 
(S$_t$), Triple (T) with a central core component and pair of conal outriders, 
and Multiple (M) with central core and two pairs of conal components. Subpulse 
drifting was expected to be phase stationary and only seen in the conal 
components of the T and M class profiles. However, the core components 
sometimes show longer periodic structures which were not classified as subpulse
drifting. 

A second classification scheme for subpulse drifting was introduced by 
\citet{wel06,wel07} for their large drifting survey involving 187 pulsars. 
These studies primarily employed the fluctuation spectrum method to quantify 
the subpulse drifting features and identified periodic phenomena in around 55\%
of the pulsars studied. Subpulse drifting was categorised based on the peak 
frequency in the fluctuation spectrum with three primary drifting classes. The 
class ``coherent'' drifters had narrow features with widths smaller than 0.05 
cycles/$P$ (where $P$ corresponds to the pulsar period). The class ``diffuse'' 
drifters had wider features compared to ``coherent'' drifters and were further 
subdivided into two categories. The subclass ``Dif'' pulsars had features which
were clearly separated from the alias boundaries of 0 and 0.5 cycles/$P$, while
the subclass ``Dif$^*$'' had features bordering either boundary. Additionally, 
a fourth class of pulsars with longitude stationary subpulse modulation were 
also observed but were not classified into the drifting population. In this 
classification scheme no information regarding the core-cone nature of the 
profile class was utilized and hence many pulsars with core components showing 
low frequency features were also identified as subpulse drifters. 

The Meterwavelength Single-pulse Polarimetric Emission Survey (MSPES) was 
conducted recently to characterise the single pulse behaviour of 123 pulsars 
\citep{mit16}. The periodic subpulse behaviour was investigated by 
\citet{bas16} with 56 pulsars showing some form of periodicity. No 
classification studies of drifting was carried out in this work. But the study
reported a clear separation between pulsars with prominent drift bands and 
periodic amplitude modulations. It was observed that the phase modulated 
drifting associated with the conal profiles showed a negative correlation 
between the drifting periodicity and the spin-down energy loss ($\dot{E}$). The
periodic amplitude modulation had large periodicities, in excess of 10$P$ and 
did not show any dependence on $\dot{E}$. It was further shown by \citet{bas17}
that the periodic amplitude modulation had similarities with the periodic 
nulling seen in certain pulsars, indicating a possible common origin for the 
two phenomena. A number of pulsars with periodic nulling have conal 
profiles which opens up the possibility that the low frequency features are not
limited to core components only. These works provide physical basis for 
the assertion made by \citet{ran86} that subpulse drifting is primarily a conal
phenomenon and the low frequency periodic modulation seen in the core 
components belongs to a different physical phenomenon. However, there are 
certain issues that have not been addressed in the MSPES studies. Firstly, 
drifting features of the conal components in core dominated profiles show 
relatively flatter phase variations. In the above works these drifting cases 
have not been clearly distinguished from periodic amplitude modulation in core 
components.
%from the periodic amplitude modulation due to the relatively flatter phase 
%variations associated with both. 
Secondly, no detailed phase variation studies associated with subpulse drifting
were reported. Finally, in order to estimate the $P_3$ dependence on $\dot{E}$,
the sparking model of drifting as suggested by \citet[][hereafter RS75]{rud75}
was used. According to the RS75 model the radio emission emerges from an 
ultra-relativistic, outflowing plasma which is generated in an Inner 
Acceleration Region (IAR) above the polar cap. The IAR maintains a 
non-stationary flow of plasma via sparking discharges that undergo non-linear 
plasma instabilities around 500 km from the stellar surface to emit 
coherent curvature radio emission \citep{ass98,mel00,gil04,mit09}. The 
subpulses are believed to be associated with each spark \citep{gil00}, and the 
drifting occurs due to movement of sparks across the magnetic field in the IAR
due to the ${\bf E \times B}$ drift, where the electric field $\mathbf{E}$
is the co-rotational electric field in the IAR. RS75 argued that in a 
pulsar where the rotation axis is anti-parallel to a star centered dipolar 
magnetic axis, as the sparks develop in the IAR, they  move slower than the 
co-rotation speed. As a consequence with every rotation the sparks lag behind. 
Since the same effect is translated to the subpulse emission, an observer sees 
this effect as subpulse drifting. In a non-aligned rotator (where the dipole 
magnetic axis and rotation axis are not coincident), assuming that the spark 
motion is primarily due to the co-rotational electric field, the sparks motion 
is around the rotation axis. This gives a preferred direction for the subpulse 
motion within the pulse window from the leading to the trailing edge of the 
profile \citep{bas16}. This justifies the assumption that negative drifting 
correspond to the sparks lagging behind co-rotation speed with a periodicity 
greater than 2$P$, while positive drifting is aliased and has periodicity 
between $P$ and 2$P$. However, it has been postulated that the magnetic field 
near the surface in the IAR is non-dipolar, while the magnetic field is 
strictly dipolar in the emission region \citep[RS75,][]{gil02,mit09,mit17b}. 
This implies that even though the sparks lag behind co-rotation speed in the 
IAR, the corresponding subpulse motion does not necessarily exhibit a 
preferential direction within the pulse window, particularly for more central 
LOS traverses of the emission beam.

As mentioned earlier the phase variations of the drifting peaks give a more 
accurate diagnostic of the subpulse motion across the pulse window compared to 
$P_2$. The possibility that the low frequency amplitude modulations can be 
associated with both core and conal emission makes the phase variation studies 
all the more important for categorising the drifting phenomenon in pulsars. 
However, only a handful of studies exist in the literature with detailed phase 
behaviour reported, for example \citet{edw03,ros11,wel16}. Phase variations
across the profile show considerable departure from linearity in a number of 
cases. A connection between non-linearity in phase variations and orthogonal 
polarization moding, especially for the outer cones, has been reported in 
\citet{ran03,ran05}. A precursor study of the detailed phase behaviour 
associated with subpulse drifting in four pulsars has been conducted by 
\citet{bas18a}. The primary objective of this paper is to extend this work and 
develop a uniform classification scheme for the subpulse drifting population. 
We have carried out extensive observations using the Giant Meterwave Radio 
Telescope (GMRT) as well as gathered archival data from the Telescope to 
characterise the subpulse drifting in the majority of pulsars where this 
phenomenon has been reported. The expanded sample is also useful for 
investigating the different physical relationships of the drifting population, 
particularly the dependence of $P_3$ on $\dot{E}$ in the presence of 
non-dipolar magnetic fields in the IAR. In section 2 we report the details of 
the observations and analysis for both archival and recently acquired data. In 
section 3 we present the details of the drifting features including the 
classification scheme introduced in this work. In section 4 we discuss the 
physical properties of drifting in detail. Finally we summarize our results in 
section 5.

\section{Observations and Analysis}
\noindent
Numerous detailed studies exist in the literature of individual pulsars 
exhibiting periodicity in their single pulse behaviour, however, there have 
been only a few major works collecting and enhancing this population. 
\citet{ran86} assembled all the available sources of periodic modulations, 
which at the time of their publication amounted to 40-50 pulsars. \citet{wel06,
wel07} considerably increased the drifting population and reported around 100 
pulsars with periodic modulation, including most of the pulsars in the 
\citet{ran86} list. More recently \citet{bas16} incremented this population by 
around 20 sources in their MSPES studies using the GMRT. In this work we 
consider primarily the pulsars exhibiting subpulse drifting and not the 
periodic amplitude modulation which is considered to be a separate phenomenon. 
As mentioned earlier drifting is restricted to the conal components of pulsar 
profiles with core components not showing any signatures of this phenomenon. 
This is most evident in the T or M type profile classes where the drifting is 
seen primarily in the leading and trailing conal components with no signatures 
seen in the central core component \citep{mit08,maa14}. On the contrary 
periodic amplitude modulation is usually seen as a low frequency feature in the
fluctuation spectrum. This phenomenon appears across both the core and the 
conal components simultaneously, at similar locations in the fluctuation 
spectrum, and is likely to be a result of changes in the conditions across the 
entire IAR. This is similar to the periodic nulling seen in certain pulsars 
which suggests a common physical origin for both \citep{bas17}. The low 
frequency feature seen in the fluctuation spectra has been interpreted in terms
of a rotating-subbeam carousel pattern of the conal emission, which is sparsely
populated and stable over several rotation periods \citep{her07,her09,for10}. 
Subpulse drifting arises due to circulation of beamlets, while the low 
frequency feature is expected to be indicative of overall circulation time. 
However, the presence of low frequency feature in core emission both in the 
form of amplitude modulation and periodic nulling strongly argues against this 
hypothesis. In certain pulsars where both these features are seen in the 
fluctuation spectra, their shapes are very different which also indicates a 
different physical origin for the two phenomena \citep{bas17}.

In order to analyze the drifting features highly sensitive single pulse 
measurements are required. We have carried out a detailed study of the subpulse
drifting population primarily using observations from the GMRT. Our objective 
was to determine the phase variations wherever possible and introduce a 
consistent classification scheme for drifting. The MSPES survey observed single
pulses at two radio frequencies, 333 and 618 MHz, from a large number of 
pulsars primarily between 0\degr~and $-$50\degr~declinations \citep{mit16}. The 
fluctuation spectral analyses for these sources were previously conducted by 
\citet{bas16}. However, detailed phase variation estimates across the profile 
were only carried out for five pulsars from this list \citep{bas18a,bas18b}. We
have used the data from these observations to carry out more detailed phase 
variation studies of drifting. Additionally, we have also assembled archival 
observations carried out in the past, at 325 MHz for 50 pulsars on three 
separate occasions, 2004 August 27, 2006 February 14 and 2007 October 26. These
observations recorded sensitive fully polarized single pulses using the now 
defunct GMRT hardware backend \citep{sir00}. Detailed studies of polarization 
characteristics for many of these pulsars have been previously reported in 
\citet{mit11}, where the observing details as well as the instrumental setup 
are explained. However, no detailed fluctuation spectral studies of the single 
pulses were then carried out. We found several pulsars with drifting 
features which have also been included in this work. There were an additional 
30-40 pulsars with periodic single pulse modulation particularly above 0\degr
declination which were outside the archival sample. We have carried out new 
observations of around 35 pulsars using the GMRT, on seven separate occasions 
between November 2017 and January 2018, for a total time of 30 hours. These 
observations were carried out as a followup to MSPES. We recorded total 
intensity signals for roughly 2000 single pulses from each pulsar, at 333 MHz, 
with an observing setup similar to \citet{bas18a}. We have compiled a list of 
61 pulsars which is possibly the most complete database of subpulse drifting 
known at present. As mentioned earlier this list does not include the periodic 
amplitude modulation and periodic nulling cases.

We have used fluctuation spectral analysis to measure the drifting features. We
employed analysis schemes similar to \citet{bas18a} in order to estimate the 
detailed phase behaviour corresponding to drifting. We estimated the Longitude 
Resolved Fluctuation Spectra (LRFS) using 256 consecutive single pulses, as 
well as their time variations by shifting the starting point by 50 pulses. The 
phase variations across the pulse window were estimated by fixing the phase 
to be zero at the longitude corresponding to peak amplitude in the average
spectrum. This process was repeated for the entire pulse sequence, as detailed
in \citet{bas18a}. The Fourier transforms were estimated for each longitude and
the phases were only calculated for significant detections (5$\sigma$ baseline 
level) of the peaks at that longitude. This implied that in certain pulsars, 
with weak peaks in the fluctuation spectra, no phase information was available.
In these cases we used the Harmonic Resolved Fluctuation Spectra 
\citep[HRFS,][]{des01,bas16} to get indications of whether the drifting is 
positive or negative or roughly phase stationary. The phase studies reported 
here were carried out for pulsars which showed long durations of systematic 
subpulse drifting. In certain pulsars there is rapid mode changing and/or 
frequent long duration nulls, and the emission switches between multiple 
states, with or without subpulse drifting, at short intervals. These require 
specialized techniques to separate emission modes and estimate drifting, and 
will be addressed in future works.

\section{The Subpulse Drifting Classes}
\noindent
Subpulse drifting has been classified in the past based on either the profile 
type or the width of the drifting feature. There are merits to both these 
schemes as although drifting is primarily a conal phenomenon there are clear 
differences seen between prominent drift bands and diffuse structures in the 
fluctuation spectra. We have classified pulsars into four groups based on the 
nature of their phase variations as well as the width of the drifting features.
The criteria for these classifications are explained in detail below. In each 
pulsar we have also specified the profile class within the core-cone model of 
the emission beam as introduced by \citet{ran90,ran93}. In some pulsars the 
profile types were not available in the literature and we have suggested 
possible profile types for the same based on their shapes and polarization 
properties.

\subsection{Coherent phase modulated drifting}

\begin{table*}
\caption{The coherent phase modulated drifting}
\label{tabphscoh}
\centering
\begin{tabular}{rccccccccc}
\hline
   & PSRJ & PSRB & $P$ & $\dot{E}$ & Mode & $P_3$ & Type & Profile &   \\
   &   &   & (s) & (10$^{30}$ erg~s$^{-1}$) &   & ($P$) &   &   &   \\
\hline
  1 & J0034$-$0721 & B0031$-$07 & 0.943 & 19.2 & A & 13$\pm$1 & ND & S$_d$ & \ref{fig_J0034_1}, \ref{fig_J0034_2} \\
    &  &  &   &   & B & 6.5$\pm$0.5 &   &   &   \\
    &  &  &   &   & C & 4.0$\pm$0.5 &   &   &   \\
    &  &  &   &   &   &   &   &   &   \\
  2 &  J0108+6608  &  B0105+65  & 1.284 & 244 & --- & 2.04$\pm$0.08 & PD & S$_d$ & \ref{fig_J0108} \\
    &   &   &   &   &   &   &   &   \\
  3 & J0151$-$0635 & B0148$-$06 & 1.465 & 5.56 & --- & 14.4$\pm$0.8 & ND & D & \ref{fig_J0151_1}, \ref{fig_J0151_2} \\
    &   &   &   &   &   &   &   &   &   \\
  4 & J0421$-$0345 & --- & 2.161 & 4.55 & --- & 3.1$\pm$0.1 & PD & *D/S$_d$ & \ref{fig_J0421}  \\
    &   &   &   &   &   &   &   &   &   \\
  5 & J0459$-$0210 & --- & 1.133 & 37.9 & --- & 2.36$\pm$0.01 & ND & *D & --- \\
    &   &   &   &   &   &   &   \\
  6 &  J0814+7429  &  B0809+74  & 1.292 & 3.08 & --- & 11.1$\pm$0.1 & ND & S$_d$ & \ref{fig_J0814}  \\
    &   &   &   &   &   &   &   &   &   \\
  7 & J0820$-$1350 & B0818$-$13 & 1.238 & 43.8 & --- & 4.7$\pm$0.1 & ND & S$_d$ & [1] \\
    &   &   &   &   &   &   &   &   &   \\
  8 & J0934$-$5249 & B0932$-$52 & 1.445 & 60.9 & --- & 3.9$\pm$0.2 & ND & S$_d$ & \ref{fig_J0934} \\
    &   &   &   &   &   &   &   &   &   \\
  9 &  J0946+0951  &  B0943+10  & 1.098 & 104 & B & 2.15$\pm$0.01 & PD & S$_d$ & [2] \\
    &   &   &   &   & Q & --- & --- & S$_d$-PC &   \\
    &   &   &   &   &   &   &   &   &   \\
 10 & J1418$-$3921 & --- & 1.097 & 26.6 & --- & 2.49$\pm$0.03 & PD & *D & \ref{fig_J1418} \\
    &   &   &   &   &   &   &   &   &   \\
 11 & J1543$-$0620 & B1540$-$06 & 0.709 & 97.4 & --- & 3.01$\pm$0.05 & PD & S$_d$ & \ref{fig_J1543} \\
    &   &   &   &   &   &   &   &   &   \\
 12 & J1555$-$3134 & B1552$-$31 & 0.518 & 17.7 & Peak1 & 17.5$\pm$3.6 & ND & *D & [1] \\
    &   &   &   &   & Peak2 & 10.2$\pm$1.0 &   &   &   \\
    &   &   &   &   &   &   &   &   &   \\
 13 & J1720$-$2933 & B1717$-$29 & 0.620 & 123 & --- & 2.452$\pm$0.006 & ND & *S$_d$ & [1] \\
    &   &   &   &   &   &   &   &   &   \\
 14 & J1727$-$2739 & --- & 1.293 & 20.1 & A & 9.7$\pm$1.6 & ND & *D & [3] \\
    &   &   &   &   & B & 5.2$\pm$0.9 &   &   &   \\
    &   &   &   &   & C & --- & --- &   &   \\
    &   &   &   &   &   &   &   &   &   \\
 15 & J1816$-$2650 & B1813$-$26 & 0.593 & 12.6 & --- & 4.1$\pm$0.2 & ND & *D & --- \\
    &   &   &   &   &   &   &   &   &   \\
 16 & J1822$-$2256 & B1819$-$22 & 1.874 & 8.12 & A & 19.6$\pm$1.6 & ND & *D & [4] \\
    &   &   &   &   & Trans A & 14.3$\pm$1.8 &   &   &   \\
    &   &   &   &   & B & --- & --- &   &   \\
    &   &   &   &   & C & 10.7$\pm$1.1 &   &   &   \\
    &   &   &   &   &   &   &   &   &   \\
 17 & J1901$-$0906 & --- & 1.782 & 11.4 & --- & 3.05$\pm$0.09 & ND & *D & \ref{fig_J1901_1}, \ref{fig_J1901_2} \\
    &   &   &   &   &   &   &   &   &   \\
 18 &  J1919+0134  & --- & 1.604 & 5.63 & --- & 6.6$\pm$0.6 & ND & *D & --- \\
    &   &   &   &   &   &   &   &   &   \\
 19 &  J1921+1948  &  B1918+19  & 0.821 & 63.9 & A & 6.1$\pm$0.3 & PD & $_c$T & \ref{fig_J1921_P1} \\
    &   &   &   &   & B & 3.8$\pm$0.1 &   &   &   \\
    &   &   &   &   & C & 2.45$\pm$0.04 &   &   &   \\
    &   &   &   &   & N & --- & --- &   &   \\
    &   &   &   &   &   &   &   &   &   \\
 20 &  J1946+1805  &  B1944+17  & 0.441 & 11.1 & A & 13.8$\pm$0.7 & ND & $_c$T & --- \\
    &   &   &   &   & B & 6.1$\pm$1.7 &   &   &   \\
    &   &   &   &   & C & --- & --- &   &   \\
    &   &   &   &   & D & --- & --- &   &   \\
    &   &   &   &   &   &   &   &   &   \\
 21 & J2046$-$0421 & B2043$-$04 & 1.547 & 15.7 & --- & 2.75$\pm$0.04 & PD & S$_d$ & \ref{fig_J2046_1}, \ref{fig_J2046_2} \\
    &   &   &   &   &   &   &   &   &   \\
 22 &  J2305+3100  &  B2303+30  & 1.576 & 29.2 & B & 2.05$\pm$0.05 & PD/ND & S$_d$ & --- \\
    &   &   &   &   & Q & $\sim$3 &   &   &   \\
    &   &   &   &   &   &   &   &   &   \\
 23 &  J2313+4253  &  B2310+42  & 0.349 & 104 & --- & 2.1$\pm$0.1 & PD & *$_c$T & \ref{fig_J2313} \\
\hline
\end{tabular}
\medskip
\\$^*$-These classifications were not available previously and suggested 
here.\\
1-\cite{bas18a}; 2-\cite{des01}; 3-\cite{wen16}; 4-\cite{bas18b}.
\end{table*}

The most prominent drifting is characterised by visible drift bands in the 
single pulse sequence. Fluctuation spectra show the presence of sharp narrow 
peaks indicating highly structured drifting. The phases corresponding to the 
peak frequency show large systematic variations across the pulse window 
signifying large scale subpulse motion. All drifting features with time 
averaged widths, measured at full width at half maximum (FWHM), less than 0.05 
cycles/$P$, and phases varying monotonically for more than 100$\degr$ across 
the majority of the pulse window have been classified as coherent phase 
modulated drifting. In Table \ref{tabphscoh} we report 23 pulsars which exhibit
this drifting effect. In addition to the basic physical parameters, $P$ and 
$\dot{E}$, the Table also shows the different emission modes, the drifting 
periodicity, the general direction of subpulse variation signified by positive 
(PD) and negative drifting (ND), and the profile classification for each 
pulsar. In many pulsars no previous profile classifications were available and 
our suggested classifications are marked with `*' in the Table. All the pulsars
in this list exhibit conal profiles with the majority showing either S$_d$ or D
type profiles. There are 7 pulsars in this group that show the presence of mode
changing with more than one stable emission state. The subpulse drifting 
changes in the different modes, with some modes showing no clear drift pattern.

In Appendix A we also show the fluctuation spectral plots for 12 pulsars, which
include the time evolution of the LRFS as well as the phase and amplitude 
variation of the peak frequency across the pulse profile. The detailed phase 
variations for three pulsars in this group, J0820$-$1350 (B0818$-$13), 
J1555$-$3134 (B1552$-$31) and J1720$-$2933 (B1717$-$29) have been shown in 
\citet{bas18a}. Pulsar J1555$-$3134 shows the presence of two distinct drifting
peaks which are not harmonically related and likely suggests fast transitions 
between two different drift states. Pulsar J1727$-$2739 nulls for around 60\% 
of the time and hence our analysis schemes were not suitable for estimating the
detailed phase behaviour. However, the single pulse behaviour was studied in 
detail by \citet{wen16} who found the presence of three distinct emission modes
during the burst phases, two of which show systematic drift motion with 
prominent drift bands. Additionally, pulsar J1822$-$2256 (B1819$-$22) has also 
been studied in detail, including the drifting behaviour, by \citet{bas18b}. 
Pulsar J0946+0951 (B0943+10) has been extensively studied \citep{des01} and 
shows the presence of large phase variations corresponding to sharp peaks in 
the fluctuation spectra. Pulsar J0459$-$0210 was part of our latest 
observations, but was affected by radio frequency interference (RFI) and could 
not be analyzed. In \citet{wel07} the fluctuation spectra showed the presence 
of a narrow peak with preferred negative drifting. Two pulsars J1816$-$2650 
(B1813$-$26) and J1919+0134 were observed in MSPES and their fluctuation 
spectra showed the presence of narrow peaks with preferred drift directions 
\citep{bas16}. However, their single pulses were not sensitive for the phase 
variation studies. In the case of PSR J1946+1805 (B1944+17) the emission is 
frequently interrupted by long nulls. Our analysis schemes were not adequate 
for estimating the detailed phase variations in this pulsar. However, the 
drifting properties have been investigated by \citet{klo10} and are consistent 
with the drifting class described here. It is also possible that the pulsar 
belongs to the switching phase-modulated drifting class described below and 
more detailed studies are required to estimate its phase variations. Finally, 
pulsar J2305+3100 (B2303+30) has prominent phase variations with a drifting 
periodicity very close to 2$P$. The temporal fluctuations of the LRFS causes 
frequent switching between positive and negative drifting which smears out the 
average phase behaviour.

\subsection{Switching phase modulated drifting}

\begin{table*}
\caption{The switching phase modulated drifting}
\centering
\begin{tabular}{rccccccccc}
\hline
   & PSRJ & PSRB & $P$ & $\dot{E}$ & Mode & $P_3$ & Profile &   \\
   &   &   & (s) & (10$^{30}$ erg~s$^{-1}$) &   & ($P$) &  &   \\
%   &   &   &   &   &   &   &   \\
\hline
 1 & J0323+3944 &  B0320+39  & 3.032 & 0.9 & --- & 8.5$\pm$0.3 & *$_c$T & \ref{fig_J0323} \\
   &   &   &   &   &   &   &   &   \\
 2 & J0815+0939 & --- & 0.645 & 20.4 & --- & 16.6$\pm$0.3 & *$_c$Q & [1] \\
   &   &   &   &   &   &   &   &   \\
 3 & J0820$-$4114 & B0818$-$41 & 0.545 & 4.60 & --- & 18.5$\pm$1.5 & $_c$Q & [2] \\
   &   &   &   &   &   &   &   &   \\
 4 & J1034$-$3224 & --- & 1.151 & 5.97 & --- & 7.2$\pm$0.5 & *$_c$Q & [3] \\
   &   &   &   &   &   &   &   &   \\
 5 & J1842$-$0359 & B1839$-$04 & 1.840 & 3.22 & --- & 12.4$\pm$0.5 & *$_c$Q & [4], \ref{fig_J1842} \\
   &   &   &   &   &   &   &   &   \\
 6 & J1921+2153 &  B1919+21  & 1.337 & 22.3 & --- & 4.2$\pm$0.2 & $_c$Q? & \ref{fig_J1921_P2} \\
   &   &   &   &   &   &   &   &   \\
 7 & J2321+6024 &  B2319+60  & 2.256 & 24.2 & A & 8$\pm$1 & $_c$Q & \ref{fig_J2321} \\
   &   &   &   &   & B & 4$\pm$1 &   &   &   \\
   &   &   &   &   & ABN & 3$\pm$0.5 &   &   &   \\
\hline
\end{tabular}
\label{tabphsw}
\medskip
\\$^*$-These classifications were not available previously and suggested
here.\\
1-\cite{cha05,sza17}; 2-\cite{bha07,bha09}; 3-\cite{bas18a}; 4-\cite{wel16}.
\end{table*}

This group of pulsars is similar to the coherent phase modulated drifters with 
sharp peaks in their fluctuation spectra and large phase variations across the 
pulse window. The primary distinguishing feature involves a sudden switch in 
the phase variation across adjacent components. The pulsars where the FWHM of 
the drifting features, in the time averaged spectra, are less than 0.05 
cycles/$P$, and the phase variations either show reversals in slope or sudden 
180\degr~jumps in adjacent components, are classified as switching 
phase-modulated drifting. The three pulsars J0815+0939, J1034$-$3224 and 
J1842$-$0359 (B1839$-$04) where detailed phase variation studies show the 
presence of the rare phenomenon of bi-drifting, i.e. the positive and negative 
drift directions are seen simultaneously at different regions of the pulse 
window, are included in this classification scheme. In Table \ref{tabphsw} we 
list the 7 pulsars which belong to this group, including their $P$, $\dot{E}$, 
different emission modes, drifting periodicity and profile classification. Once
again we have marked with `*' our suggestions for the pulsars without previous 
profile classifications. As shown in the Table most of the pulsars in this 
group belong to $_c$Q profile class, with four conal components, the only 
exception being J0323+3944 (B0320+39) which likely has $_c$T profile shape. 
Additionally, pulsar J1034$-$3224 also has a low level preceding pre-cursor 
component which does not exhibit any detectable drifting \citep{bas15,bas18a}. 
Incidentally, all three bi-drifting pulsars as well as PSR J0820$-$4114 
(B0818$-$41) have large profile widths of 100\degr~longitude or more, 
suggesting small inclination angles between the rotation and magnetic axes.

Pulsar J0815+0939 shows positive drifting only in the second component while 
the remaining components show negative drifting \citep{cha05,sza17}. In PSR 
J1034$-$3224 alternate components have opposite direction of phase variation, 
with the first and third components showing negative drifting and the second 
and fourth showing positive \citep{bas18a}. In case of J1842$-$0359 the leading
part shows phase variation corresponding to negative drifting which reverses 
direction towards the middle into positive drifting \citep{wel16}. Our analysis
at 618 MHz is shown in Appendix B and also indicates a reversal in phase 
between the leading and trailing components. The phase behaviour of pulsar 
J0323+3944 has been shown in detail in Appendix B. An 180\degr~phase jump has 
been previously reported \citep{edw03} for this pulsar near the center of the 
profile which is also seen in our plots. Pulsar J0820$-$4114 was observed as 
part of MSPES, but the detected single pulses were not sensitive for estimating
drifting behaviour. However, the pulsar has been studied in detail by 
\citet{bha07,bha09} who report that the central region shows a different 
subpulse motion compared to the outer components. It appears that no reversal 
in the drift direction is seen in the different components but more detailed 
phase variation studies would be required. Pulsar J1921+2153 (B1919+21) was 
observed in MSPES at 618 MHz and we have carried out detailed phase variation 
studies as shown in Appendix B. A jump in the phase variation is seen towards 
the center signifying different subpulse motion between the inner and outer 
parts of the profile. Finally, pulsar J2321+6024 (B2319+60) has been reported 
to exhibit three distinct drift modes by \citet{wri81}. Our drifting analysis 
could only estimate the phase variations corresponding to the most dominant 
mode with $P_3$ = 8$P$ as shown in Appendix B. The phase variations show 
positive drifting, and no drift reversals are seen in any of the components. 
But a clear shift in the phase variations is seen between the adjacent second 
and third components which justifies the classification.

\subsection{Diffuse phase modulated drifting}

\begin{table*}
\caption{The diffuse phase modulated drifting}
\centering
\begin{tabular}{rcccccccc}
\hline
   & PSRJ & PSRB & $P$ & $\dot{E}$ & $P_3$ & Type & Profile &   \\
   &   &   & (s) & (10$^{30}$erg~s$^{-1}$) & ($P$) &   &   &   \\
\hline
  1 & J0152$-$1637 & B0149$-$16 & 0.833 & 88.8 & 5.9$\pm$1.0 & ND & D & [1] \\
    &   &   &   &   &   &   &   &  \\
  2 &  J0304+1932  &  B0301+19  & 1.388 & 19.1 & 6.4$\pm$1.7 & ND & D & [1] \\
    &   &   &   &   &   &   &   &  \\
  3 &  J0525+1115  &  B0523+11  & 0.354 & 65.3 & 3.2$\pm$0.5 & PD & *D & [1] \\
    &   &   &   &   &   &   &   &  \\
  4 & J0630$-$2834 & B0628$-$28 & 1.244 & 146 & 6.9$\pm$1.5 & PD & S$_d$ & [2] \\
    &   &   &   &   &   &   &   &  \\
  5 &  J0823+0159  &  B0820+02  & 0.865 & 6.38 & 4.7$\pm$0.6 & PD & *S$_d$ & \ref{fig_J0823} \\
    &   &   &   &   &   &   &   &  \\
  6 & J0944$-$1354 & B0942$-$13 & 0.570 & 9.63 & 6.4$\pm$0.3 & ND & *$_c$T & [1] \\
    &   &   &   &   &   &   &   &  \\
  7 & J0959$-$4809 & B0957$-$47 & 0.670 & 10.8 & 5.6$\pm$1.3 & ND & D & [1] \\
    &   &   &   &   &   &   &   &  \\
  8 & J1041$-$1942 & B1039$-$19 & 1.386 & 14.0 & 4.3$\pm$0.4 & PD & *$_c$T & [1] \\
    &   &   &   &   &   &   &   &  \\
  9 & J1703$-$1846 & B1700$-$18 & 0.804 & 131 & 3.6$\pm$0.2 & ND & *S$_d$ & [2] \\
    &   &   &   &   &   &   &   &  \\
 10 & J1720$-$0212 & B1718$-$02 & 0.478 & 30.0 & 5.4$\pm$0.1 & PD & *D & [2] \\
    &   &   &   &   &   &   &   &  \\
 11 & J1741$-$0840 & B1738$-$08 & 2.043 & 10.5 & 4.6$\pm$0.6 & PD & $_c$Q? & [1] \\
    &   &   &   &   &   &   &   &  \\
 12 & J1840$-$0840 & --- & 5.309 & 6.25 & 15.0$\pm$0.8 & PD & *D & [3] \\
    &   &   &   &   &   &   &   &  \\
 13 &  J2018+2839  &  B2016+28  & 0.558 & 33.7 & 4.0$\pm$0.2 & ND & S$_d$ & \ref{fig_J2018} \\
    &   &   &   &   &   &   &   &  \\
 14 &  J2046+1540  &  B2044+15  & 1.138 & 4.88 & 23.0$\pm$6.1 & ND & D & [1] \\
    &   &   &   &   &   &   &   &  \\
 15 &  J2317+2149  &  B2315+21  & 1.445 & 13.7 & 5.2$\pm$0.4 & ND & $_c$T & [1] \\
\hline
\end{tabular}
\label{tabphsdif}
\medskip
\\$^*$-These classifications were not available previously and suggested
here.\\
1-\cite{bas16}; 2-\cite{wel07}; 3-\cite{gaj17};
\end{table*}

This group of pulsars exhibits wide structures in their fluctuation spectra 
which indicate the presence of multiple drift bands in the emission. Pulsars 
with FWHM of drifting features greater than 0.5 cycles/$P$ in the time average 
fluctuation spectra and no clearly measurable phase variations, but a clear 
preference for a drift direction, are identified as diffuse phase modulated 
drifters. A notable example of this phenomenon has been recently seen in the B 
mode of the pulsar J1822$-$2256 by \citet{bas18b}. The single pulse behaviour 
shows the presence of multiple short lived subpulse tracks with the fluctuation
spectra showing a wide structure without any distinct peaks. In Table 
\ref{tabphsdif} we have assembled 15 pulsars which exhibit wide structures in 
their fluctuation spectra. The individual drift bands usually have short 
durations, so have no significant peaks in their average spectra. As a result 
the average phase behaviour cannot be determined using the techniques employed 
in this work. However, as described in \citet{bas16} the harmonic spectrum 
gives an indication of the general nature of the drift direction. Negative 
drifting is seen in the 0-0.5 cycles/$P$ region of the HRFS while for positive
drifting the peak structure is shifted into the 0.5-1 cycles/$P$ region. In 
addition to $P$ and $\dot{E}$ the Table also lists mean drifting periodicity, 
drifting type and classification of the profile, including `*' identifiers for 
newly suggested classifications.  

As indicated in the Table the fluctuation spectral studies have been reported 
earlier for 12 pulsars, 9 pulsars in \citet{bas16}, PSR J0630$-$2834 
(B0628$-$28), J1703$-$1846 (B1700$-$18) and J1720$-$0212 (B1718$-$02) in 
\citet{wel07} and J1840$-$0840~in \citet{gaj17}. For two pulsars J0823+0159 
(B0820+02) and J2018+2839 (B2016+28) we have carried out new measurements for 
the fluctuation spectra as shown in Appendix C. Pulsar J2018+2839 is 
particularly interesting since the wide peak corresponding to drifting fills up
the entire window between 0-0.5 cycles/$P$ in the average LRFS. Short duration 
drift bands are also prominently seen in its single pulse sequence. It should 
be noted that the wide structures seen in these pulsars are different from the 
temporal fluctuations of the single drifting peaks reported in \citet{bas18a}. 
The closest counterpart is the two peaks seen in the fluctuation spectra of PSR
J1555$-$3134, where the drifting is likely to alternate between two different 
states at rapid intervals.

\subsection{Low-mixed phase-modulated drifting}

\begin{table*}
\caption{The low-mixed phase-modulated drifting}
\centering
\begin{tabular}{rcccccccc}
\hline
   & PSRJ & PSRB & $P$ & $\dot{E}$ & $P_3$ & Profile \\
   &   &   & (s) & (10$^{30}$erg~s$^{-1}$) & ($P$) &   \\
\hline
  1 & J0624$-$0424 & B0621$-$04 & 1.039 & 29.2 & 2.05$\pm$0.01 & M & \ref{fig_J0624} \\
    &   &   &   &   &   &   &   \\
  2 &  J0837+0610  &  B0834+06  & 1.274 & 130 & 2.17$\pm$0.03 & D & \ref{fig_J0837_1}, \ref{fig_J0837_2}  \\
    &   &   &   &   &   &   &   \\
  3 & J0846$-$3533 & B0844$-$35 & 1.116 & 45.5 & 2.03$\pm$0.02 & *M & \ref{fig_J0846_1}, \ref{fig_J0846_2} \\
    &   &   &   &   &   &   &   \\
  4 &  J1239+2453  &  B1237+25  & 1.382 & 14.3 & 2.8$\pm$0.1 & M & \ref{fig_J1239_1}, \ref{fig_J1239_2} \\
    &   &   &   &   &   &   &   \\
  5 & J1328$-$4921 & B1325$-$49 & 1.479 & 7.45 & 3.4$\pm$0.2 & *M/$_c$Q & \ref{fig_J1328_1}, \ref{fig_J1328_2} \\
    &   &   &   &   &   &   &   \\
  6 & J1625$-$4048 & --- & 2.355 & 1.34 & 61$\pm$30 & *T/$_c$T & \ref{fig_J1625} \\
    &   &   &   &   &   &   &   \\
  7 & J1650$-$1654 & --- & 1.750 & 23.6 & 2.6$\pm$0.1 & *D/T & \ref{fig_J1650} \\
    &   &   &   &   &   &   &   \\
  8 & J1700$-$3312 & --- & 1.358 & 74.2 & 2.2$\pm$0.1 & *M & --- \\
    &   &   &   &   &   &   &   \\
  9 & J1703$-$3241 & B1700$-$32 & 1.212 & 14.6 & 4.7$\pm$0.5 & T & --- \\
    &   &   &   &   &   &   &   \\
 10 & J1733$-$2228 & B1730$-$22 & 0.872 & 2.55 & 23$\pm$13 & *T & \ref{fig_J1733} \\
    &   &   &   &   &   &   &   \\
 11 &  J1740+1311  &  B1737+13  & 0.803 & 110 & 9$\pm$1 & M & --- \\
    &   &   &   &   &   &   &   \\
 12 & J1801$-$2920 & B1758$-$29 & 1.082 & 103 & 2.48$\pm$0.08 & *T & --- \\
    &   &   &   &   &   &   &   \\
 13 & J1900$-$2600 & B1857$-$26 & 0.612 & 35.2 & 7.6$\pm$0.8 & M & \ref{fig_J1900_1}, \ref{fig_J1900_2} \\
    &   &   &   &   &   &   &   \\
 14 &  J1912+2104  &  B1910+20  & 2.233 & 36.1 & 2.70$\pm$0.04 & M/T? & --- \\
    &   &   &   &   &   &   &   \\
 15 & J2006$-$0807 & B2003$-$08 & 0.581 & 9.27 & 15.2$\pm$2.5 & M & --- \\
    &   &   &   &   &   &   &   \\
 16 & J2048$-$1616 & B2045$-$16 & 1.962 & 57.3 & 3.23$\pm$0.03 & T & --- \\
\hline
\end{tabular}
\label{tabphstat}
\medskip
\\$^*$-These classifications were not available previously and suggested
here.\\
\end{table*}

\noindent
This group comprises pulsars which have been previously identified as 
exhibiting phase-stationary drifting, where the subpulses are not expected to 
move across the pulse window but show periodic fluctuations in intensity. These
pulsars usually show the presence of a central core component in their profiles
which does not exhibit any drifting feature. However, our detailed studies 
reveal that in many of these cases the phases are not stationary across 
the profile but show significant variations, though these variations are 
lower compared to the coherent phase-modulated pulsars. We have classified 
pulsars which show relatively shallow phase variations of less than 100\degr
across different components of the profile as low-mixed phase-modulated 
drifting. The mixed nature reflects the fact that in certain components the 
phase variations have opposite slopes. However, the extent of these variations 
is significantly lower than the switching phase-modulated drifters. The FWHM of
the drifting feature in the fluctuation spectra varies for different pulsars in
this group as well, similar to the coherent and diffuse classes described 
above, and possibly can be subdivided into two groups. However, the presence of
the core component, along with lower phase variations identify this group. In 
Table \ref{tabphstat} we list 16 pulsars including their $P$, $\dot{E}$, 
drifting periodicity and profile classification. Pulsars with periodic 
behaviour of the core component are believed to exhibit a different phenomenon 
and are not included in this list. 

In Appendix D we show the detailed phase variations across the profile for 9 
pulsars in this list. As mentioned above the phase variations are only seen in 
the conal components, with the variations much flatter than the coherent and 
switching phase modulation cases. However, even in these cases the phase 
variations are not zero but show curved trajectories. This implies that 
even for central LOS traverses through the emission beams the conal components 
show subpulse motion across the pulse window, which is smaller than the more 
peripheral traverse. Pulsar J0837+0610 (B0834+06) has a D profile and is the 
only one in this list without the presence of any clear core component. 
However, the presence of relatively shallow phase variations of less than 
50\degr~across its two components prompted its drifting classification. In 
pulsar J1700$-$3312 we were not able to estimate the phase behaviour due to the
lower sensitivity of the single pulses. However, the presence of a possible 
core component and the low drifting periodicity around 2$P$ justified its 
classification. More sensitive single pulse studies are required to estimate 
the phase variations in this pulsar. In the case of three pulsars J1703$-$3241 
(B1700$-$32), J1740+1311 (B1737+13) and J2048$-$1616 (B2045$-$16) the drifting 
exhibited wide features, with FWHM in excess of 0.05 cycles/$P$, without any 
prominent single drifting periodicity, similar to the diffuse phase modulation 
case. Hence, no clear phase variation studies could be carried out for these 
pulsars, but detailed fluctuation spectra for these sources have been reported 
in \citet{bas16}. In each of these pulsars it is clear that the drifting is 
restricted to the conal regions only. We did not have access to single pulses 
from the pulsar J1912+2104 (B1910+20) and could not estimate the phase 
variations. The pulsar has been studied by \citet{wel07} where a low 
periodicity drifting feature is seen for only the leading conal component. 
Finally, two pulsars J1801$-$2920 (B1758$-$29) and J2006$-$0807 (B2003$-$08) 
also show the presence of drifting features in the fluctuation spectra 
associated with only the conal components \citep{bas16}. However, their single 
pulse behaviour is also characterized by frequent nulls which renders our 
analysis techniques inadequate for phase variation studies. These would require
a separate analysis scheme where the burst regions are separated out and 
drifting analysis is carried out on individual segments.

\section{Discussion}
\subsection{The effect of Aliasing and dependence on Spin-down energy loss}

\begin{figure*}
\begin{tabular}{@{}lr@{}}
{\mbox{\includegraphics[scale=1.0,angle=0.]{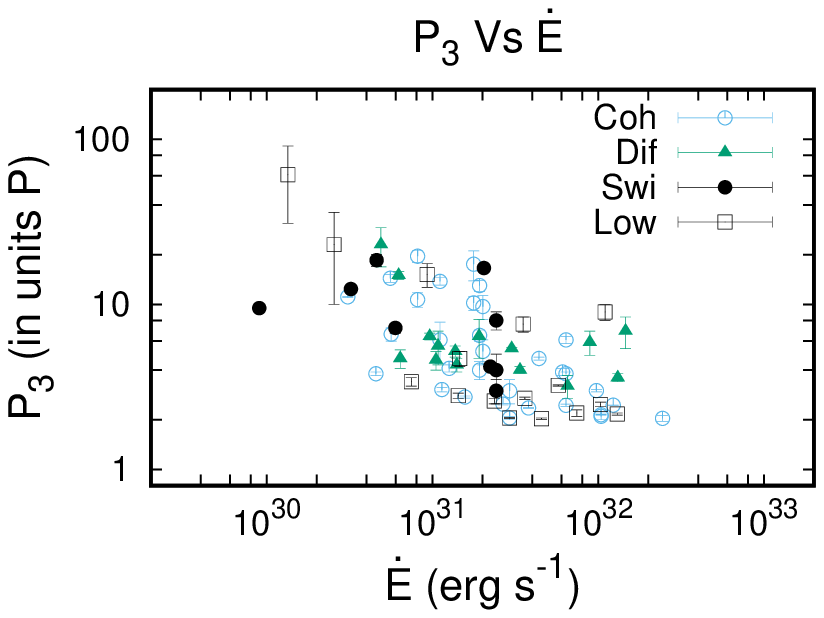}}} &
{\mbox{\includegraphics[scale=1.0,angle=0.]{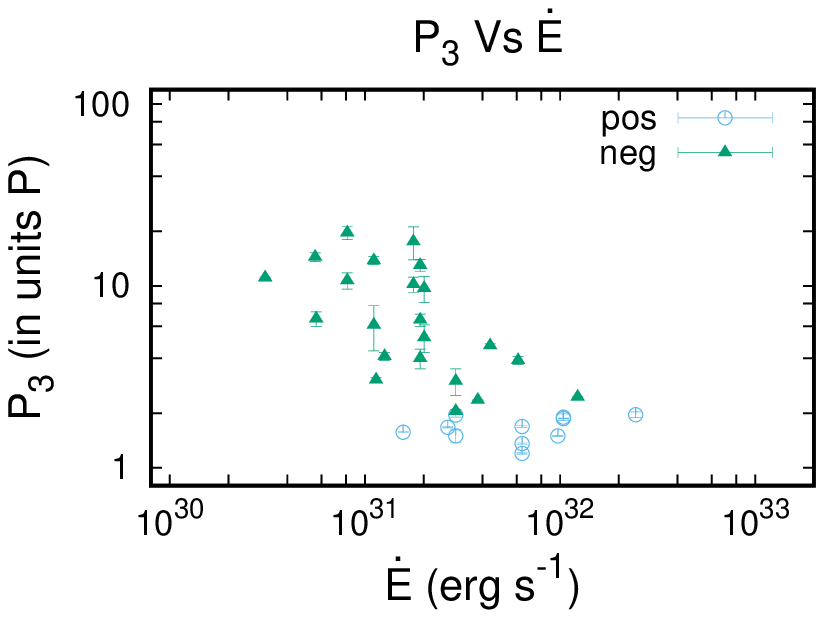}}} \\
\end{tabular}
\caption{The Figure shows the variation of the drifting periodicity $P_3$ as a
function of the spin-down energy loss ($\dot{E}$). The left panel shows all 
pulsars with subpulse drifting with the four different classes of drifting 
separately indicated in the Figure. The right panel only shows the coherent 
drifting class associated with peripheral line of sight cuts of the emission 
beam. The alias around 2$P$ is resolved by assuming $P_3>$2$P$ for negative
drift directions and $P_3<$2$P$ for positive drifting.}
\label{fig_p3edot}
\end{figure*}

\noindent
As discussed in the introduction, the RS75 model, which postulates that the 
sparking discharges in the IAR lag behind co-rotation speed, has been used to 
unravel the aliasing associated with subpulse drifting. This gives a natural 
solution for estimating the alias where negative drifting has periodicities in 
excess of 2$P$, and positive drifting has periodicities less than 2$P$. This 
was used to determine a dependence between the drifting periodicity and 
spin-down energy loss ($\dot{E}$) which exhibits a negative correlation 
\citep{bas16}. However, there are some issues with the above assumptions. 
Firstly, a number of pulsars like the switching phase-modulated and low-mixed 
phase-modulated drifters do not show any specific drift direction. Secondly, 
the magnetic field structure in the IAR is highly non-dipolar in nature, while 
the magnetic field in the emission region is dipolar \citep{mit17b}. This 
ensures that the spark tracks in the non-dipolar IAR which lag behind 
co-rotation speed will have more convoluted subpulse motion in the emission 
region due to the transition of the magnetic fields from the non-dipolar to 
dipolar. This still does not justify a carousel like motion of subpulses as 
proposed in certain models \citep{gil00,des01}, as the subpulse motion in this 
model requires the sparks to lag behind co-rotation speed in certain instances 
and lead co-rotation speed in others. A more detailed model of subpulse 
drifting needs to be explored to understand the observational results. But, we 
once again take a detailed look into the dependence of $P_3$ on $\dot{E}$ for
the extended sample studied here. In Figure \ref{fig_p3edot}, left panel, we 
have plotted $P_3$ as a function of $\dot{E}$ for all 61 pulsars which are 
expected to exhibit subpulse drifting. The four different drifting classes are 
represented separately in the Figure. We have not resolved the alias using the 
drift direction and have used periodicities $>$2$P$ for all measurements. On 
the right panel we have isolated only the coherent drifting pulsars which are 
mainly associated with LOS cuts of the emission beam towards the edge of the 
profile. These are less likely to be affected by non-dipolar effects of the 
inner field lines. Similar to \citet{bas16} we have resolved the alias around 
2$P$ by assuming that negative drifting has $P_3>$2$P$ and positive drifting 
has $P_3<$2$P$. As seen in the left panel of the Figure, despite a large 
scatter there exists a dependence between the two quantities with $P_3$
increasing with decreasing $\dot{E}$. The dependence flattens out between 
10$^{31}$ and 10$^{32}$ erg~s$^{-1}$ where $P_3$s are likely to be aliased. 
This is further highlighted by the fact that most $P_3$ values close to 2$P$ 
boundary appear in this range. The correlation becomes even more prominent on 
the right panel where we are dealing with cleaner examples of systematic 
subpulse drifting.

\subsection{The Pulsars without subpulse drifting}

\begin{table*}
\caption{List of Pulsars with prominent conal components but no detectable 
drifting}
\centering
\begin{tabular}{cccc|cccc|cccc}
\hline
   & Pulsar & $\dot{E}$ & Profile &   & Pulsar & $\dot{E}$ & Profile &   & Pulsar & $\dot{E}$ & Profile  \\
   &   & (10$^{30}$erg~s$^{-1}$) &   &   &   & (10$^{30}$erg~s$^{-1}$) &   &   &   & (10$^{30}$erg~s$^{-1}$) &  \\
\hline
%   &   &   &   &   &   &   &   &   &   &   &  \\
 1 & B0052+51 & 39.8 & *D & 26 & B1530+27 & 21.6 & S$_d$/D & 51 & B1845$-$01 & 723 & $_c$T \\
 2 & J0134$-$2937 & 1.20$\times10^3$ & *$_c$T & 27 & B1541+09 & 40.7 & T & 52 & J1850+0026 & 11.2 & *M \\
 3 & B0138+59 & 8.44 & M & 28 & B1558$-$50 & 4.26$\times10^3$ & T & 53 & J1857$-$1027 & 8.31 & *T/$_c$T \\
 4 & B0144+59 & 1.34$\times10^3$ & *T & 29 & B1601$-$52 & 35.5 & D & 54 & B1905+39 & 11.3 & M \\
 5 & B0329+54 & 222 & T/M & 30 & B1604$-$00 & 161 & T & 55 & B1907+03 & 13.9 & T/M \\
 6 & B0402+61 & 1.05$\times10^3$ & T/M & 31 & B1612+07 & 53.0 & S$_d$ & 56 & B1914+09 & 5.04$\times10^3$ & T$_{1/2}$ \\
 7 & B0447$-$12 & 48.2 & *M & 32 & B1633+24 & 39.9 & $_c$T & 57 & B1917+00 & 147 & T \\
 8 & B0450$-$18 & 1.37$\times10^3$ & T & 33 & B1648$-$17 & 130 & *D & 58 & B1923+04 & 78.3 & S$_d$ \\
 9 & B0450+55 & 2.37$\times10^3$ & T & 34 & B1649$-$23 & 25.2 & *T$_{1/2}$ & 59 & B1929+10 & 3.93$\times10^3$ & T$_{1/2}$ \\
10 & B0458+46 & 846 & T & 35 & J1652+2651 & 33.6 & *D & 60 & B2021+51 & 816 & D/S$_d$ \\
11 & J0520$-$2553 & 84.2 & *D & 36 & B1717$-$16 & 59.6 & *D & 61 & B2044+15 & 4.88 & D \\
12 & B0525+21 & 30.1 & D & 37 & B1718$-$32 & 235 & *T$_{1/2}$ & 62 & B2110+27 & 59.5 & S$_d$ \\
13 & B0559$-$05 & 828 & *T & 38 & B1727$-$47 & 1.13$\times10^4$ & T & 63 & B2148+63 & 121 & S$_d$ \\
14 & B0727$-$18 & 5.64$\times10^3$ & *T & 39 & B1742$-$30 & 8.49$\times10^3$ & T & 64 & B2154+40 & 38.2 & $_c$T/D \\
15 & B0736$-$40 & 1.21$\times10^3$ & T & 40 & B1745$-$12 & 782 & T/M & 65 & B2227+61 & 1.02$\times10^3$ & *$_c$Q/M \\
16 & B0740$-$28 & 1.43$\times10^5$ & T/M & 41 & B1747$-$46 & 125 & T/M & 66 & B2306+55 & 73.5 & D \\
17 & B0751+32 & 14.2 & D & 42 & B1753+52 & 4.52 & *D & 67 & B2323+63 & 37.6 & D/$_c$Q \\
18 & B0905$-$51 & 4.43$\times10^3$ & *$_c$Q/M & 43 & B1754$-$24 & 4.00$\times10^4$ & *T & 68 & B2327$-$20 & 41.2 & T \\
19 & B0919+06 & 6.79$\times10^3$ & T & 44 & B1758$-$29 & 103 & *T & 69 & J2346$-$0609 & 32.6 & *D \\
20 & B0950+08 & 560 & S$_d$ & 45 & B1804$-$08 & 259 & T &  &  &  &  \\
21 & B1133+16 & 87.9 & D & 46 & J1808$-$0813 & 72.8 & *S$_d$ &  &  &  &  \\
22 & B1254$-$10 & 60.9 & *D & 47 & B1821+05 & 21.0 & T &  &  &  &  \\
23 & B1322+83 & 74.3 & S$_d$ & 48 & B1826$-$17 & 7.57$\times10^3$ & T &  &  &  &  \\
24 & B1508+55 & 488 & T & 49 & B1831$-$04 & 116 & M &  &  &  &  \\
25 & B1524$-$39 & 53.3 & D & 50 & B1845$-$19 & 11.5 & *T &  &  &  &  \\
\hline
\end{tabular}
\label{tabnodrift}
\medskip
\\$^*$-These classifications were not available previously and suggested
here.\\
\end{table*}

As noted earlier drifting is primarily a conal phenomenon and hence core 
dominated pulsars belonging to the S$_t$ and T profile classes, with weak or 
absent conal emission, do not show any drifting. However, the low frequency 
features associated with periodic amplitude modulation or nulling are seen in a
number of these pulsars. Additionally, there are also many pulsars with 
prominent conal emission where detailed studies have not revealed any 
measurable drifting features. In Table \ref{tabnodrift} we have listed 69 
pulsars with distinct conal components which do not show any drifting, along 
with their $\dot{E}$ and profile classifications. There are 23 pulsars in this 
list with $\dot{E} >$ 5$\times$10$^{32}$ erg~s$^{-1}$ and the remaining 46 are 
below this limit. There are also around 100 pulsars with profile 
classifications where the single pulses are too weak to carry out detailed 
analysis.

One primary feature of drifting studies reported in this work is that the 
majority of measurements are carried out at a single frequency of 325 MHz. This
however makes it difficult to estimate frequency evolution of the drifting 
phenomenon particularly at higher frequencies. It has been observed that the 
conal emission in core dominated pulsars becomes more prominent at higher 
frequencies \citep{ran83,mit17}. It is possible that some core dominated 
pulsars show subpulse drifting in their conal components at higher frequencies.
But a majority of such pulsars have $\dot{E}$ greater than 10$^{33}$ 
erg~s$^{-1}$ making it unlikely for drifting to be seen. Detailed single pulse 
observations at frequencies in excess of 2 GHz would be required to address 
this issue. There are 46 pulsars where sensitive single pulse studies show no 
drifting. This implies that drifting is seen in 57\% of pulsars with conal 
emission and $\dot{E} <$ 5$\times$10$^{32}$ erg~s$^{-1}$. As seen in the Table 
non-drifting pulsars are not restricted to any specific profile class. It is 
not clear at present why a large fraction of conal pulsars do not show presence
of drifting. One possibility is that the absence of drifting is an extreme 
example of diffuse drifting category where the pulsar switches rapidly between 
multiple subpulse tracks. In absence of any apparent physical characteristics 
distinguishing between the two populations, more detailed modelling is 
essential to better understand the conditions affecting the drifting 
phenomenon.

\subsection{The Conal emission in Pulsars}

\begin{figure*}
\begin{tabular}{@{}lr@{}}
{\mbox{\includegraphics[scale=0.68,angle=0.]{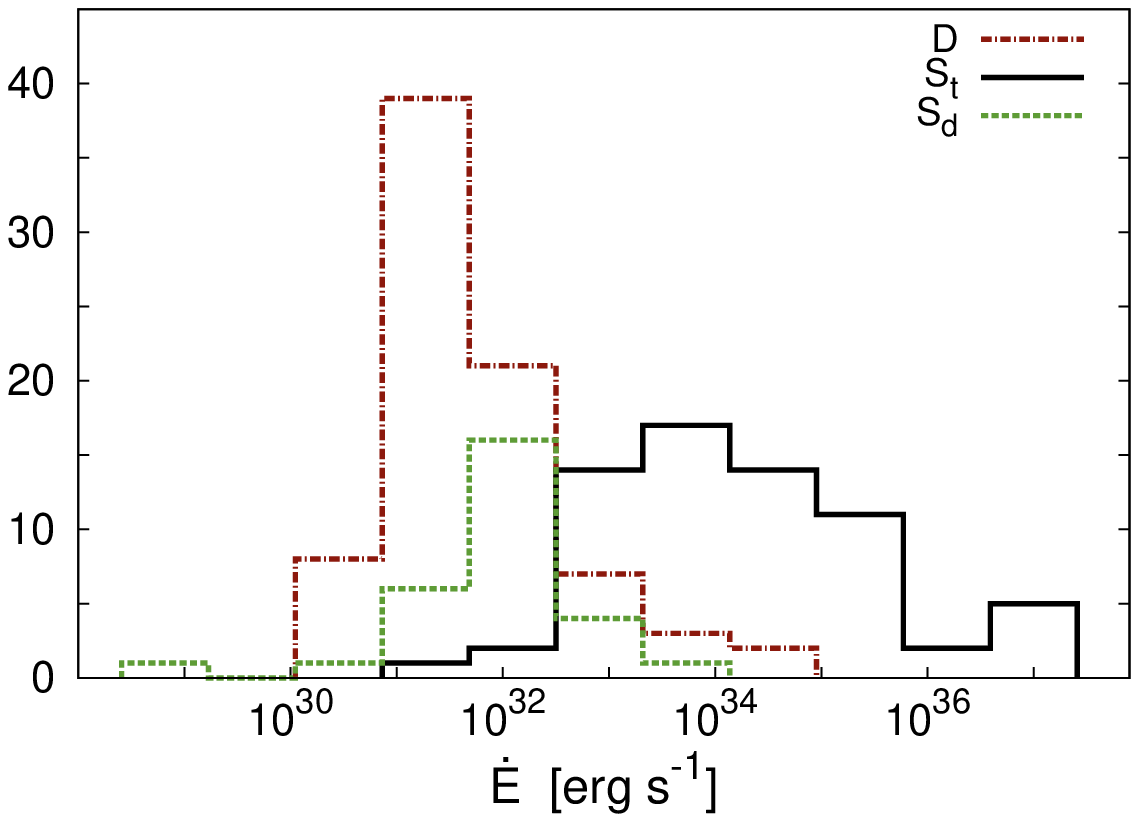}}} &
{\mbox{\includegraphics[scale=0.68,angle=0.]{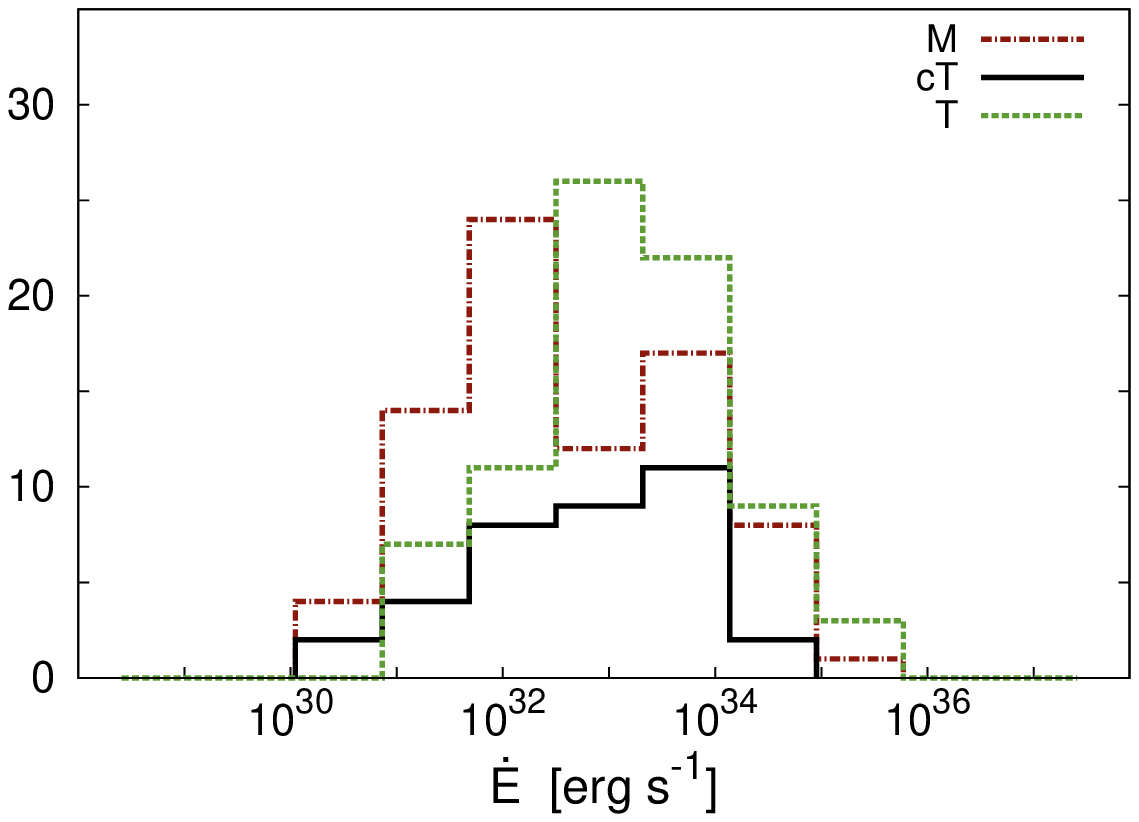}}} \\
\end{tabular}
\caption{The Figure shows the distribution of the different profile classes as 
a function of the spin-down energy loss ($\dot{E}$). The left panel shows the 
distribution for the three profile classes, conal single (S$_d$), conal double 
(D) and core single (S$_t$). The right panel shows the distribution for the
Multiple (M), Triple (T) and conal Triple ($_c$T) profile classes. The $_c$T 
class in this distribution is a combination of the conal Triple and conal 
Quadruple ($_c$Q) classes.}
\label{fig_prof_edot}
\end{figure*}

\noindent
There are no cases of drifting with $\dot{E} >$ 5$\times$10$^{32}$ erg~s$^{-1}$
which has also been noted previously in \citet{bas16}. It has also been 
established that drifting is primarily a conal phenomenon. This raises the 
question about whether the different profile classes have similar dependencies 
on $\dot{E}$. As mentioned earlier profile classifications were introduced in 
the works of \citet{ran90,ran93}. Detailed profile classifications for the 
MSPES pulsars were carried out by \citet{skr17}. There are around 300 pulsars 
with classifications, and their distribution as a function of $\dot{E}$ for 6 
different classes S$_d$, S$_t$, D, $_c$T, T and M are shown in Figure 
\ref{fig_prof_edot}. The $_c$T distribution also includes the $_c$Q class since
they both represent conal profiles with multiple components. The Figure shows a
clear demarcation between core dominated profiles and conal pulsars. The 
$\dot{E}$ of majority of conal pulsars belonging to the S$_d$ and D classes lie
below 10$^{33}$ erg~s$^{-1}$ while the core single profiles mostly have higher 
$\dot{E}$ values. The differences become less clear in more complicated profile
types (T, $_c$T and M), though their numbers become lower at higher $\dot{E}$ 
range. This gives a possible indication about absence of drifting in the higher
$\dot{E}$ range. These results highlight that the distinction between core and 
conal dominated profiles are not just a geometrical effect, related to the line
of sight traverse through the emission beam, but also a likely outcome of other
physical processes. In previous studies there have been indications of physical
differences between profile classes. For example \citet{ran93} used 
$B_{12}/P^2$ as an indicator for core and conal species with the cores having
larger values. Similarly, \citet{gil00} introduced the complexity parameter to 
distinguish core and conal profile classes. It has been recently shown by 
\citet{skr18} that the underlying widths of core and conal components in 
profiles are similar and has a $P^{-0.5}$ dependence. In order to understand 
the physical processes leading to different profile classes more detailed 
modelling is required. The dependence seen between $P_3$ and $\dot{E}$ reported
in this work should serve as an input into these models. One example is the 
partially screened gap (PSG) model where the IAR is screened by thermally 
emitted ions from the stellar surface \citep{gil03,sza15}. The drifting 
periodicity in this model is inversely proportional to screening factor of the 
gap and hence a inverse dependence of $P_3$ on $\dot{E}$ can be established 
\citep{bas16}. However, this model needs to be extended to explain variations 
seen in the physical properties of core and conal pulsars.

\subsection{The association between Drifting and Profile Classifications}

\begin{figure*}
{\mbox{\includegraphics[scale=0.42,angle=0.]{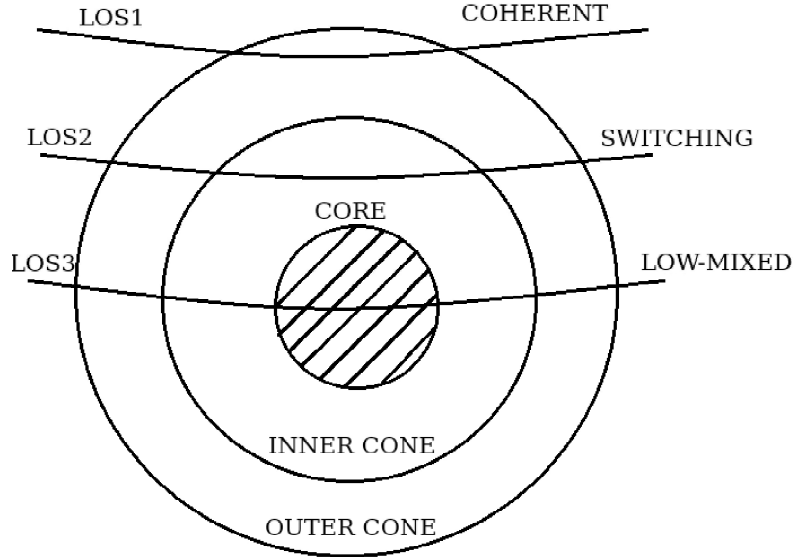}}} 
\caption{The Figure shows a schematic of the association of the different 
drifting classes with the line-of-sight (LOS) traverse of the emission beam. 
Coherent drifting with regular phase variations are associated with outermost 
LOS1. Switching phase-modulated drifting is usually associated with more 
interior LOS2. The central core region does not show any systematic subpulse 
drifting. Low-mixed phase-modulated drifting is associated with conal regions 
of central LOS3.}
\label{fig_drift_schem}
\end{figure*}

\noindent
A distinct association between the nature of subpulse drifting and profile 
class was noted by \citet{ran86}. 
%Although no detailed phase behaviour was 
%reported in this work, the tendency of the phase variation were noted and a 
%general association was made between the nature of subpulse drift variation 
%and the profile class. 
It was seen that the most ordered drifting was associated with the S$_d$ class 
and barely resolved D profiles where the LOS was supposed to traverse the 
emission beam tangentially. On the contrary well resolved D profiles as well as
conal components of the T and M profiles were not expected to drift but show 
pulse-to-pulse phase-stationary modulation. This behaviour was interpreted as 
the LOS traversing emission beam more centrally, with conal radio emission 
circulating around magnetic axis. The presence of detailed phase variation 
measurements associated with drifting makes it appropriate to revisit this 
suggested association.

As shown in Table \ref{tabphscoh} the majority of pulsars which exhibit large
and systematic phase variations across the profile are associated with S$_d$ 
and barely resolved D profiles. The phases monotonically increase or decrease 
across the profile for several hundred degrees. 
%This is entirely consistent with the expectations of \citet{ran86}. 
However, as profile shapes become more complicated the drifting behaviour 
also show large diversity. Well resolved conal D profiles do not necessarily 
exhibit phase-stationary behaviour. For example in PSR J0837+0610 (B0834+06) 
the phase variations, particularly in the trailing component, are relatively 
flat with less than 20\degr~variation. The pulsar has been classified in the 
low-mixed phase-modulated drifting category. In contrast a number of pulsars 
with D profiles show large phase variations across the components and have been 
classified in the coherent phase-modulated drifting class. The most prominent D
class pulsars with clearly separated components are J0151$-$0635 (B0148$-$06), 
where the leading component exhibits between 100-150\degr~phase variations and 
the trailing component between 50-100\degr; and J1901$-$0906 where the drifting
is prominent primarily in the trailing component with more than 200\degr~phase 
variation across the component. 
%{\bf The polarization modal analysis would be 
%instructive in understanding the phase variations in these cases\citep{ran03}.}
In three pulsars J1921+1948 (B1918+19), J1946+1805 (B1944+17) and J2313+4253 
(B2310+42) belonging to the $_c$T profile class, the phases show more than 
several hundred degree variations across the profile and are classified as 
coherent phase-modulated drifters. Pulsar J0323+3944 (B0320+39) also shows 
similar large phase variations, but a 180\degr~phase jump is seen towards the 
center of the profile. Due to its complex phase behaviour we have classified 
the profile as $_c$T. For six $_c$Q pulsars showing subpulse drifting, once 
again large phase variations exceeding several hundred degrees across the 
profiles are seen. However, in each of these pulsars phase variations show 
sudden jumps between components, which include the three bi-drifting pulsars 
where slope of phase variations changes sign due to drift reversals. These 
pulsars along with J0323+3944 (B0320+39) have been identified as a separate 
class of switching phase-modulated drifting. 
%This category of drifting has been
%identified much later, owing to the improved sensitivities of observations, and
%greatly deviate from the simplistic associations proposed in \citet{ran86}. 
Finally, the drifting in the conal components of T and M class profiles, in 
addition to PSR J0837+0610 (B0834+06), is identified as low-mixed 
phase-modulated drifting which is different from phase stationary behaviour.
%expected in \citet{ran86}. 
The phase variations are not strictly zero but show variations of 
50\degr~across certain components. The most prominent variations are seen in 
the inner cone of PSR J1239+2453 (B1237+25), as well as J1733$-$2228 (B1730-22)
in its strong leading component. 

In summary a general evolution of the phase variations corresponding to 
drifting is indeed seen with the pulsar profile class. The more conal profiles,
S$_d$, D and $_c$T are associated with tangential LOS cuts of the emission beam
and show a systematic change in the phase variations across the profile 
components. In case of more inward LOS traverses, corresponding to $_c$Q 
profile type and certain $_c$T pulsars, the phase jumps and phase reversals 
become more prominent. Finally, for central LOS traverse corresponding to the M
and T class profiles, the drifting is absent in the central core component 
while the conal components show relatively smaller phase variations, which 
however still show the complexity of the switching class. A schematic 
explaining the association of the different drifting categories with the LOS 
traverse of the emission beam is shown in the Figure \ref{fig_drift_schem}. But
it should be noted that the schematic is only representative of the generalized
association between drifting behaviour, with profile class with clear 
deviations seen in individual pulsars. Estimation of the detailed emission
geometry in each case would likely provide a better understanding of this 
association. There are a few possibilities for the deviation of the phase 
variations from linearity. A flattening of phases towards the peripheral 
longitudes is seen in a number of S$_d$ pulsars. This can develop due to 
curvature of the LOS in more tangential traverses of the emission beam. In 
certain pulsars it has been shown that phase jumps are correlated with 
variations in the orthogonal polarization modes \citep{ran03,ran05}, 
particularly for the outer conal components. This phenomenon needs to be 
explored in more detail by carrying out sensitive polarization modal studies in
drifting pulsars. Finally, presence of non-dipolar surface fields in the IAR, 
as proposed by RS75, is another possible source of deviation from linear phase 
variations. The sparks, responsible for the outflowing plasma, undergo 
{\bf E}x{\bf B} drift in the IAR in presence of highly curved non-dipolar 
magnetic fields. On the other hand the subpulses corresponding to the observed 
emission originate at regions associated with dipolar field lines 
\citep{mit17b}. The transition from non-dipolar fields in the IAR to dipolar 
fields in the emission region is likely to lead to non-linearity in drifting 
phases. However, detailed modelling is required to understand the spark motion 
in the IAR.

\section{Summary}
\noindent
We have carried out a detailed analysis of subpulse drifting in pulsars and 
assembled the most complete sample exhibiting this phenomenon. We have made 
careful distinctions between drifting and other phenomena like periodic 
amplitude modulations and periodic nulling. We have used fluctuation spectral 
analysis to characterise the drifting features, particularly the nature of 
frequency peaks and phase variations across the profile associated with the 
peaks. Based on these measurements the drifting population was divided into 
four main categories: coherent phase-modulated drifting with narrow frequency 
features and regular phase variations across the profile; switching 
phase-modulated drifting with narrow frequency features but exhibiting sudden 
jumps and reversals in phase variations; diffuse phase-modulated drifting with 
relatively wide frequency features and indication of a drift direction; and 
low-mixed phase-modulated drifting associated with pulsars with central core 
components, which do not show drifting, and corresponding conal components, 
showing relatively lower but complex phase variations. The classification 
scheme introduced in this work differs from the previous schemes of 
\citet{ran86} and \citet{wel06,wel07}. In contrast to \citet{ran86} the 
classification is based not just on profile class but also the nature of 
subpulse drifting seen in each profile class. However, our classification uses 
the suggestion by \citet{ran86} that drifting is restricted to conal 
components, which was not utilized in the works of \citet{wel06,wel07}. 
Drifting periodicities are anti-correlated with spin-down energy loss 
($\dot{E}$). It is also observed that drifting is restricted to a narrow range 
of $\dot{E}$ $<$ 5$\times$10$^{32}$ erg~s$^{-1}$. The drifting 
classification shows an association with the profile types and line-of-sight 
(LOS) geometry. The coherent phase-modulated drifting is associated with S$_d$ 
and barely resolved D profiles with tangential LOS traverses of the emission 
beam. The switching phase-modulated drifting is seen in more interior LOS 
traverses of $_c$T and $_c$Q pulsars. While the low-mixed phase-modulated 
drifting usually appears in T and M profiles with central LOS traverse.
%A detailed estimation
%of LOS geometry in each pulsar will facilitate better understanding of the 
%association seen between drifting phase behaviour and profile class.}
%This is likely to be a manifestation of the conal pulsars to be restricted in 
%this range, and requires more detailed understanding of the evolution of 
%pulsars between the core dominated and conal profile classes. 

\section*{Acknowledgments}
\noindent
We thank the referee Prof. Joanna Rankin for her comments which helped to 
improve the paper. Dipanjan Mitra acknowledges funding from the grant 
``Indo-French Centre for the Promotion of Advanced Research - CEFIPRA". We 
thank the staff of the GMRT who have made these observations possible. The GMRT
is run by the National Centre for Radio Astrophysics of the Tata Institute of 
Fundamental Research.

\appendix
\section{The coherent phase modulated drifting : Fluctuation Spectral Plots}

% J0034-0721 - 333 MHz
\begin{figure*}
\begin{tabular}{@{}lr@{}}
{\mbox{\includegraphics[scale=0.4,angle=0.]{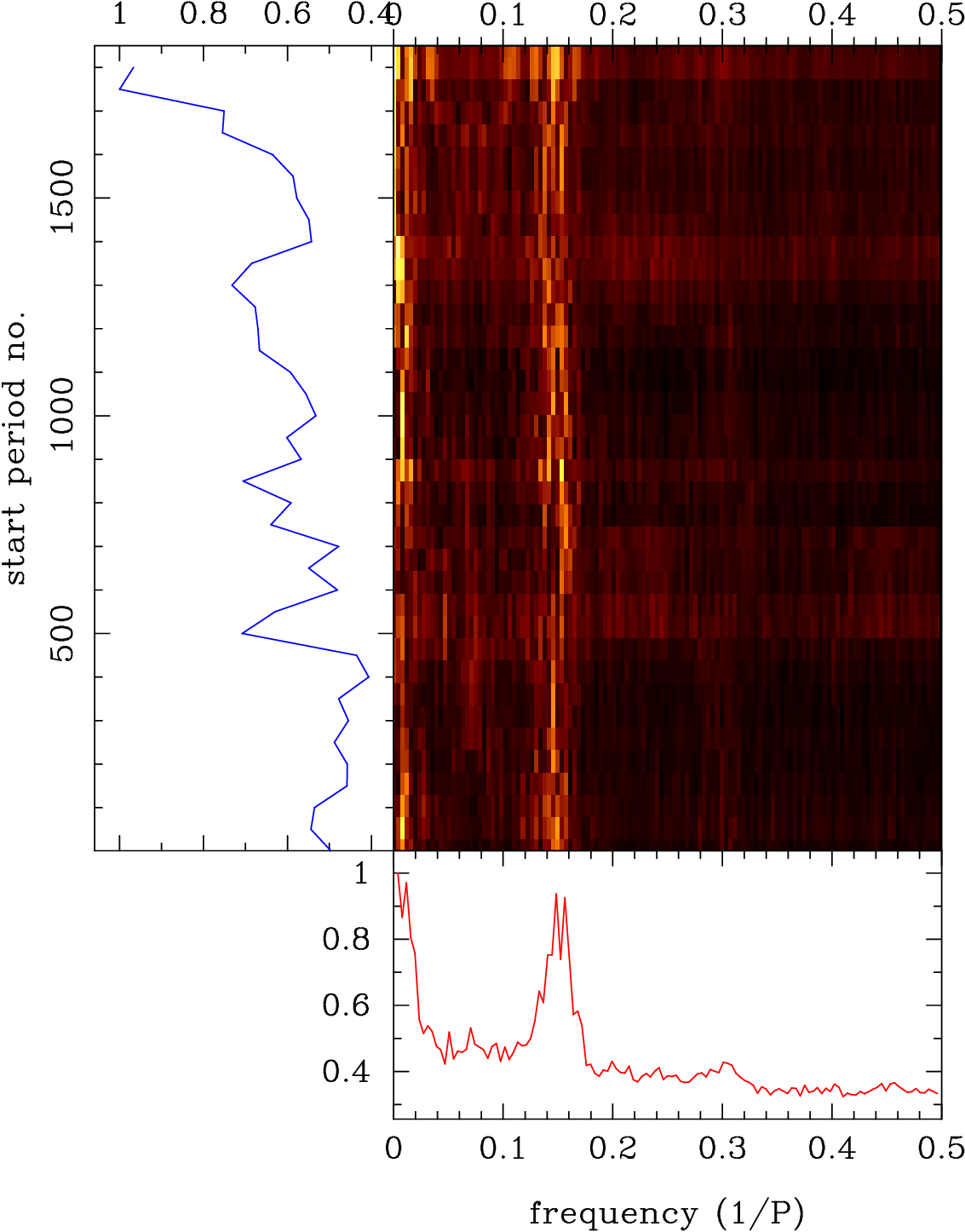}}} &
\hspace{50px}
{\mbox{\includegraphics[scale=0.4,angle=0.]{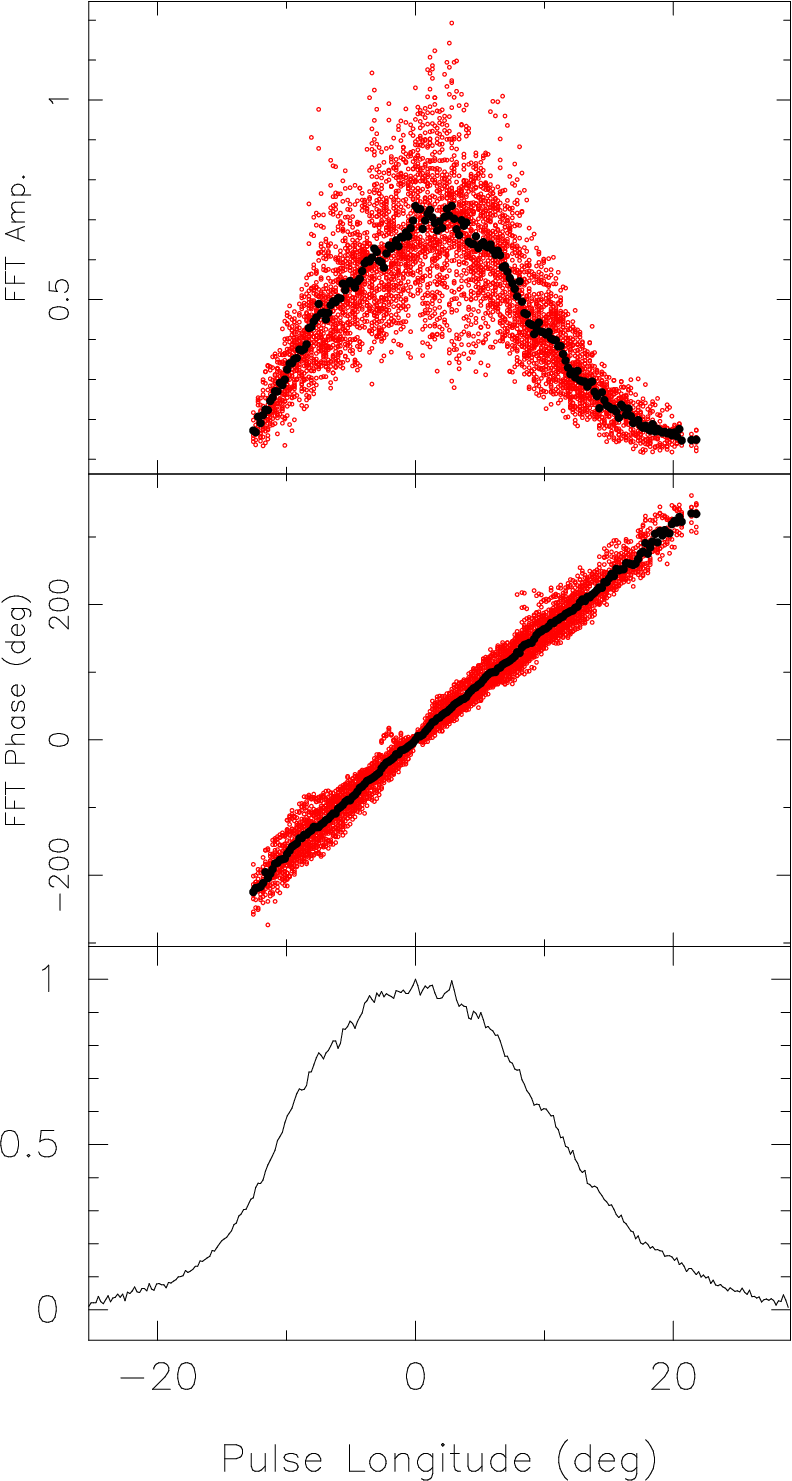}}} \\
\end{tabular}
\caption{The fluctuation spectral analysis for PSR J0034$-$0721 at 333 MHz. The
left panel shows time variation of LRFS. The right panel shows the variation of
the peak frequency across the pulse window; peak amplitude (top window), phase
variations (middle window), profile (bottom window).}
\label{fig_J0034_1}
\end{figure*}

% J0034-0721 - 618 MHz
\begin{figure*}
\begin{tabular}{@{}lr@{}}
{\mbox{\includegraphics[scale=0.4,angle=0.]{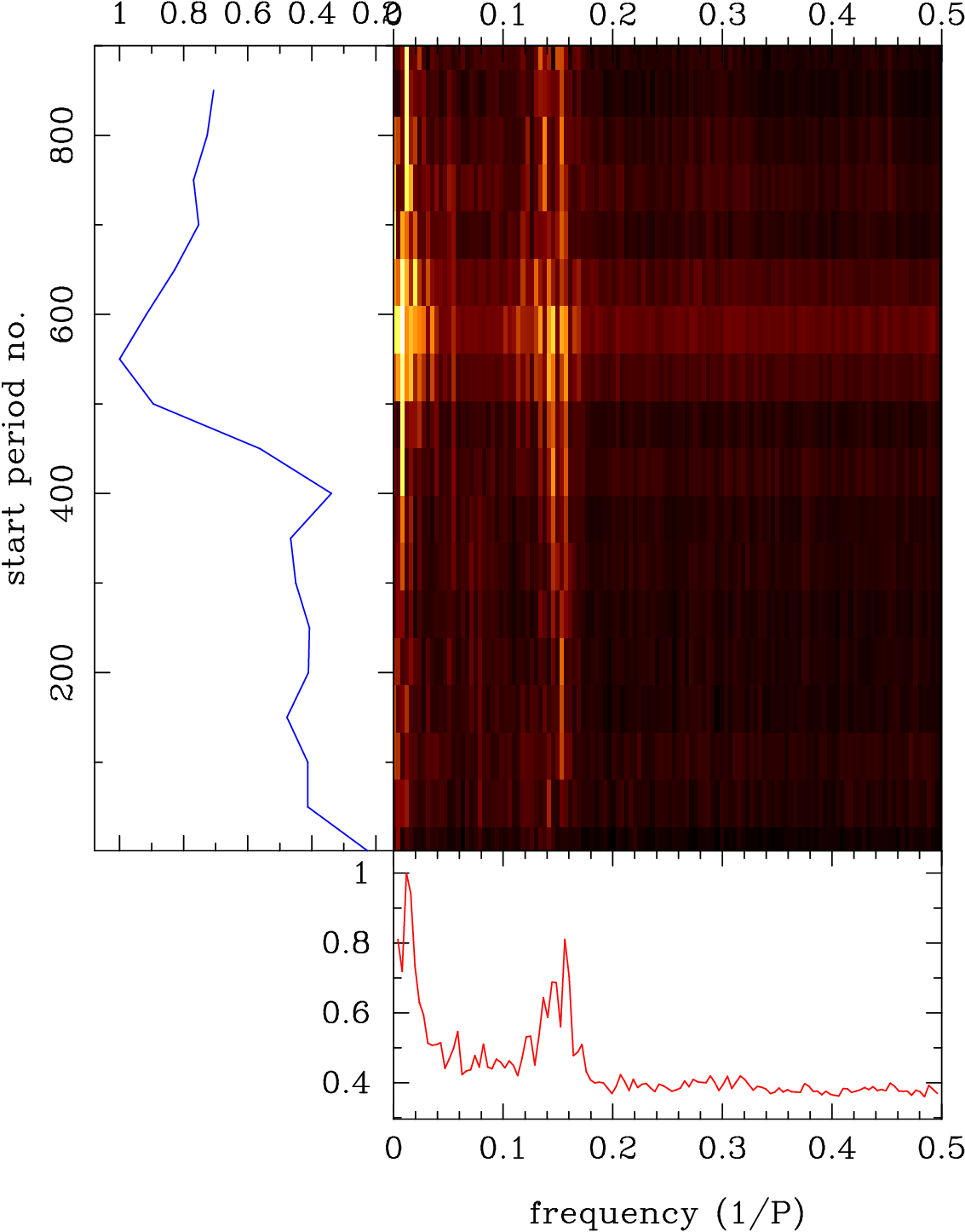}}} &
\hspace{50px}
{\mbox{\includegraphics[scale=0.4,angle=0.]{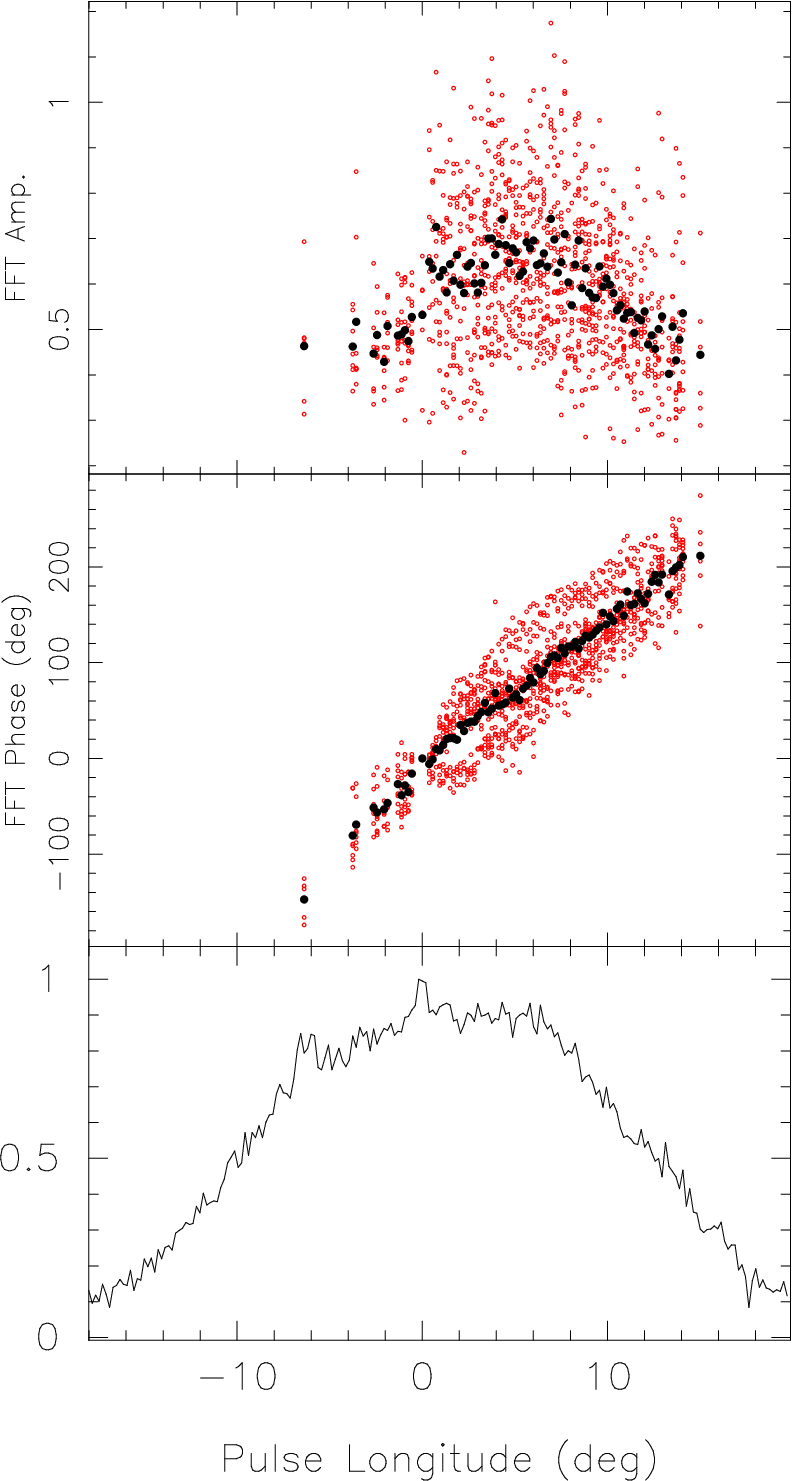}}} \\
\end{tabular}
\caption{The fluctuation spectral analysis for PSR J0034$-$0721 at 618 MHz. The
left panel shows time variation of LRFS. The right panel shows the variation of
the peak frequency across the pulse window; peak amplitude (top window), phase
variations (middle window), profile (bottom window).}
\label{fig_J0034_2}
\end{figure*}

% J0108+6608 - 339 MHz
\begin{figure*}
\begin{tabular}{@{}lr@{}}
{\mbox{\includegraphics[scale=0.4,angle=0.]{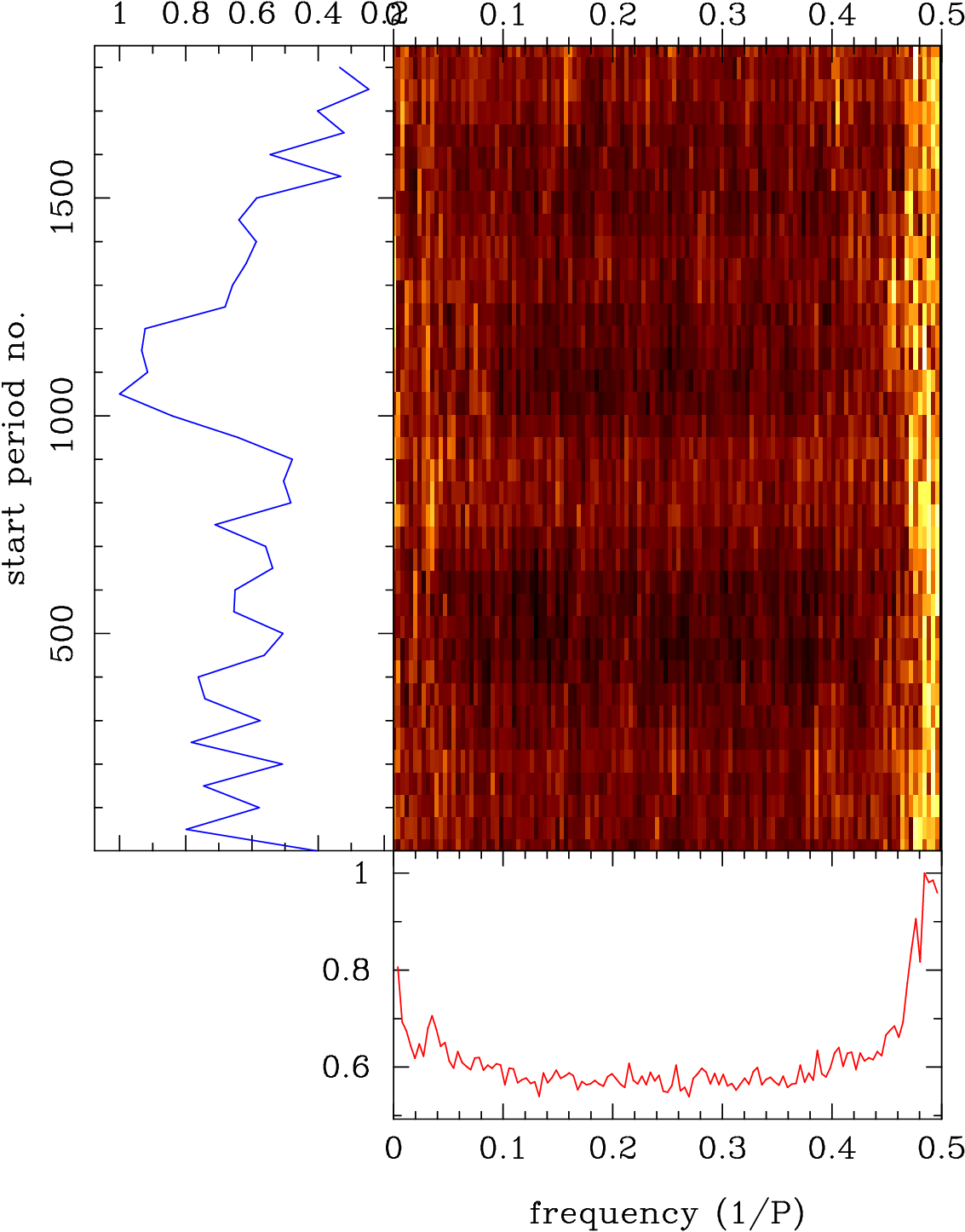}}} &
\hspace{50px}
{\mbox{\includegraphics[scale=0.4,angle=0.]{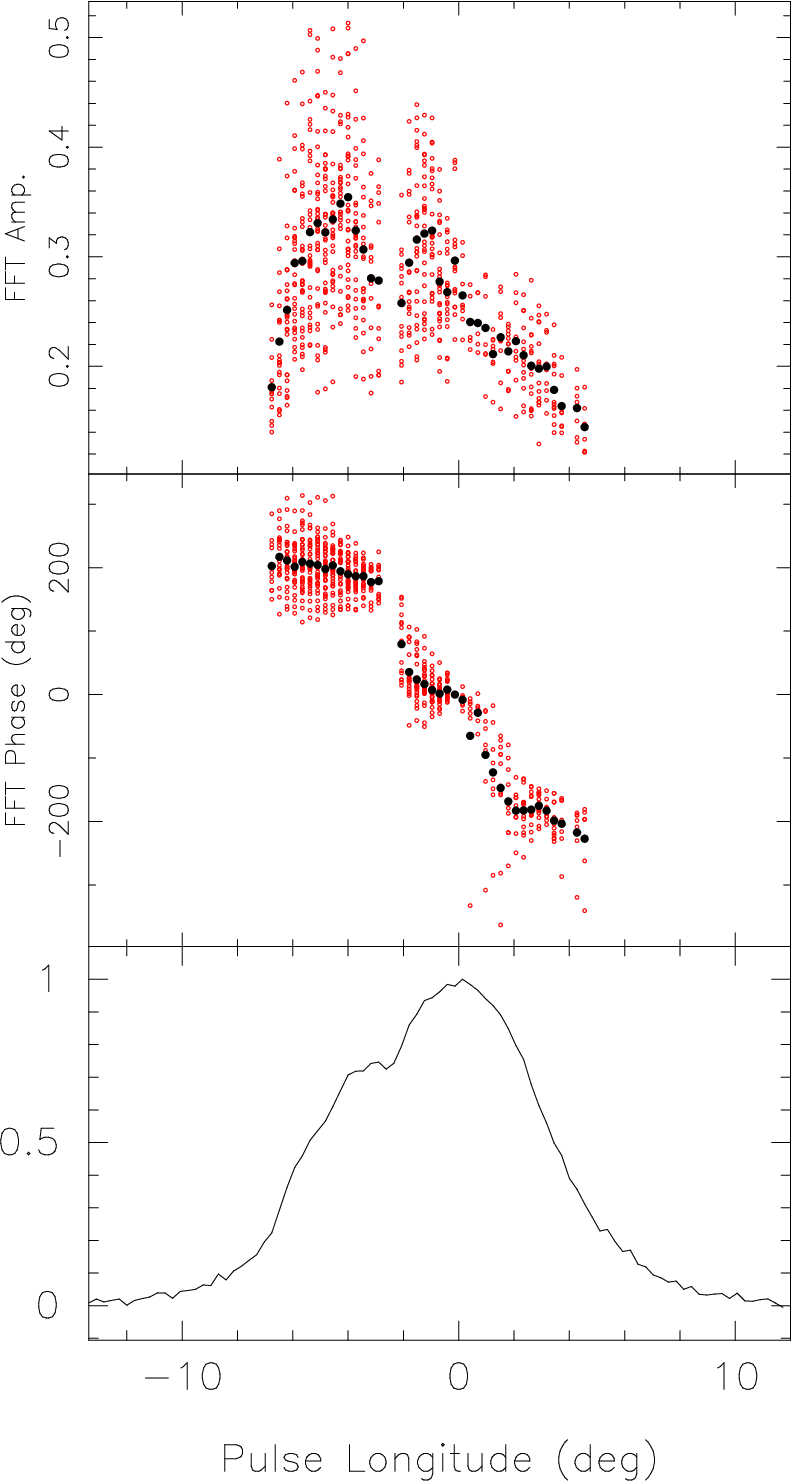}}} \\
\end{tabular}
\caption{The fluctuation spectral analysis for PSR J0108+6608 at 339 MHz. The
left panel shows time variation of LRFS. The right panel shows the variation of
the peak frequency across the pulse window; peak amplitude (top window), phase
variations (middle window), profile (bottom window).}
\label{fig_J0108}
\end{figure*}

% J0151-0635 - 333 MHz
\begin{figure*}
\begin{tabular}{@{}lr@{}}
{\mbox{\includegraphics[scale=0.4,angle=0.]{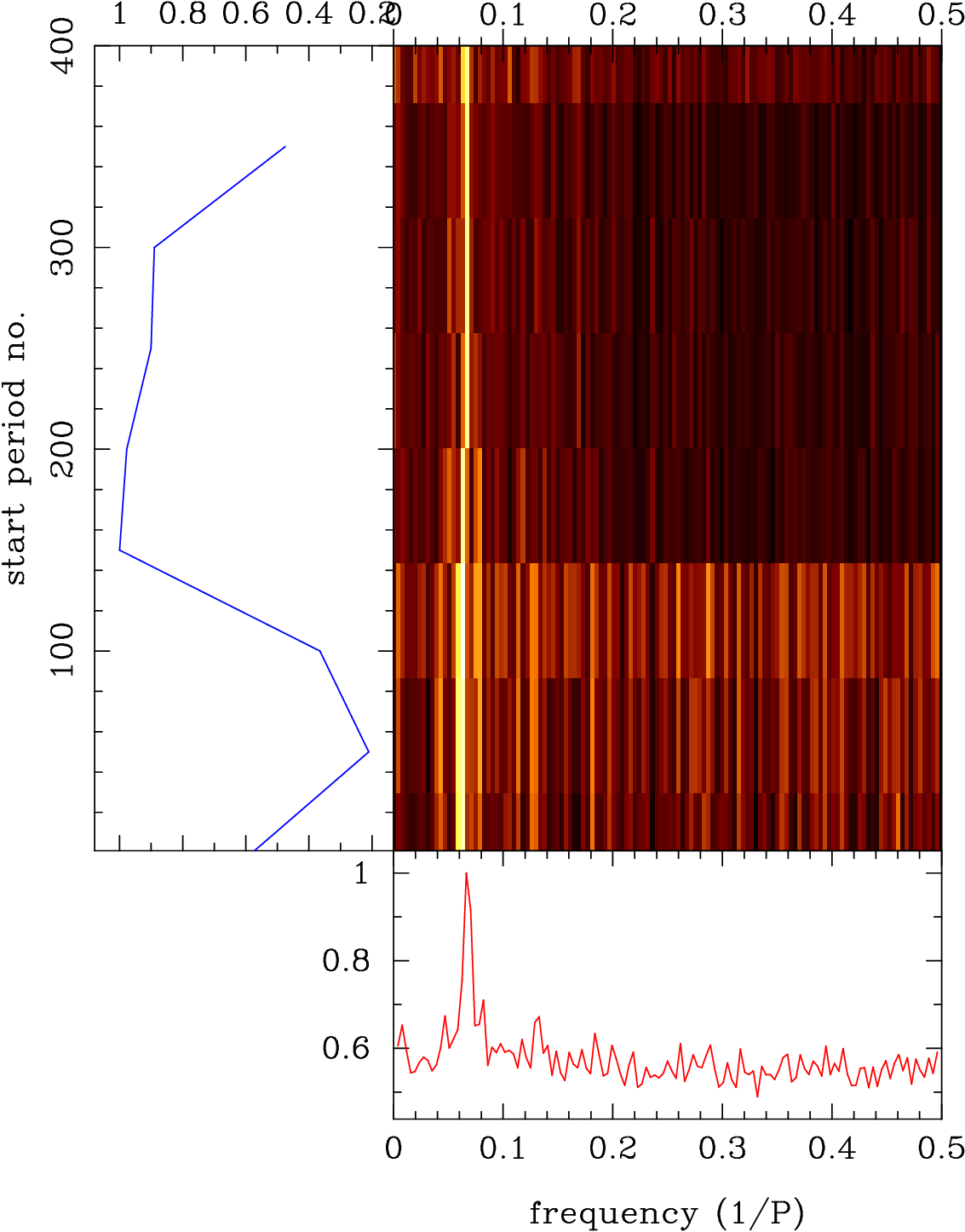}}} &
\hspace{50px}
{\mbox{\includegraphics[scale=0.4,angle=0.]{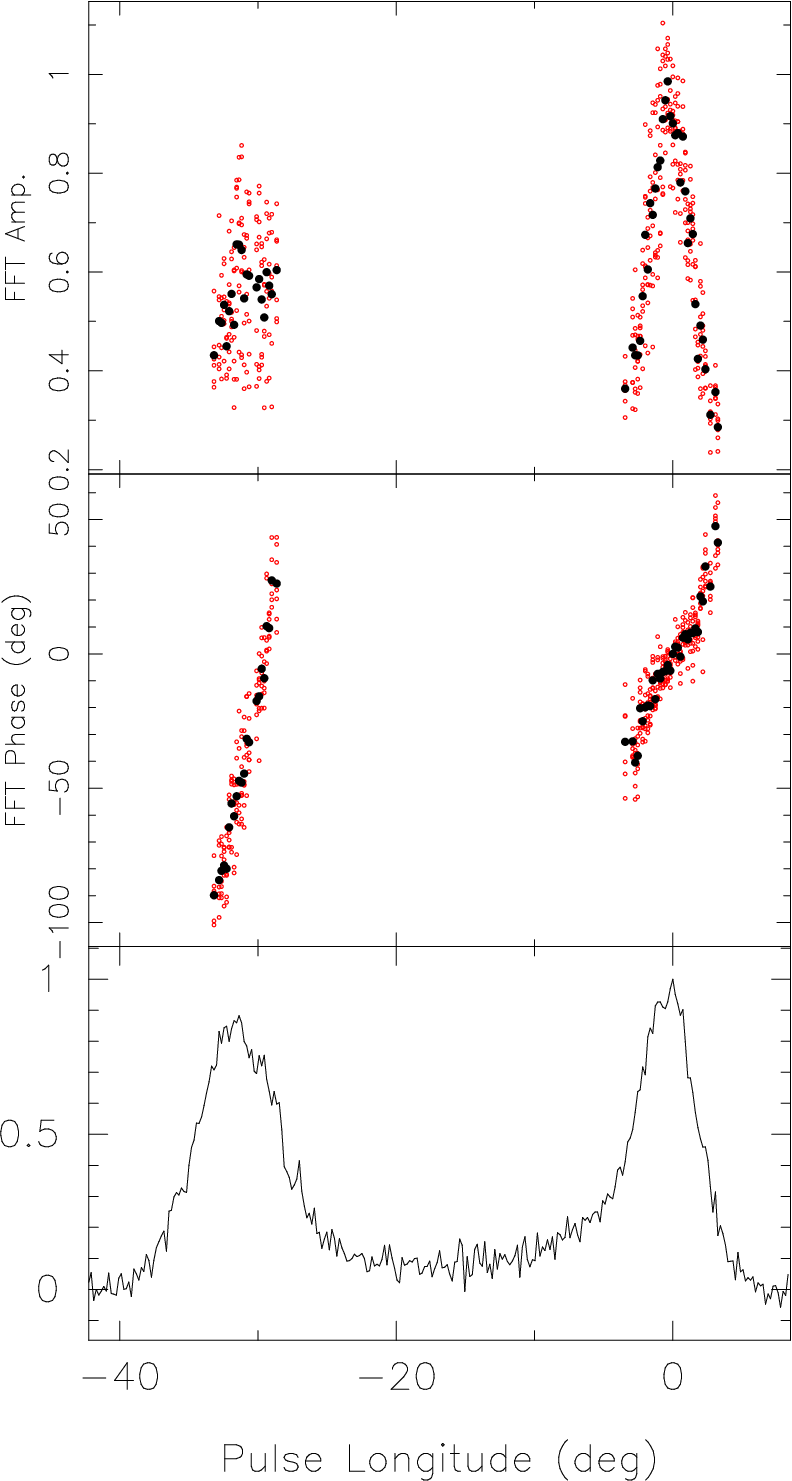}}} \\
\end{tabular}
\caption{The fluctuation spectral analysis for PSR J0151$-$0635 at 333 MHz. The
left panel shows time variation of LRFS. The right panel shows the variation of
the peak frequency across the pulse window; peak amplitude (top window), phase 
variations (middle window), profile (bottom window).}
\label{fig_J0151_1}
\end{figure*}

% J0151-0635 - 618 MHz
\begin{figure*}
\begin{tabular}{@{}lr@{}}
{\mbox{\includegraphics[scale=0.4,angle=0.]{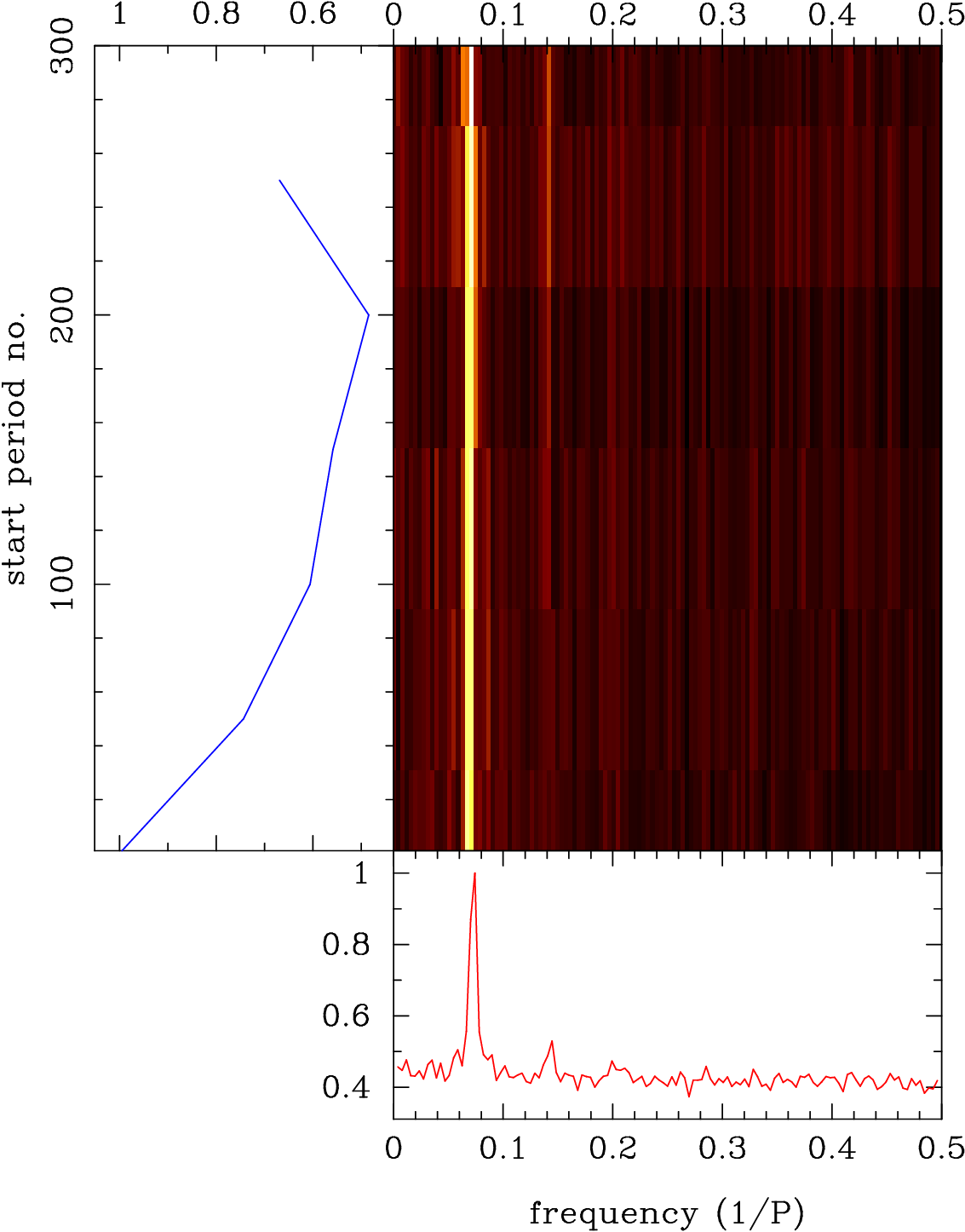}}} &
\hspace{50px} 
{\mbox{\includegraphics[scale=0.4,angle=0.]{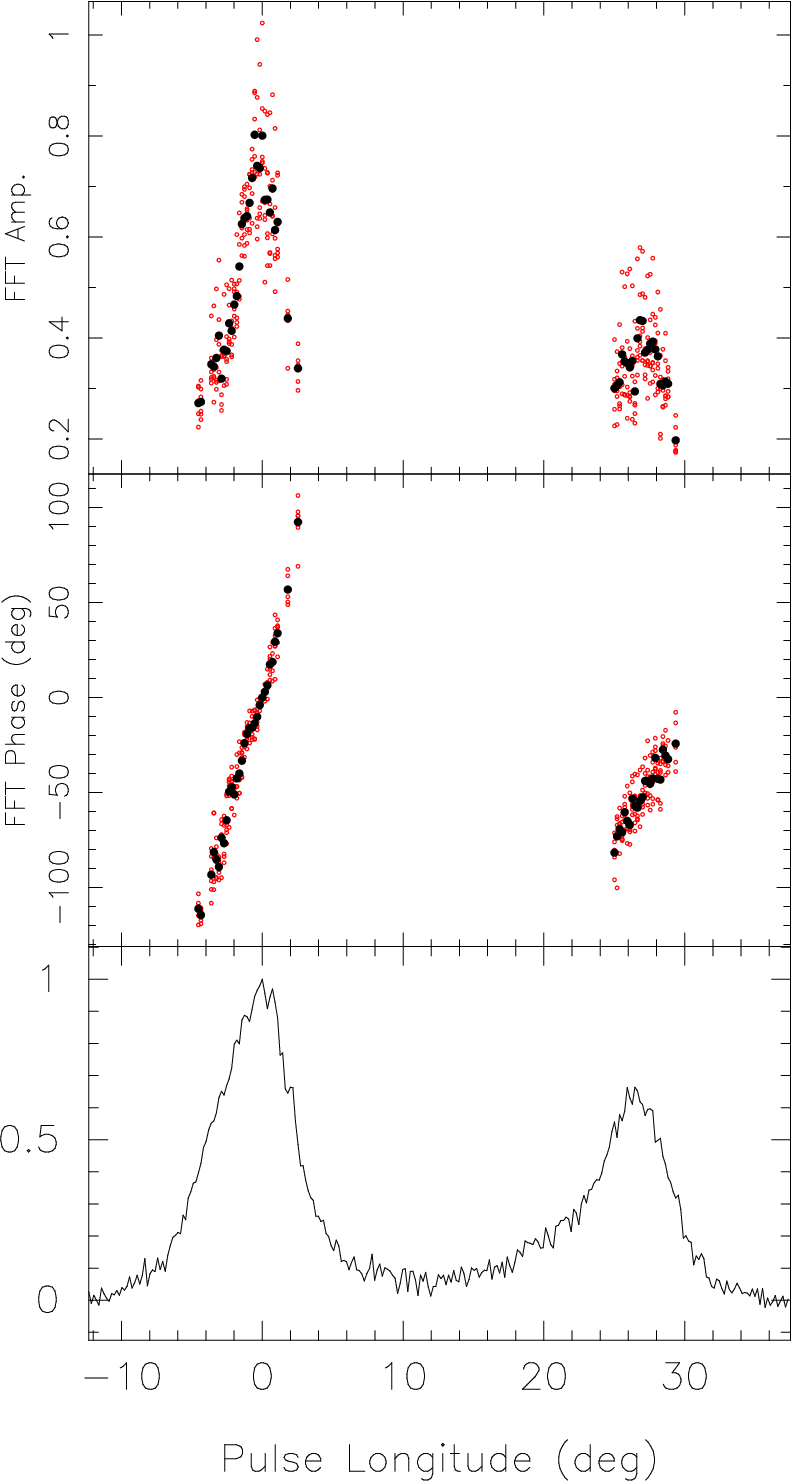}}} \\
\end{tabular}
\caption{The fluctuation spectral analysis for PSR J0151$-$0635 at 618 MHz. The
left panel shows time variation of LRFS. The right panel shows the variation of
the peak frequency across the pulse window; peak amplitude (top window), phase 
variations (middle window), profile (bottom window).}
\label{fig_J0151_2}
\end{figure*}

% J0421-0345 - 339 MHz
\begin{figure*}
\begin{tabular}{@{}lr@{}}
{\mbox{\includegraphics[scale=0.4,angle=0.]{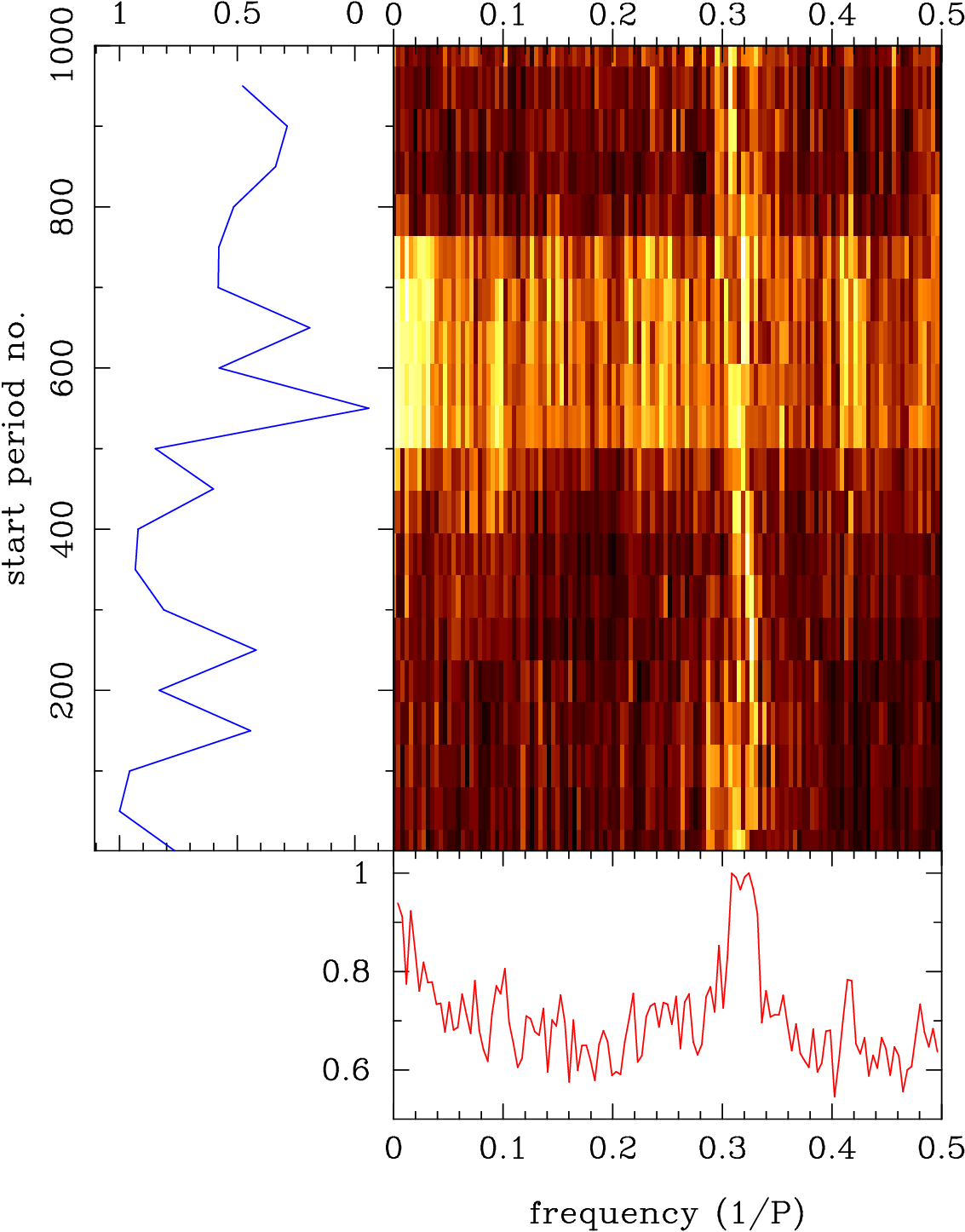}}} &
\hspace{50px}
{\mbox{\includegraphics[scale=0.4,angle=0.]{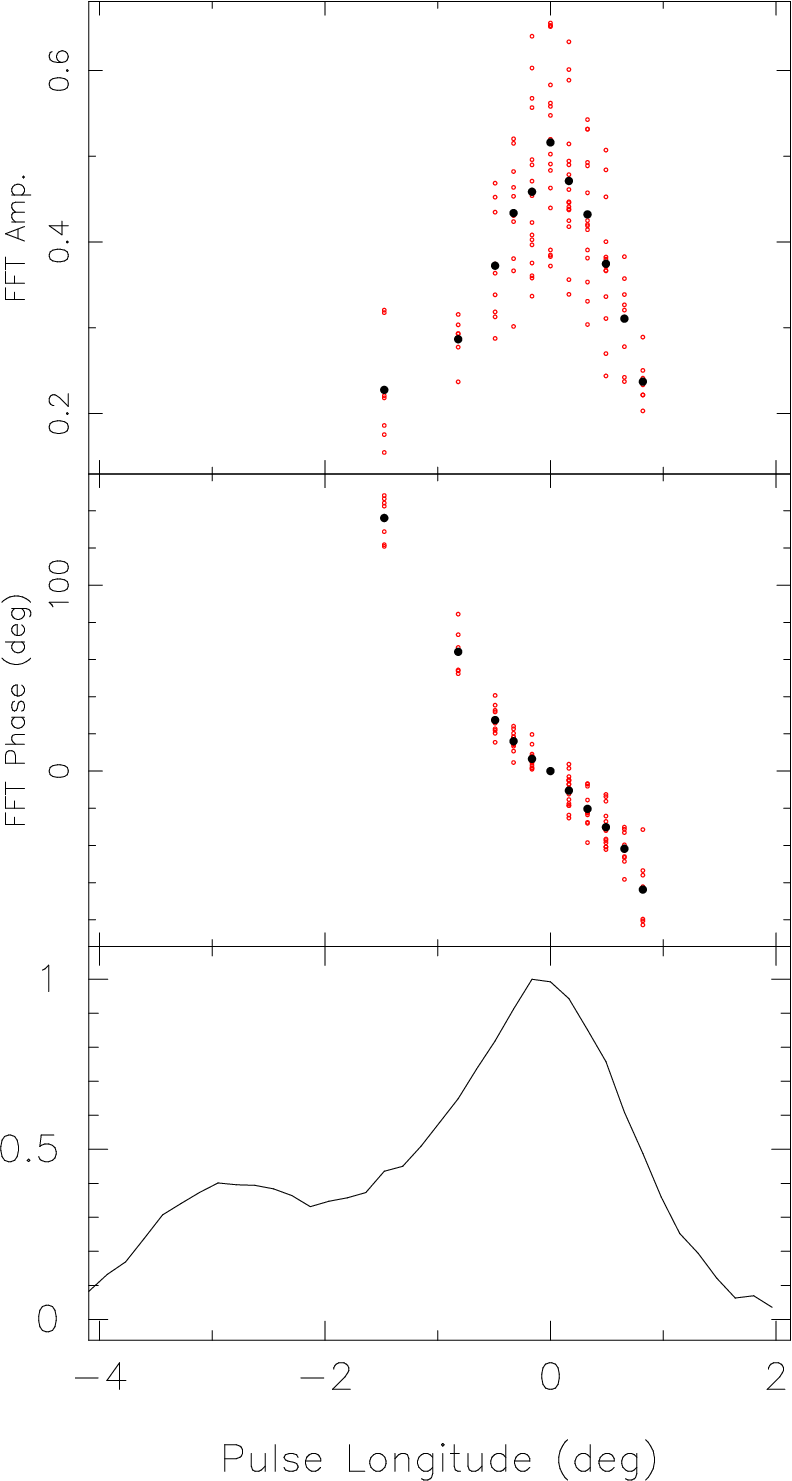}}} \\
\end{tabular}
\caption{The fluctuation spectral analysis for PSR J0421$-$0345 at 339 MHz. The 
left panel shows time variation of LRFS. The right panel shows the variation of
the peak frequency across the pulse window; peak amplitude (top window), phase 
variations (middle window), profile (bottom window).}
\label{fig_J0421}
\end{figure*}

% J0814+7429 - 325 MHz
\begin{figure*}
\begin{tabular}{@{}lr@{}}
{\mbox{\includegraphics[scale=0.4,angle=0.]{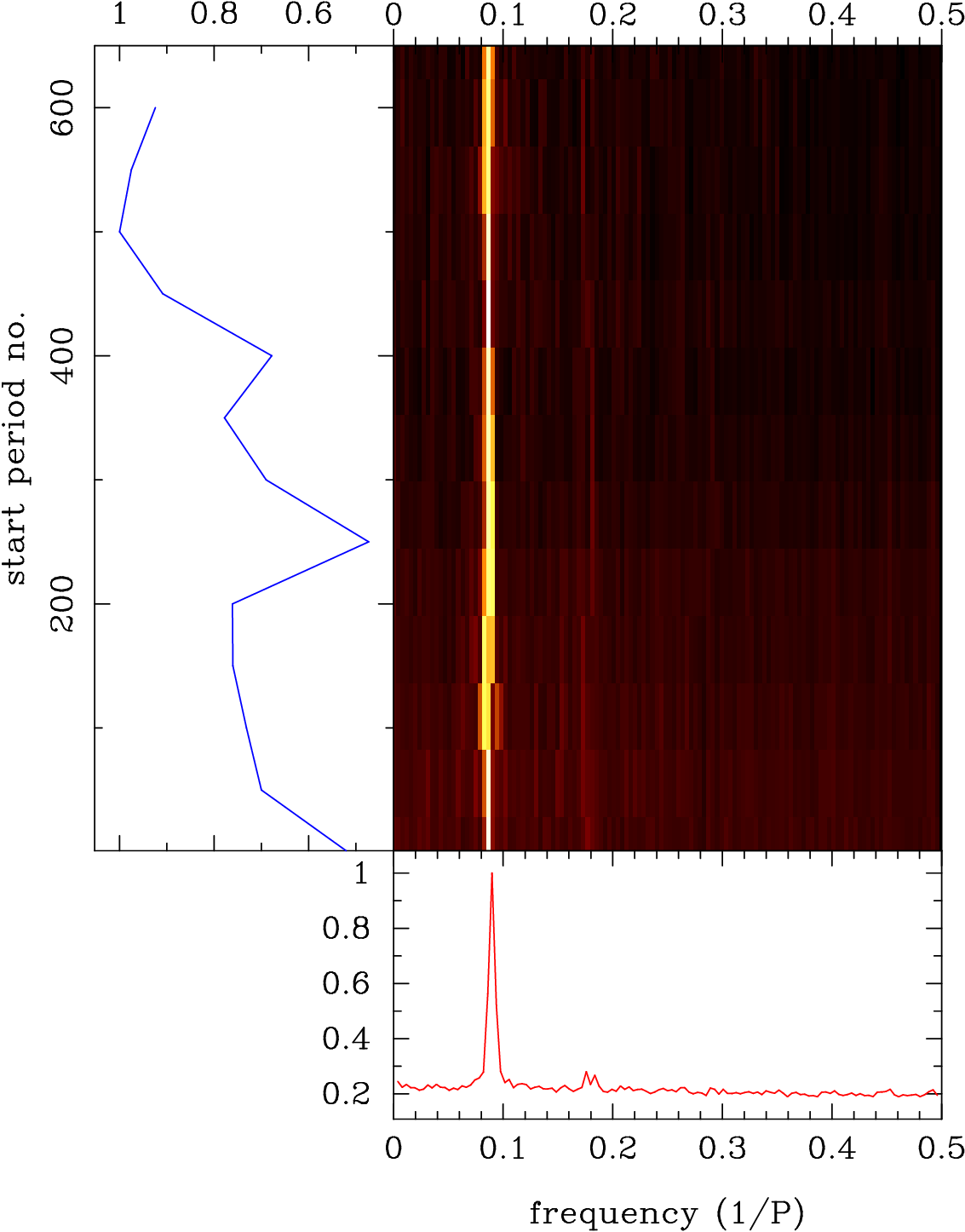}}} &
\hspace{50px}
{\mbox{\includegraphics[scale=0.4,angle=0.]{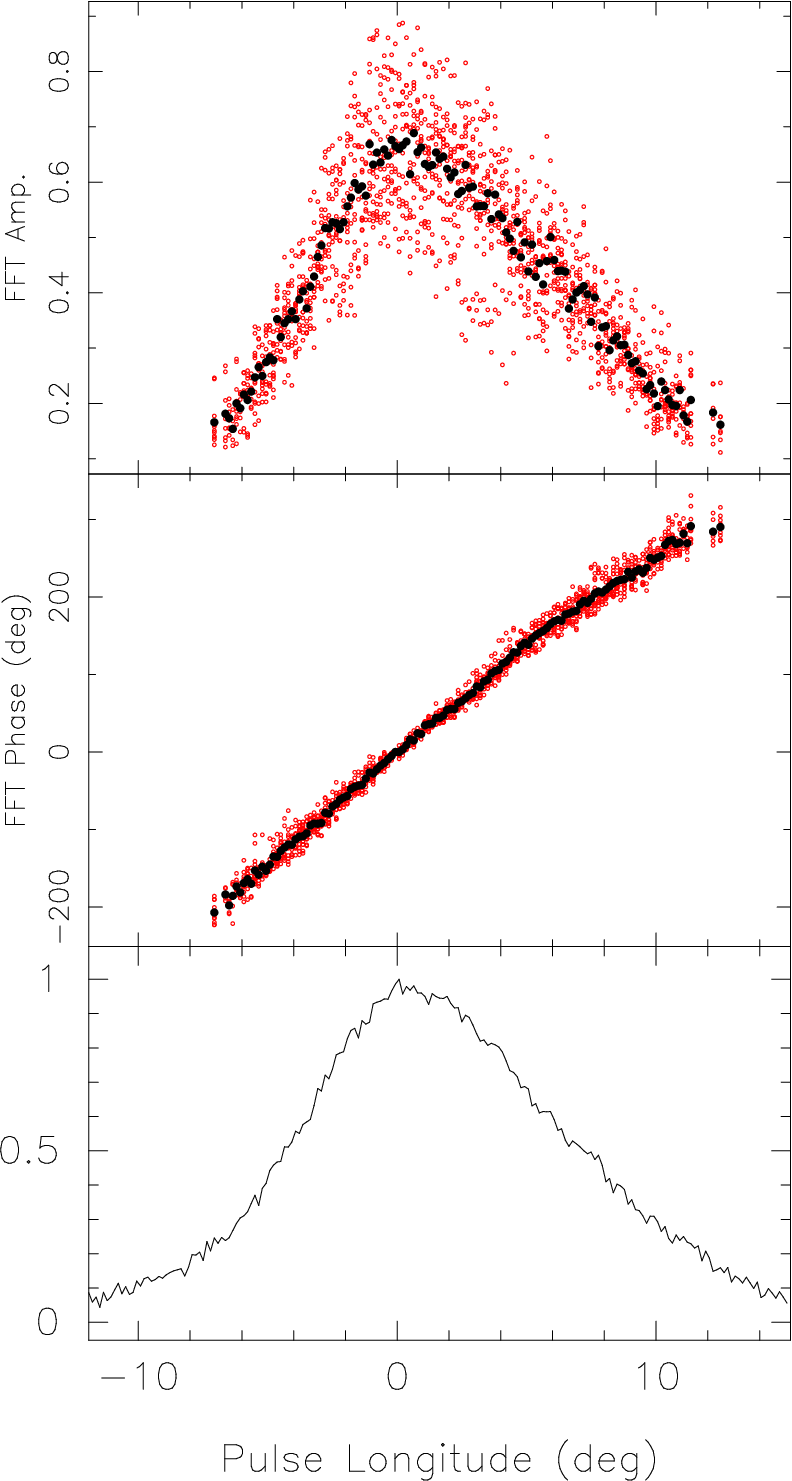}}} \\
\end{tabular}
\caption{The fluctuation spectral analysis for PSR J0814+7429 at 325 MHz. The 
left panel shows time variation of LRFS. The right panel shows the variation of
the peak frequency across the pulse window; peak amplitude (top window), phase 
variations (middle window), profile (bottom window).}
\label{fig_J0814}
\end{figure*}

% J0934-5249 - 339 MHz
\begin{figure*}
\begin{tabular}{@{}lr@{}}
{\mbox{\includegraphics[scale=0.4,angle=0.]{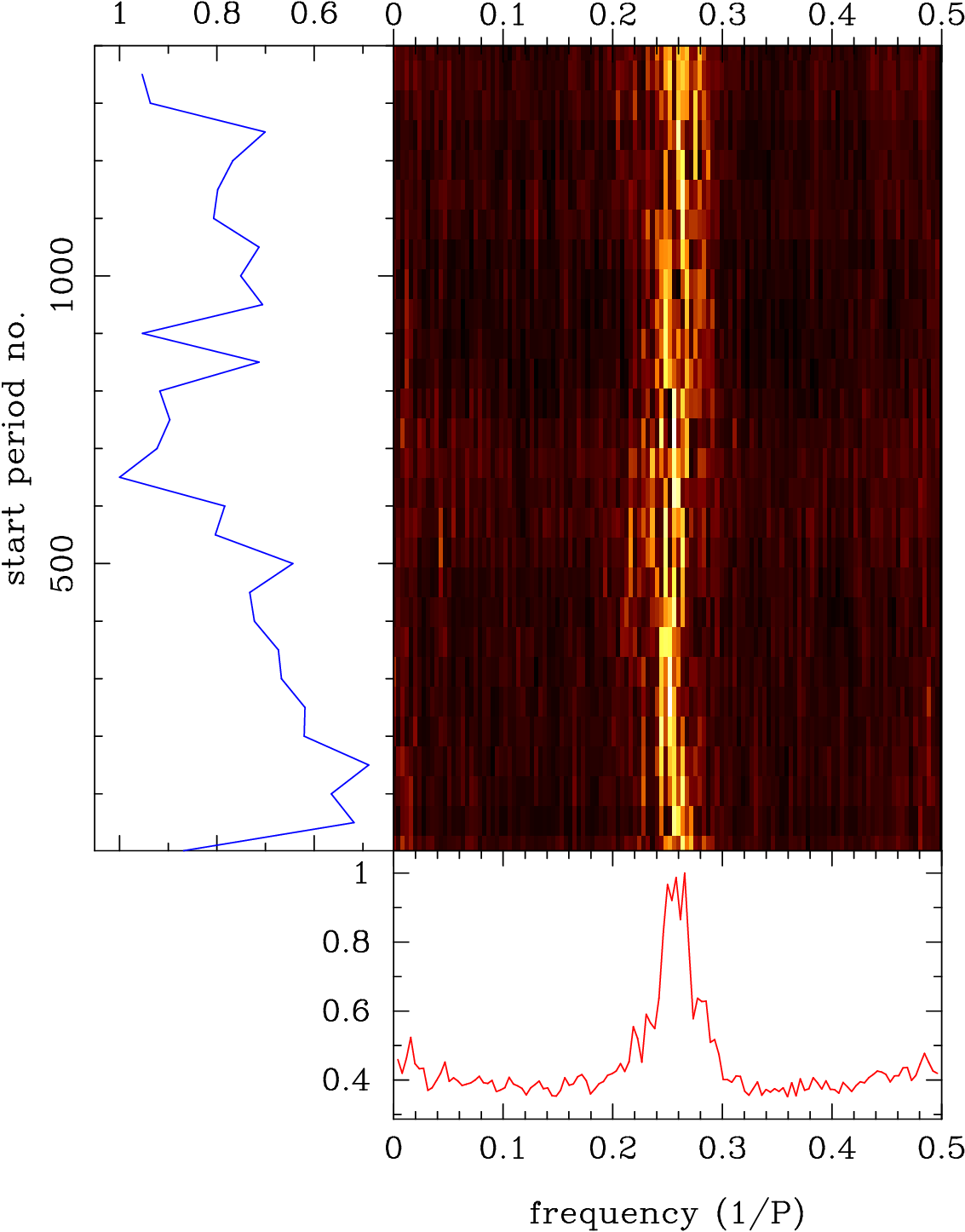}}} &
\hspace{50px}
{\mbox{\includegraphics[scale=0.4,angle=0.]{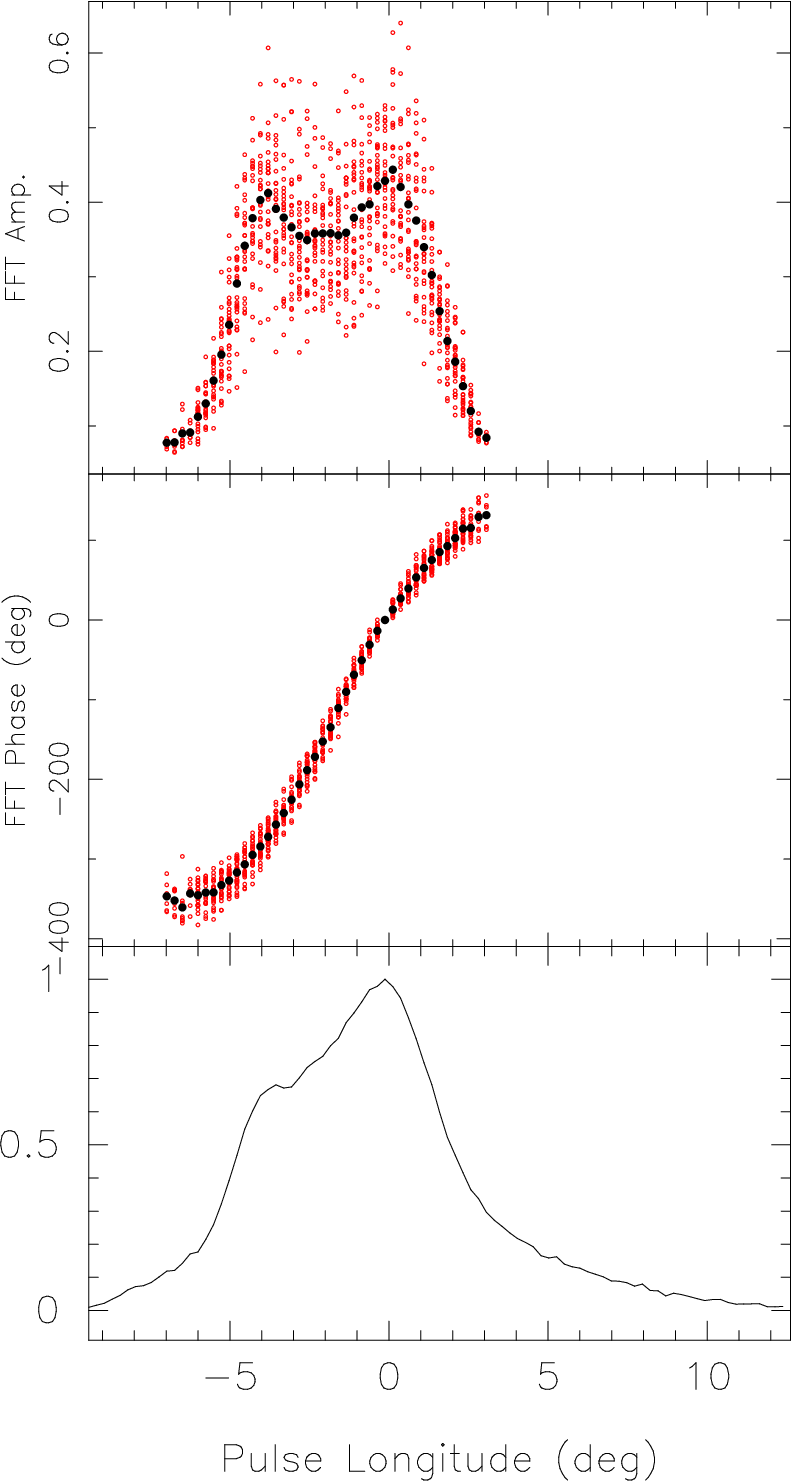}}} \\
\end{tabular}
\caption{The fluctuation spectral analysis for PSR J0934$-$5249 at 339 MHz. The 
left panel shows time variation of LRFS. The right panel shows the variation of
the peak frequency across the pulse window; peak amplitude (top window), phase 
variations (middle window), profile (bottom window).}
\label{fig_J0934}
\end{figure*}

% J1418-3921 - 618 MHz
\begin{figure*}
\begin{tabular}{@{}lr@{}}
{\mbox{\includegraphics[scale=0.4,angle=0.]{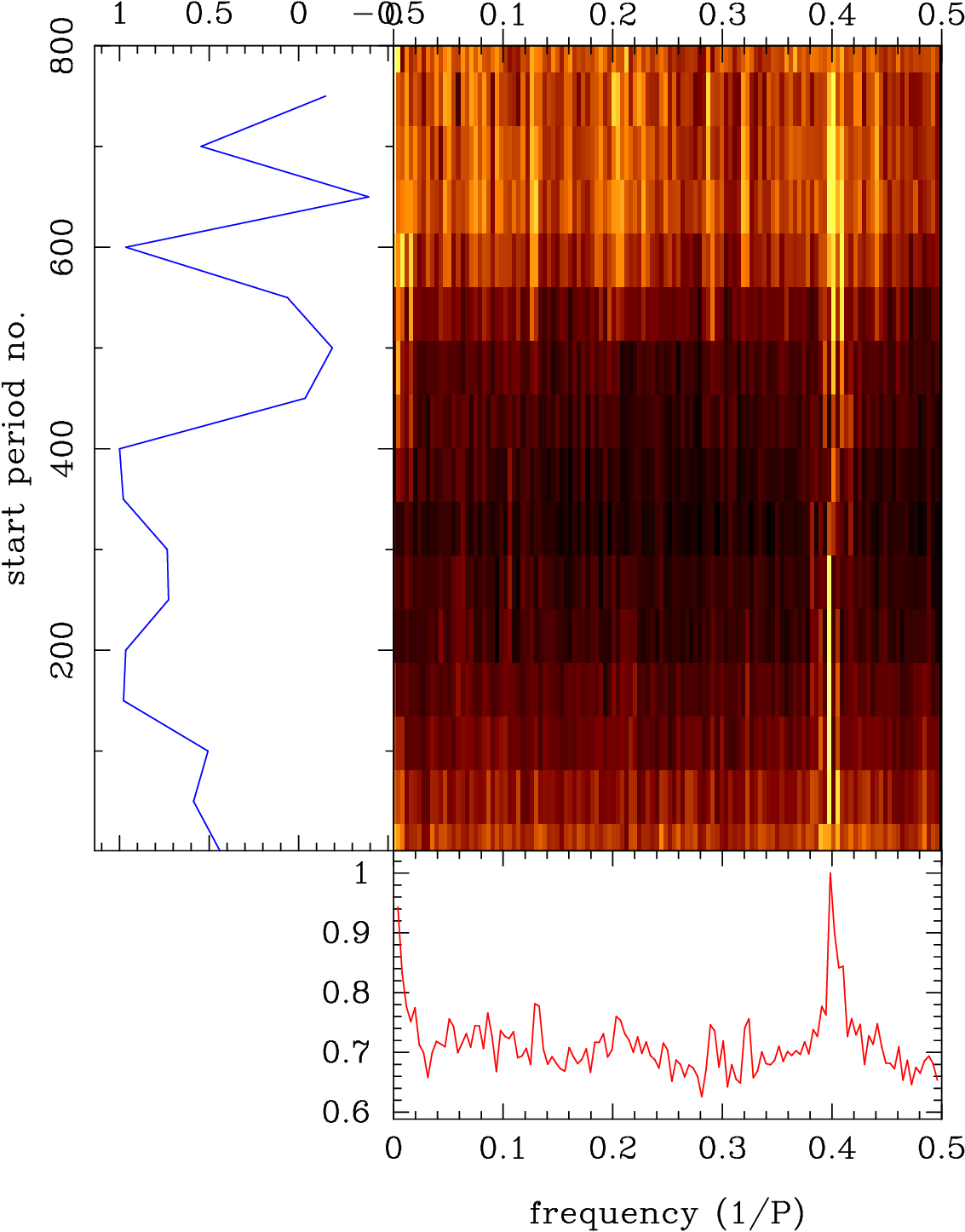}}} &
\hspace{50px}
{\mbox{\includegraphics[scale=0.4,angle=0.]{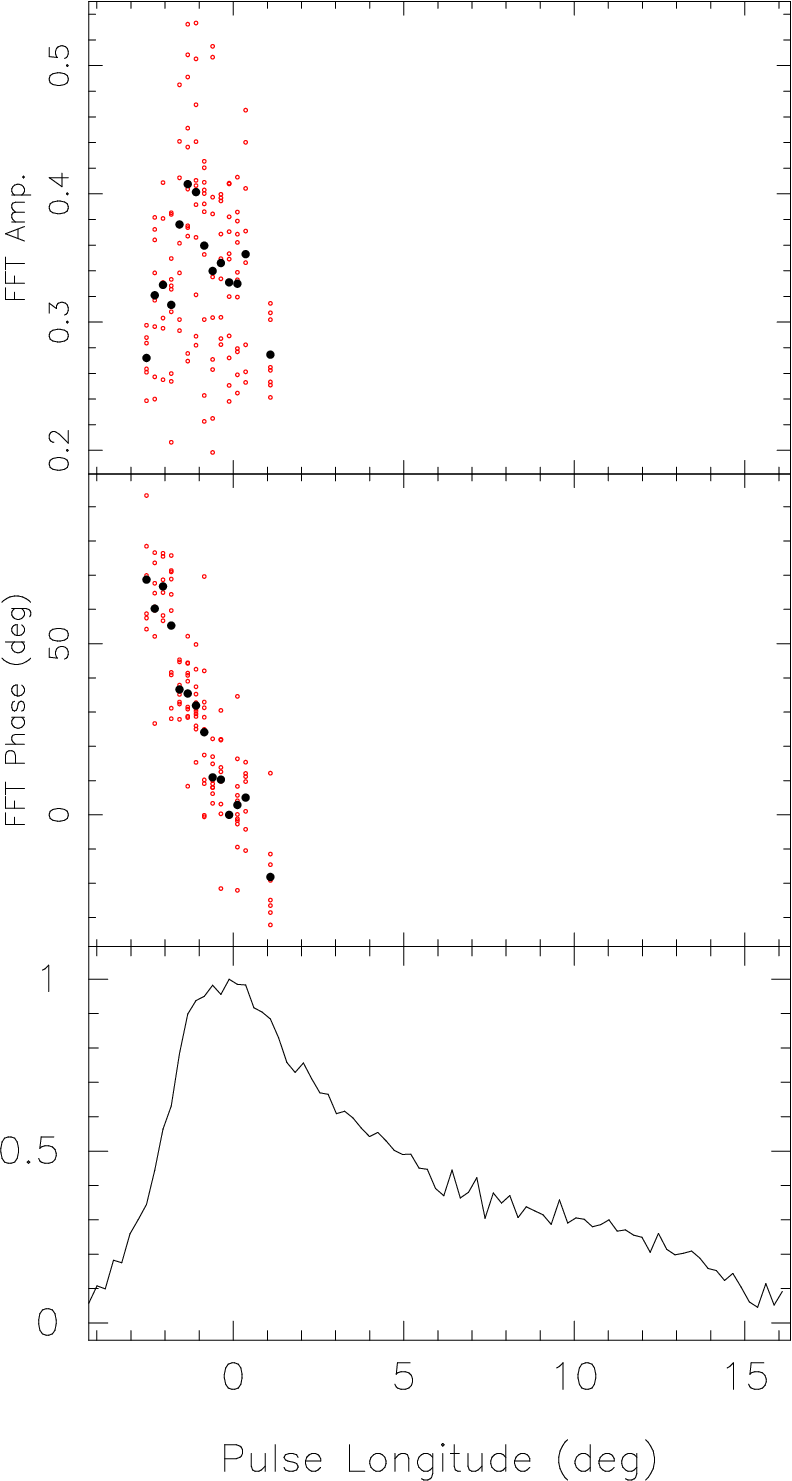}}} \\
\end{tabular}
\caption{The fluctuation spectral analysis for PSR J1418$-$3921 at 618 MHz. The 
left panel shows time variation of LRFS. The right panel shows the variation of
the peak frequency across the pulse window; peak amplitude (top window), phase 
variations (middle window), profile (bottom window).}
\label{fig_J1418}
\end{figure*}

% J1543-0620 - 339 MHz
\begin{figure*}
\begin{tabular}{@{}lr@{}}
{\mbox{\includegraphics[scale=0.4,angle=0.]{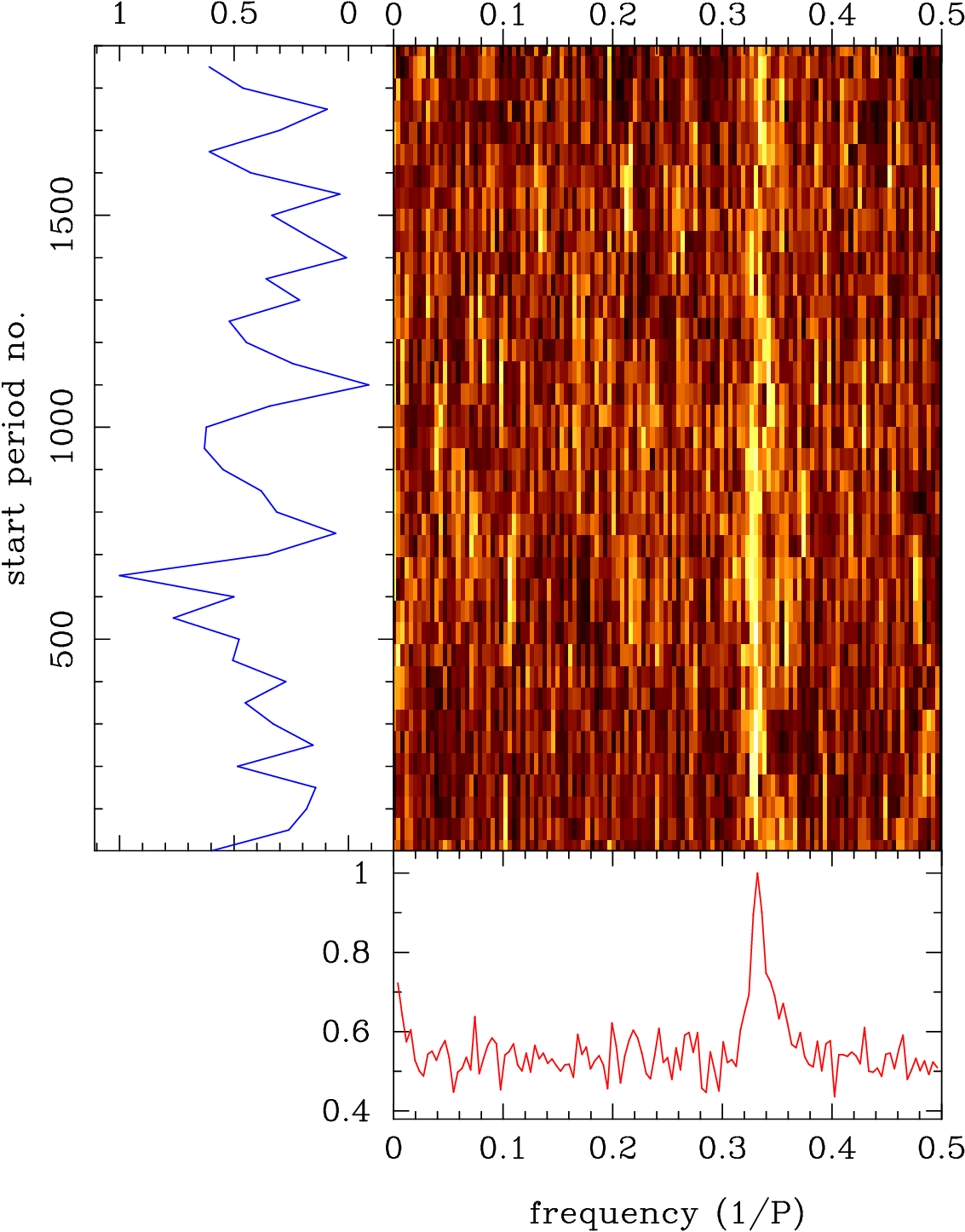}}} &
\hspace{50px}
{\mbox{\includegraphics[scale=0.4,angle=0.]{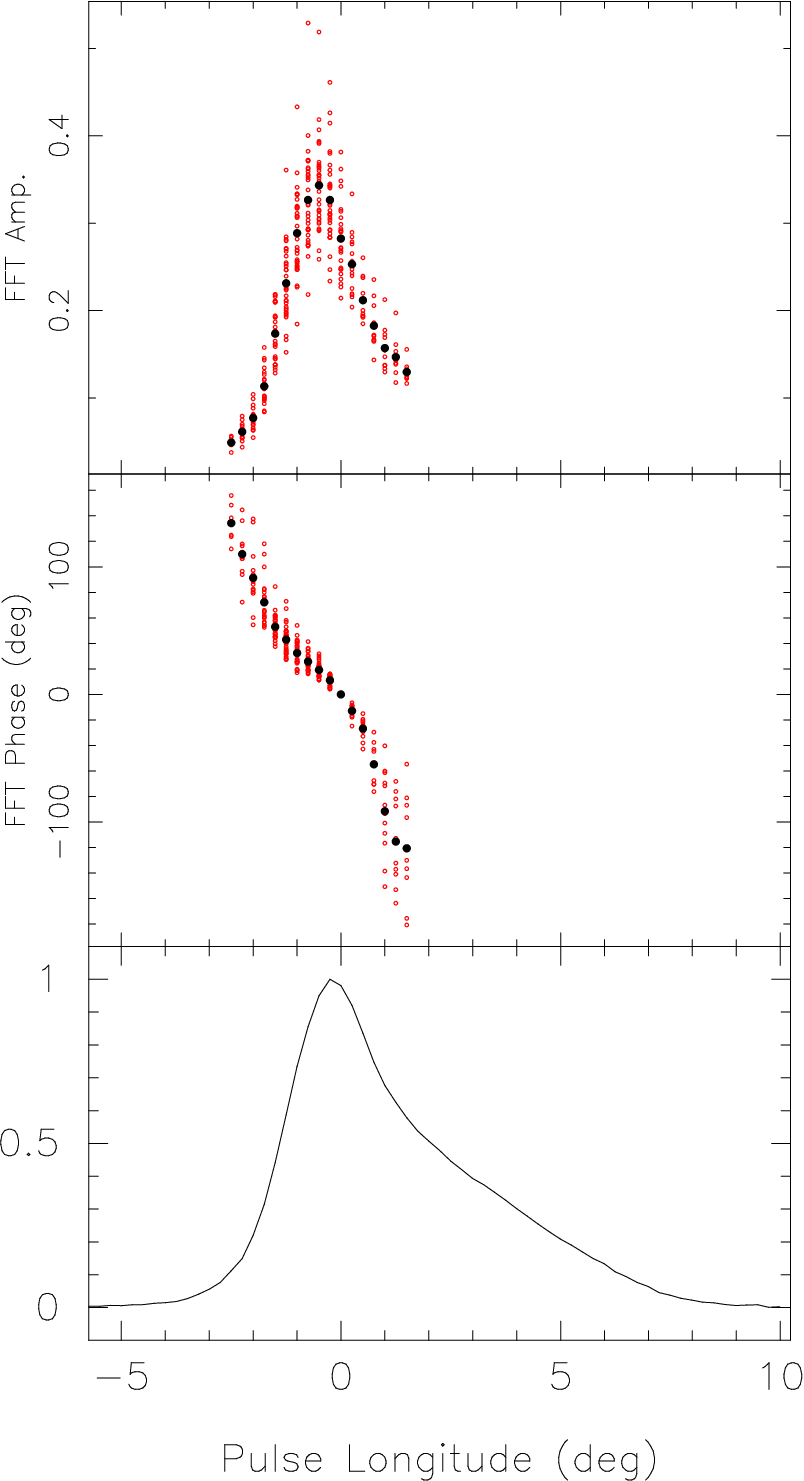}}} \\
\end{tabular}
\caption{The fluctuation spectral analysis for PSR J1543$-$0620 at 339 MHz. The
left panel shows time variation of LRFS. The right panel shows the variation of
the peak frequency across the pulse window; peak amplitude (top window), phase 
variations (middle window), profile (bottom window).}
\label{fig_J1543}
\end{figure*}

% J1901-0906 - 333 MHz
\begin{figure*}
\begin{tabular}{@{}lr@{}}
{\mbox{\includegraphics[scale=0.4,angle=0.]{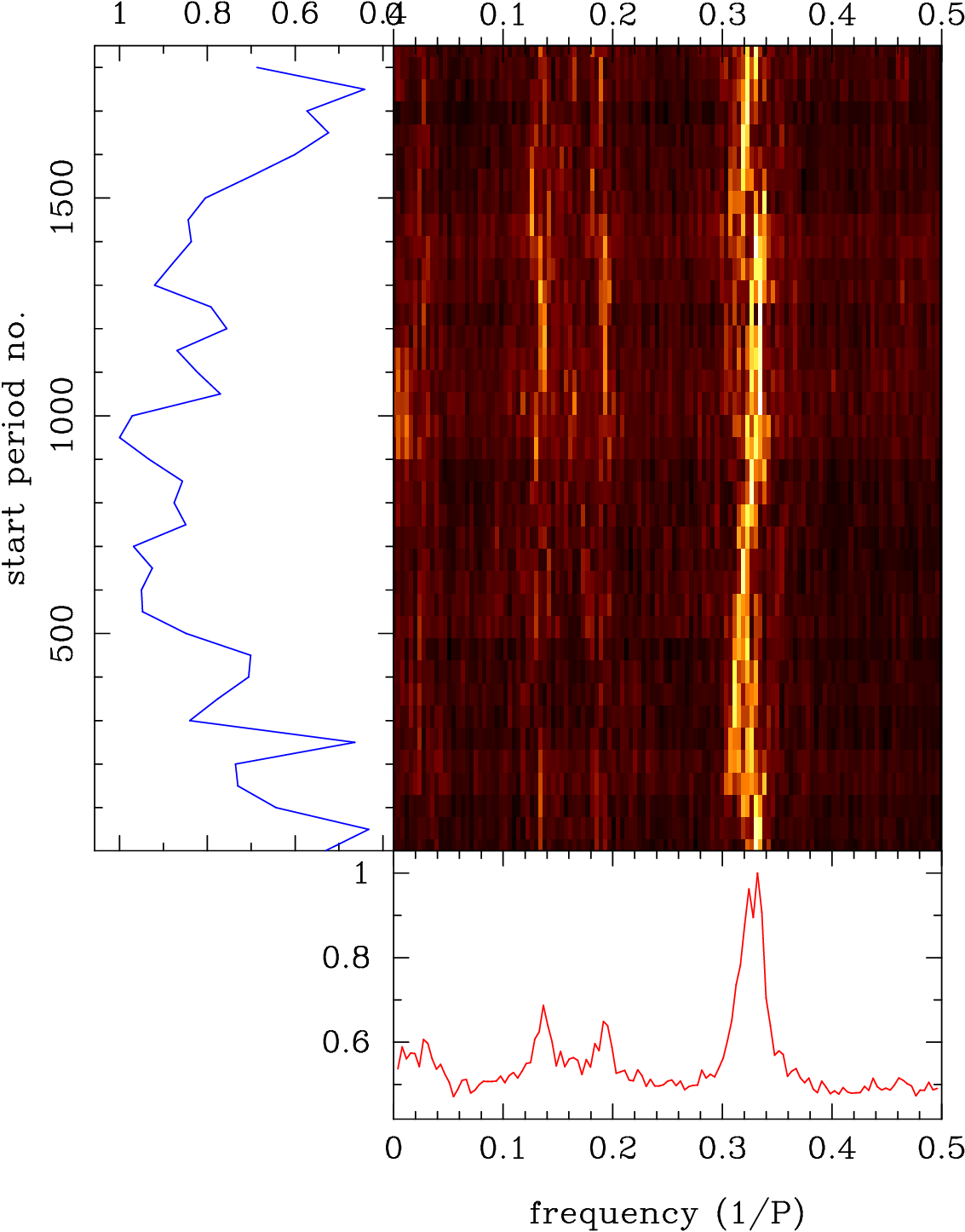}}} &
\hspace{50px}
{\mbox{\includegraphics[scale=0.4,angle=0.]{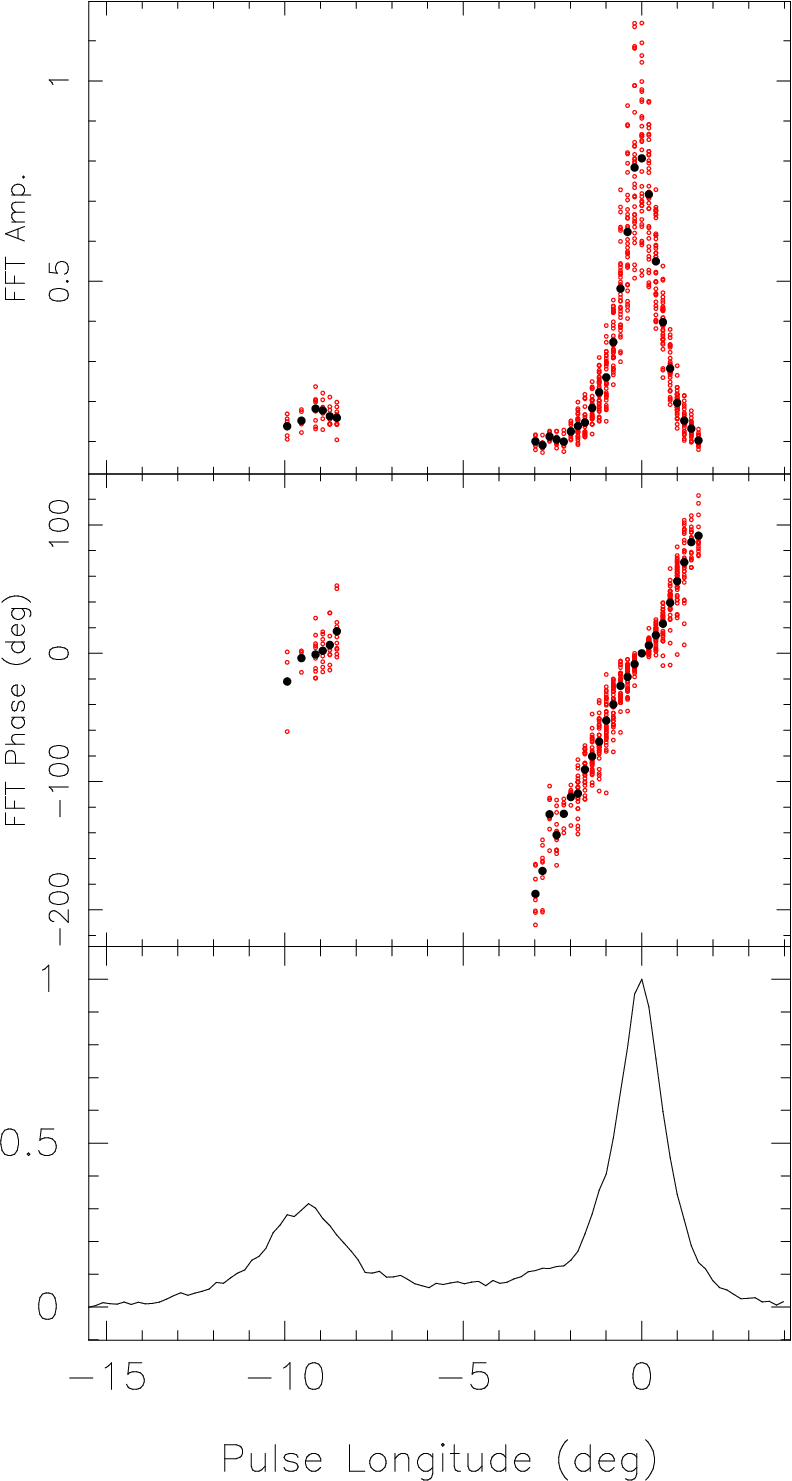}}} \\
\end{tabular}
\caption{The fluctuation spectral analysis for PSR J1901$-$0906 at 333 MHz. The
left panel shows time variation of LRFS. The right panel shows the variation of
the peak frequency across the pulse window; peak amplitude (top window), phase 
variations (middle window), profile (bottom window).}
\label{fig_J1901_1}
\end{figure*}

% J1901-0906 - 618 MHz
\begin{figure*}
\begin{tabular}{@{}lr@{}}
{\mbox{\includegraphics[scale=0.4,angle=0.]{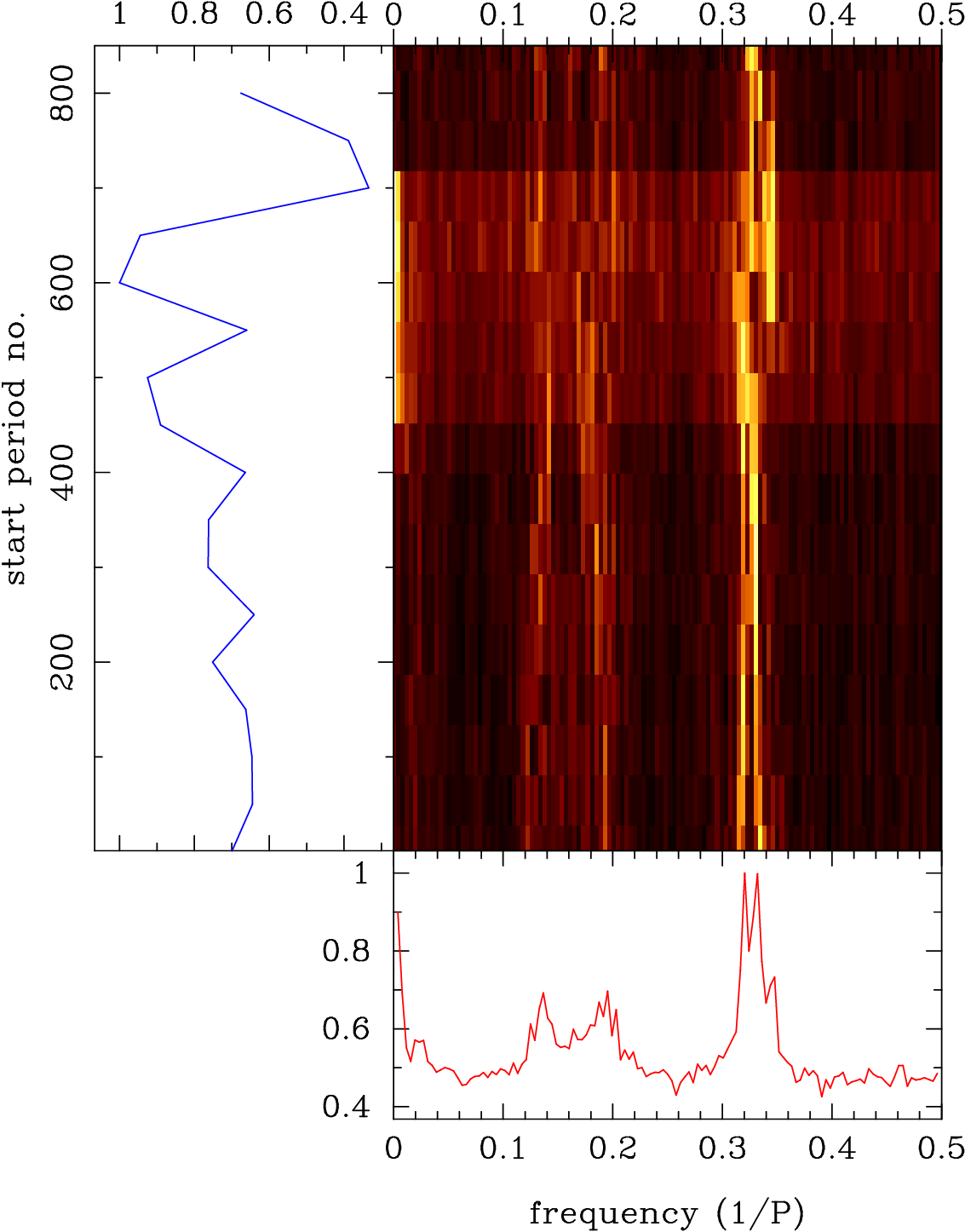}}} &
\hspace{50px}
{\mbox{\includegraphics[scale=0.4,angle=0.]{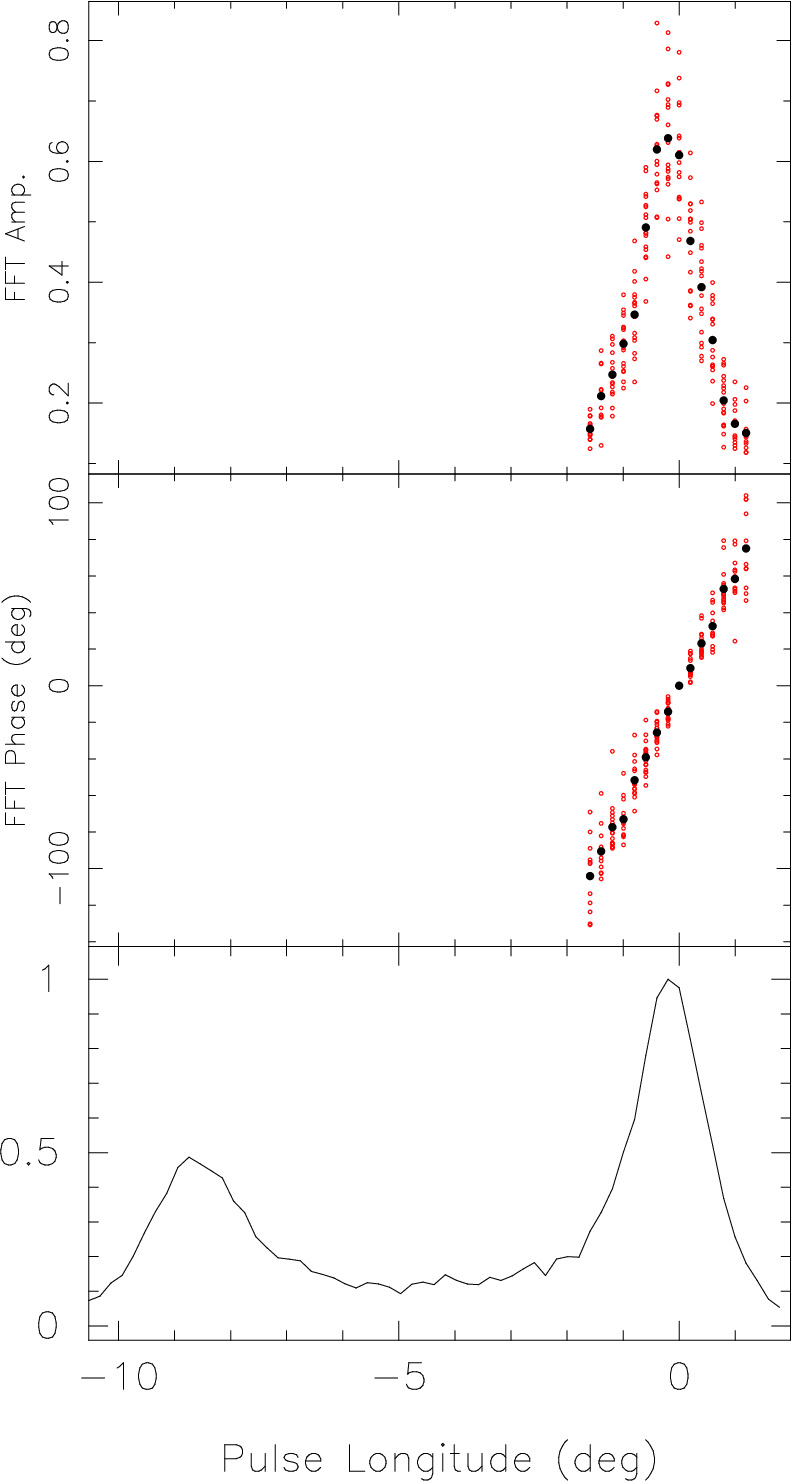}}} \\
\end{tabular}
\caption{The fluctuation spectral analysis for PSR J1901$-$0906 at 618 MHz. The
left panel shows time variation of LRFS. The right panel shows the variation of
the peak frequency across the pulse window; peak amplitude (top window), phase 
variations (middle window), profile (bottom window).}
\label{fig_J1901_2}
\end{figure*}

% J1921+1948 - 618 MHz
\begin{figure*}
\begin{tabular}{@{}lr@{}}
{\mbox{\includegraphics[scale=0.4,angle=0.]{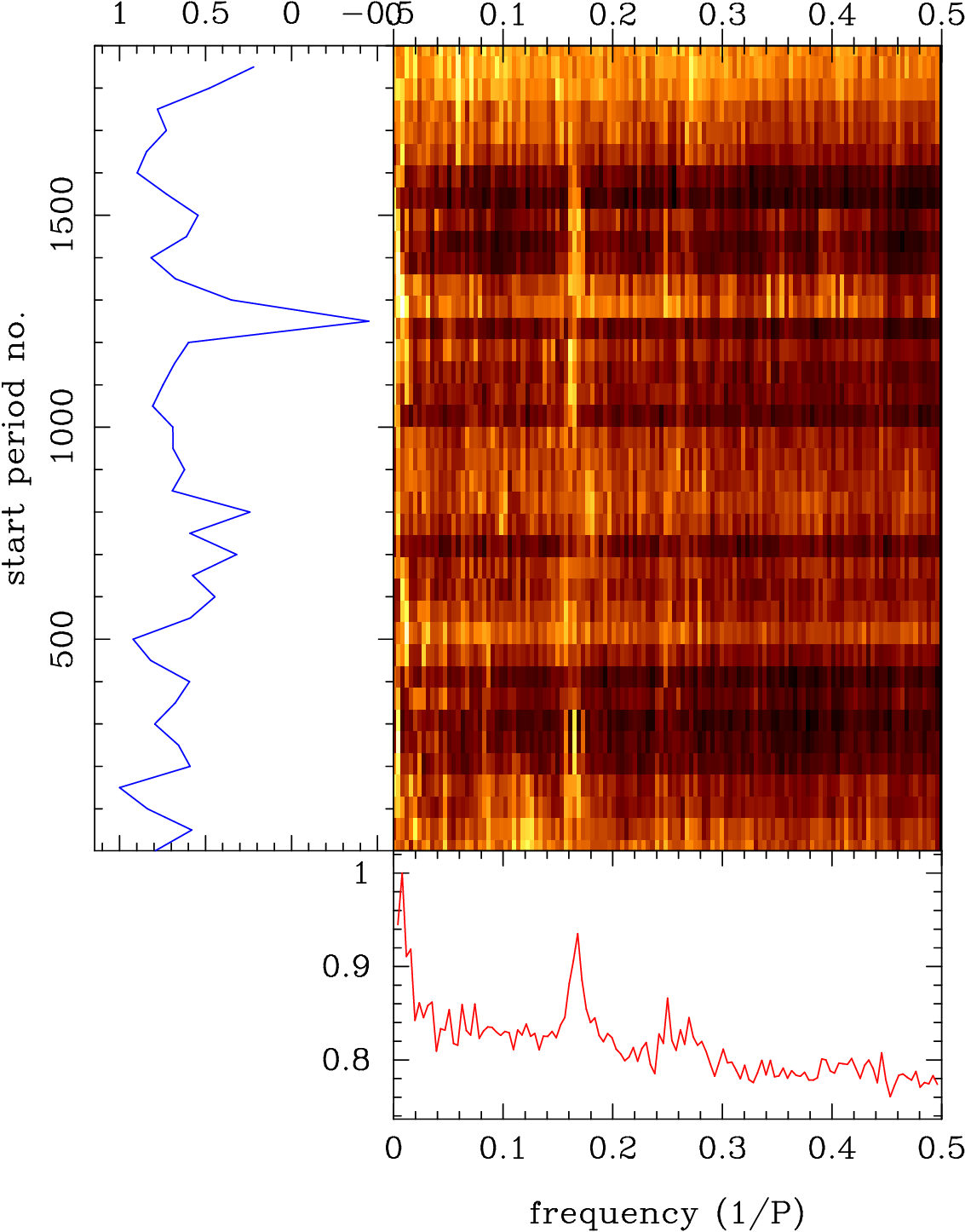}}} &
\hspace{50px}
{\mbox{\includegraphics[scale=0.4,angle=0.]{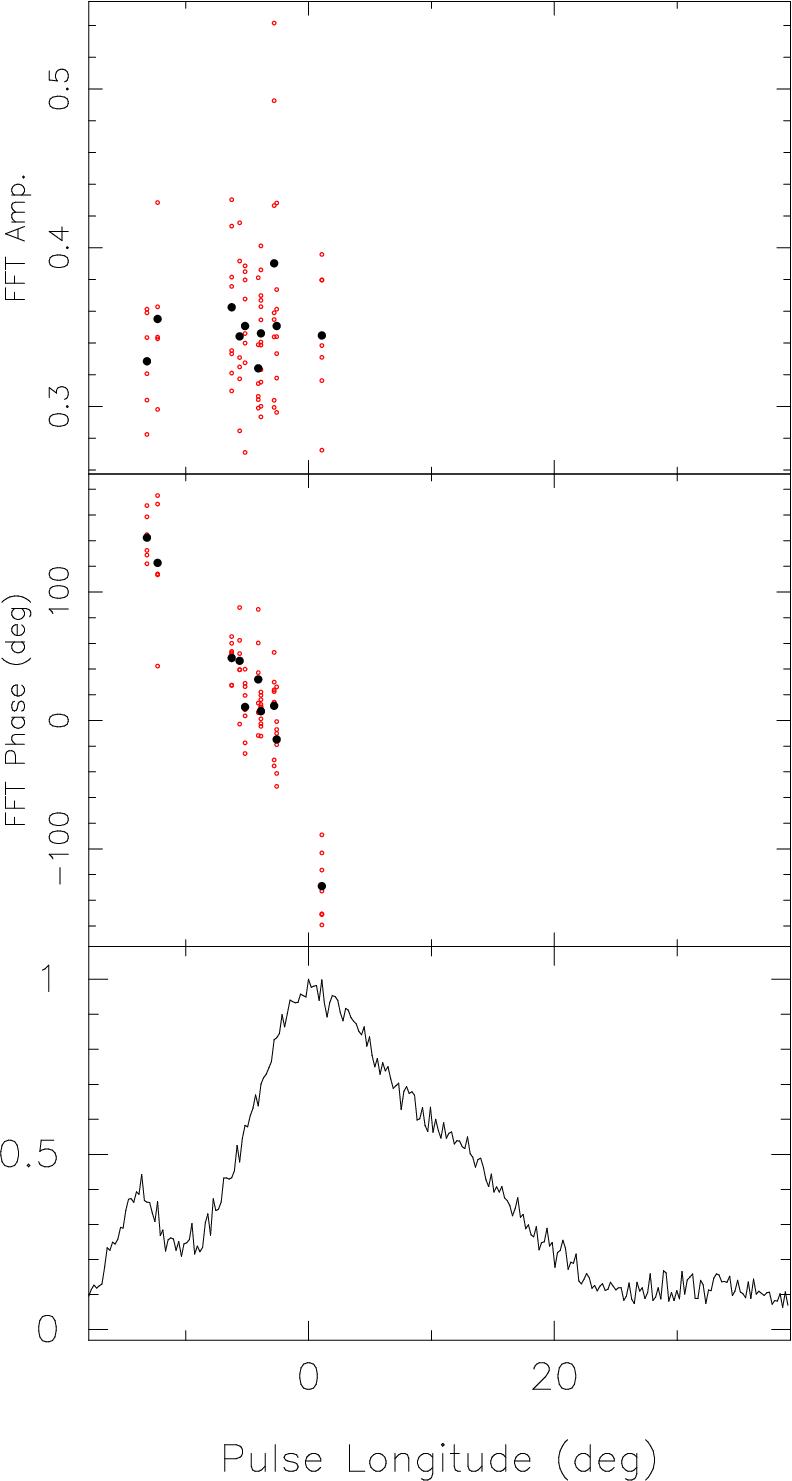}}} \\
\end{tabular}
\caption{The fluctuation spectral analysis for PSR J1921+1948 at 333 MHz. The
left panel shows time variation of LRFS. The right panel shows the variation of
the peak frequency across the pulse window; peak amplitude (top window), phase 
variations (middle window), profile (bottom window).}
\label{fig_J1921_P1}
\end{figure*}

% J2046-0421 - 333 MHz
\begin{figure*}
\begin{tabular}{@{}lr@{}}
{\mbox{\includegraphics[scale=0.4,angle=0.]{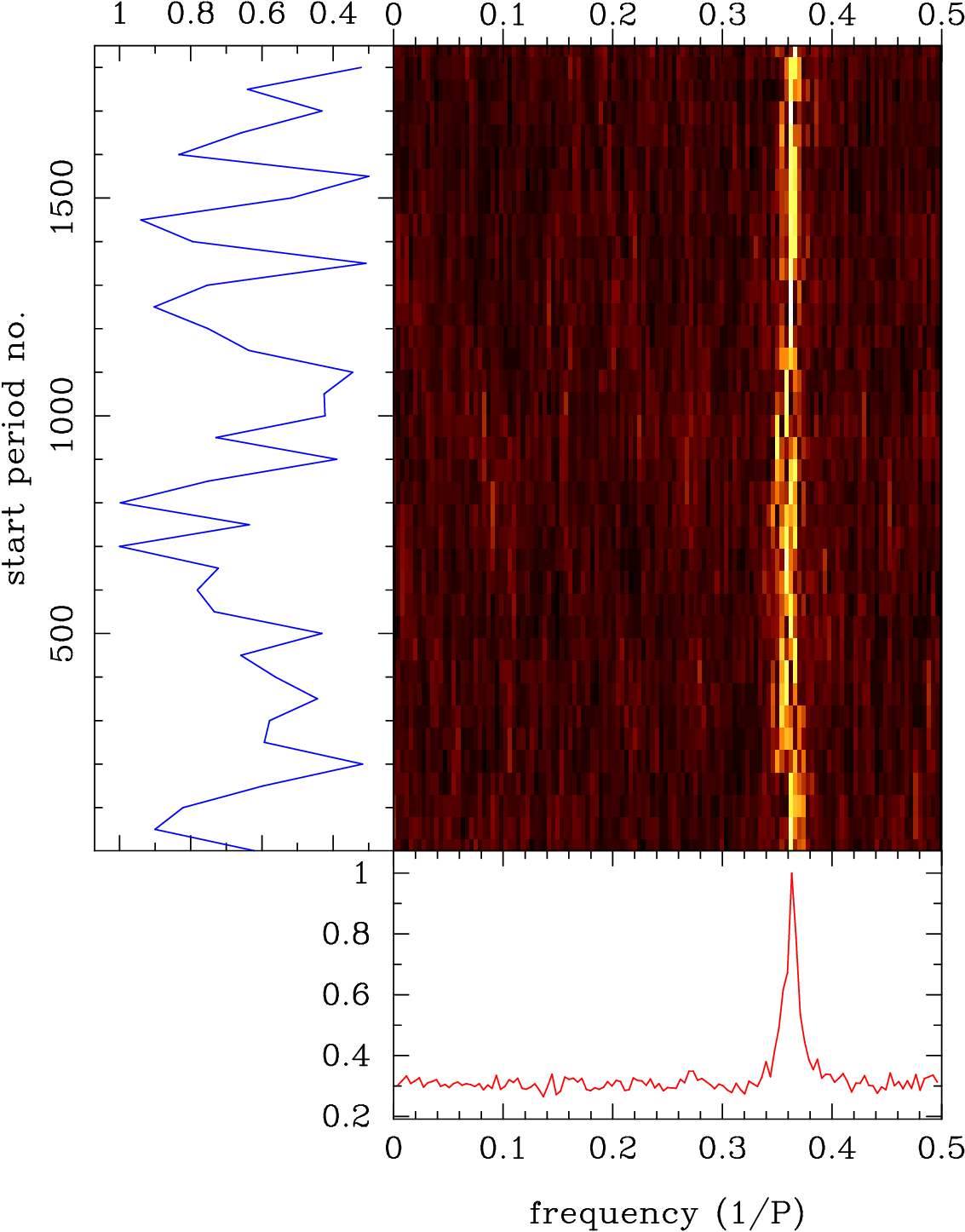}}} &
\hspace{50px}
{\mbox{\includegraphics[scale=0.4,angle=0.]{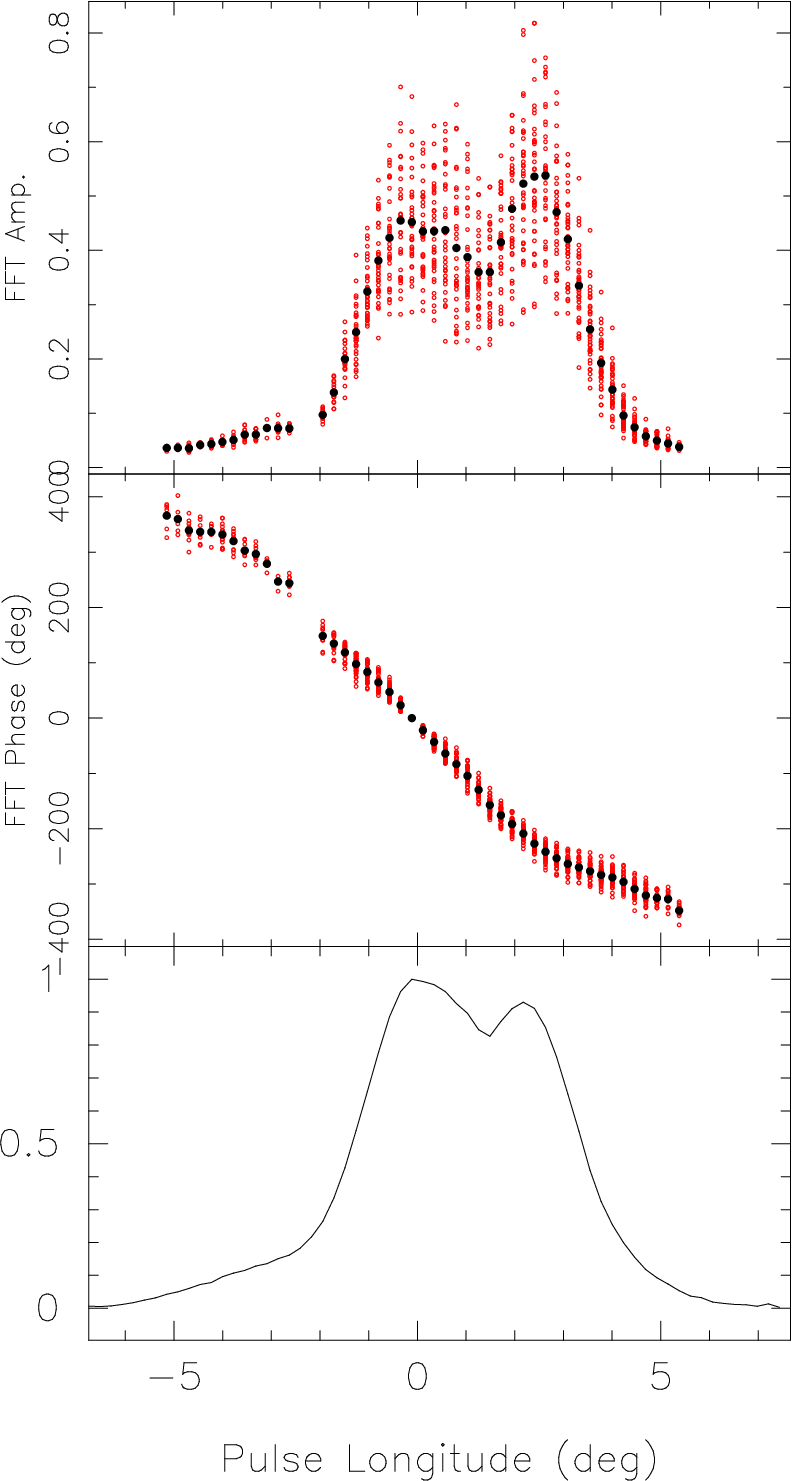}}} \\
\end{tabular}
\caption{The fluctuation spectral analysis for PSR J2046$-$0421 at 333 MHz. The
left panel shows time variation of LRFS. The right panel shows the variation of
the peak frequency across the pulse window; peak amplitude (top window), phase 
variations (middle window), profile (bottom window).}
\label{fig_J2046_1}
\end{figure*}

% J2046-0421 - 618 MHz
\begin{figure*}
\begin{tabular}{@{}lr@{}}
{\mbox{\includegraphics[scale=0.4,angle=0.]{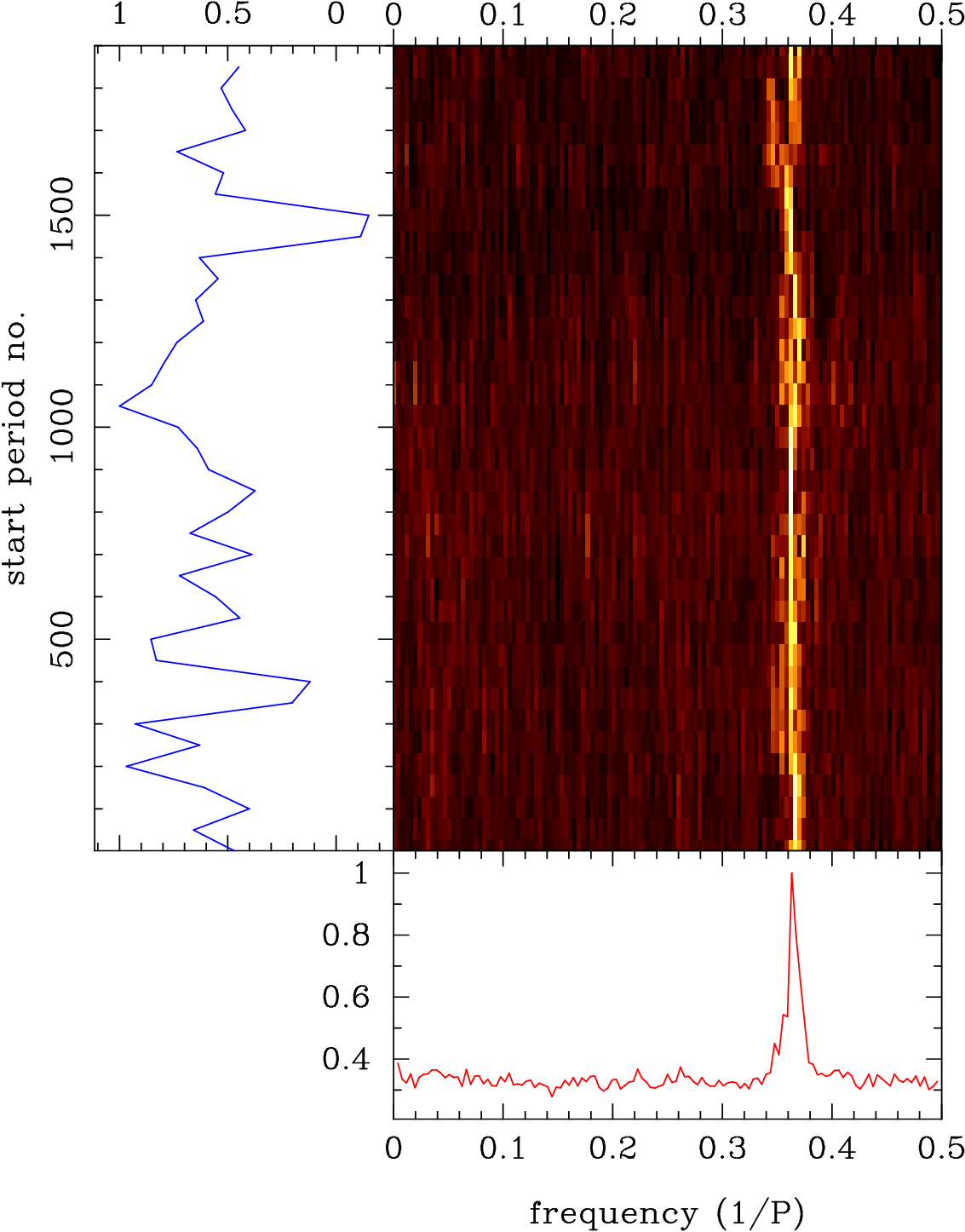}}} &
\hspace{50px}
{\mbox{\includegraphics[scale=0.4,angle=0.]{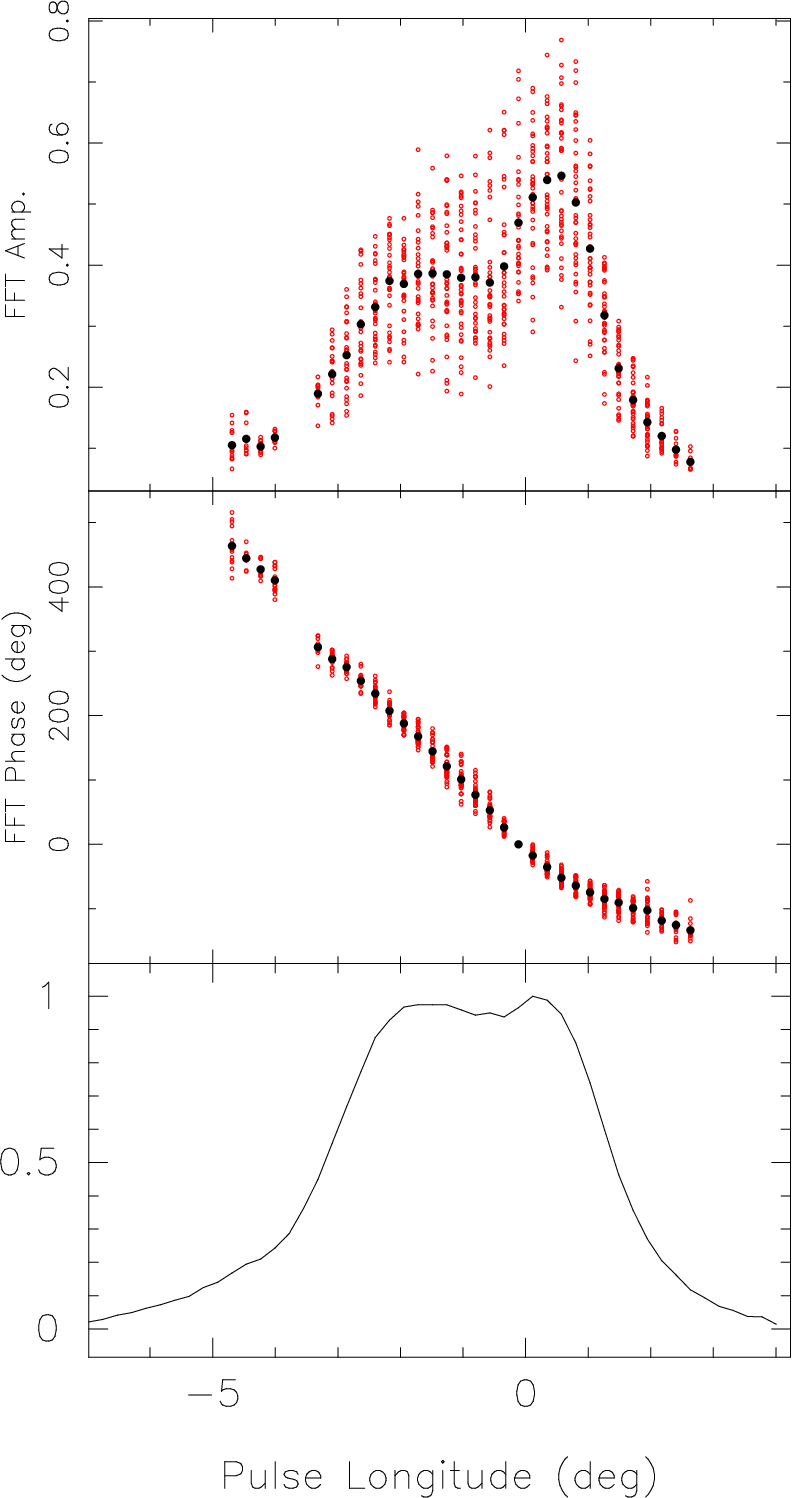}}} \\
\end{tabular}
\caption{The fluctuation spectral analysis for PSR J2046$-$0421 at 618 MHz. The
left panel shows time variation of LRFS. The right panel shows the variation of
the peak frequency across the pulse window; peak amplitude (top window), phase 
variations (middle window), profile (bottom window).}
\label{fig_J2046_2}
\end{figure*}

% J2313+4253 - 339 MHz
\begin{figure*}
\begin{tabular}{@{}lr@{}}
{\mbox{\includegraphics[scale=0.4,angle=0.]{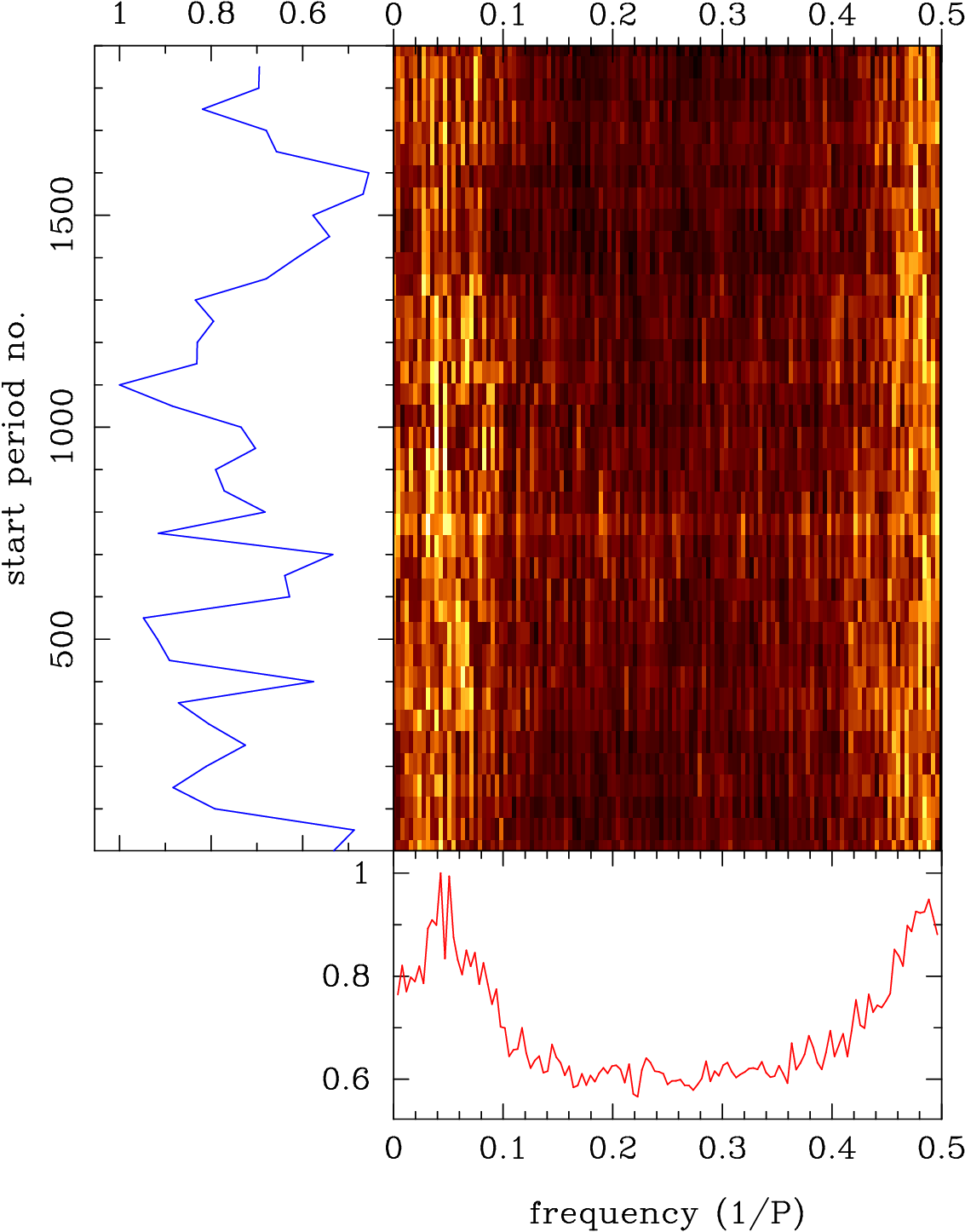}}} &
\hspace{50px}
{\mbox{\includegraphics[scale=0.4,angle=0.]{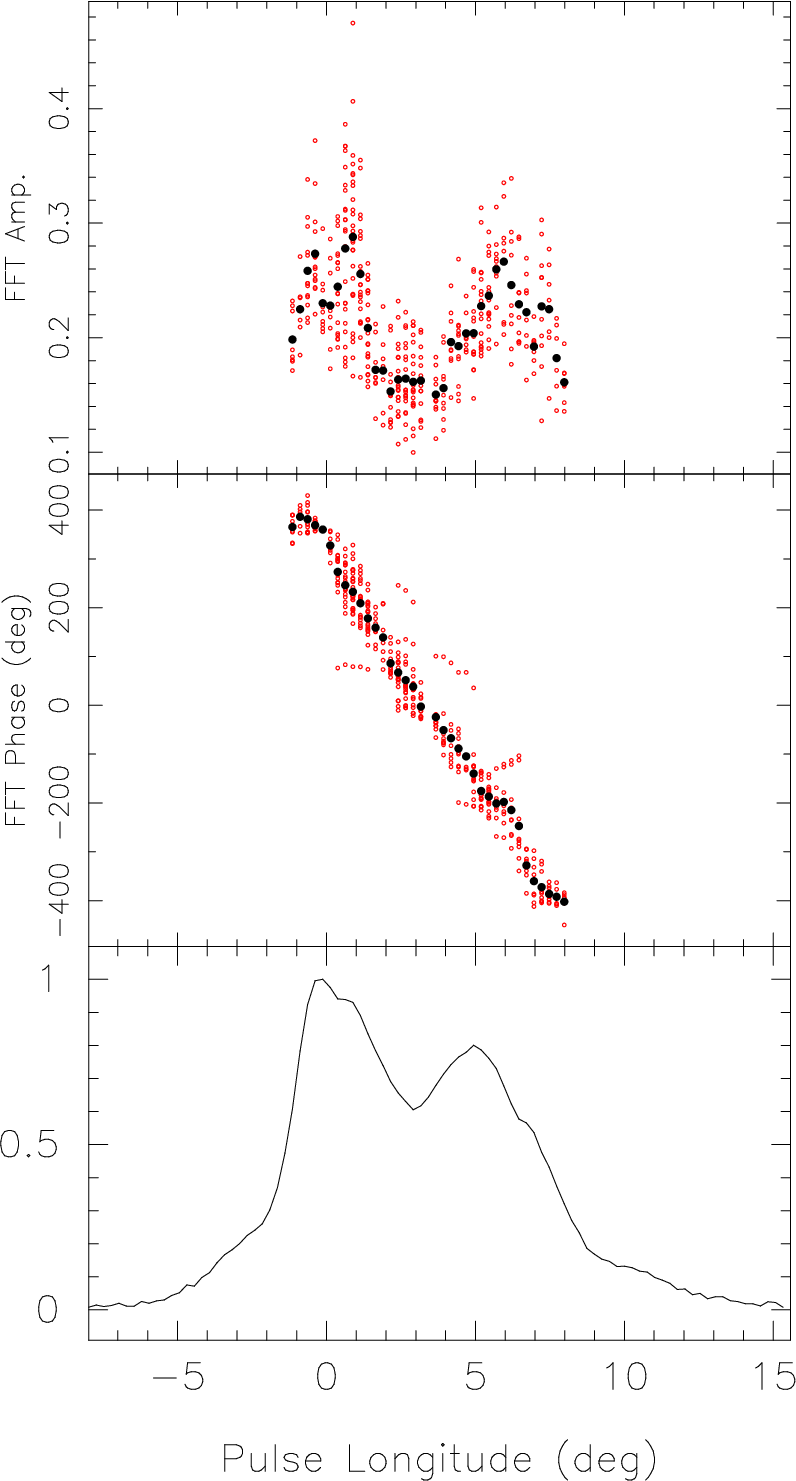}}} \\
\end{tabular}
\caption{The fluctuation spectral analysis for PSR J2313+4253 at 339 MHz. The
left panel shows time variation of LRFS. The right panel shows the variation of
the peak frequency across the pulse window; peak amplitude (top window), phase
variations (middle window), profile (bottom window).}
\label{fig_J2313}
\end{figure*}

\section{The switching phase modulated drifting : Fluctuation Spectral Plots}

% J0323+3944 - 339 MHz
\begin{figure*}
\begin{tabular}{@{}lr@{}}
{\mbox{\includegraphics[scale=0.4,angle=0.]{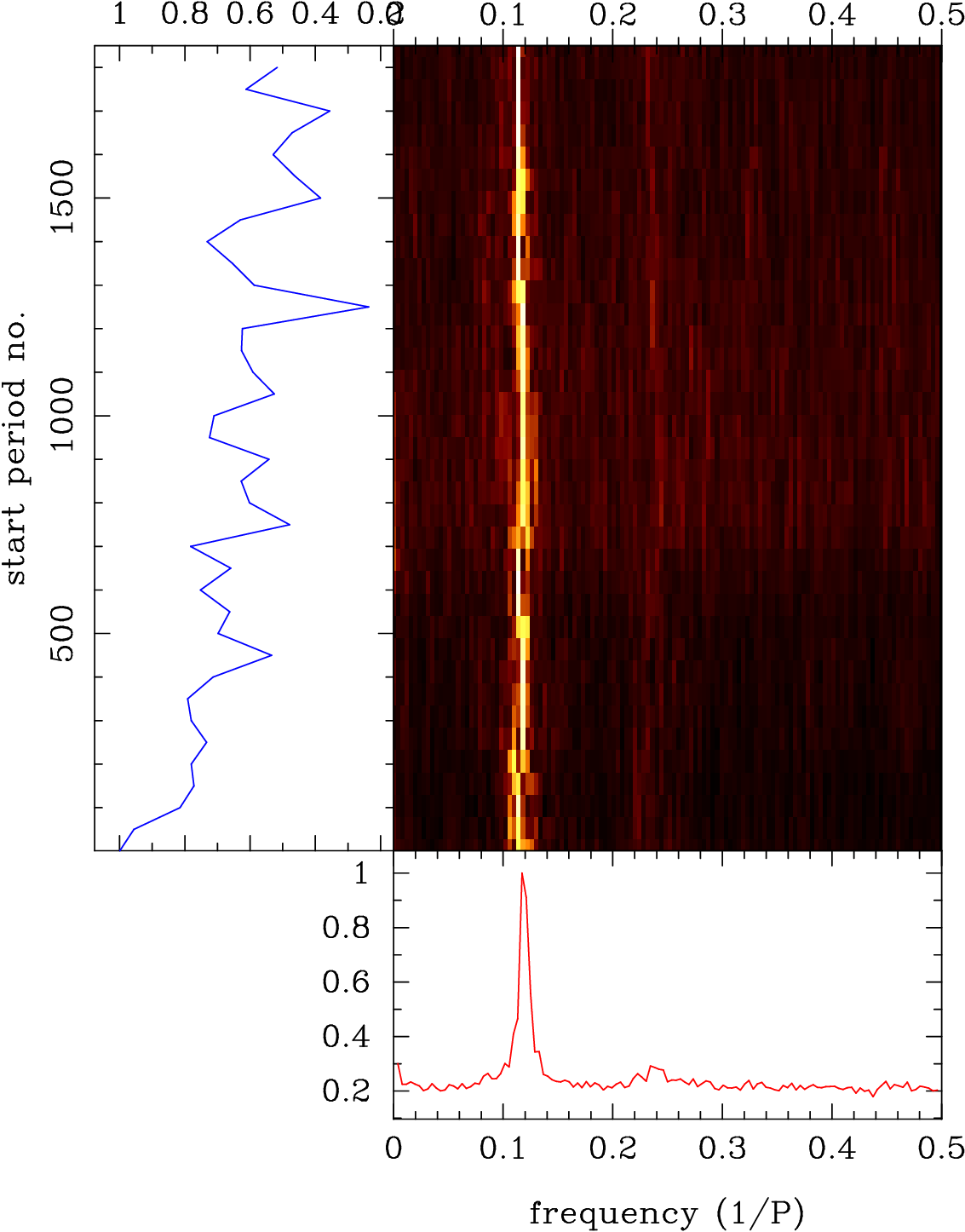}}} &
\hspace{50px}
{\mbox{\includegraphics[scale=0.4,angle=0.]{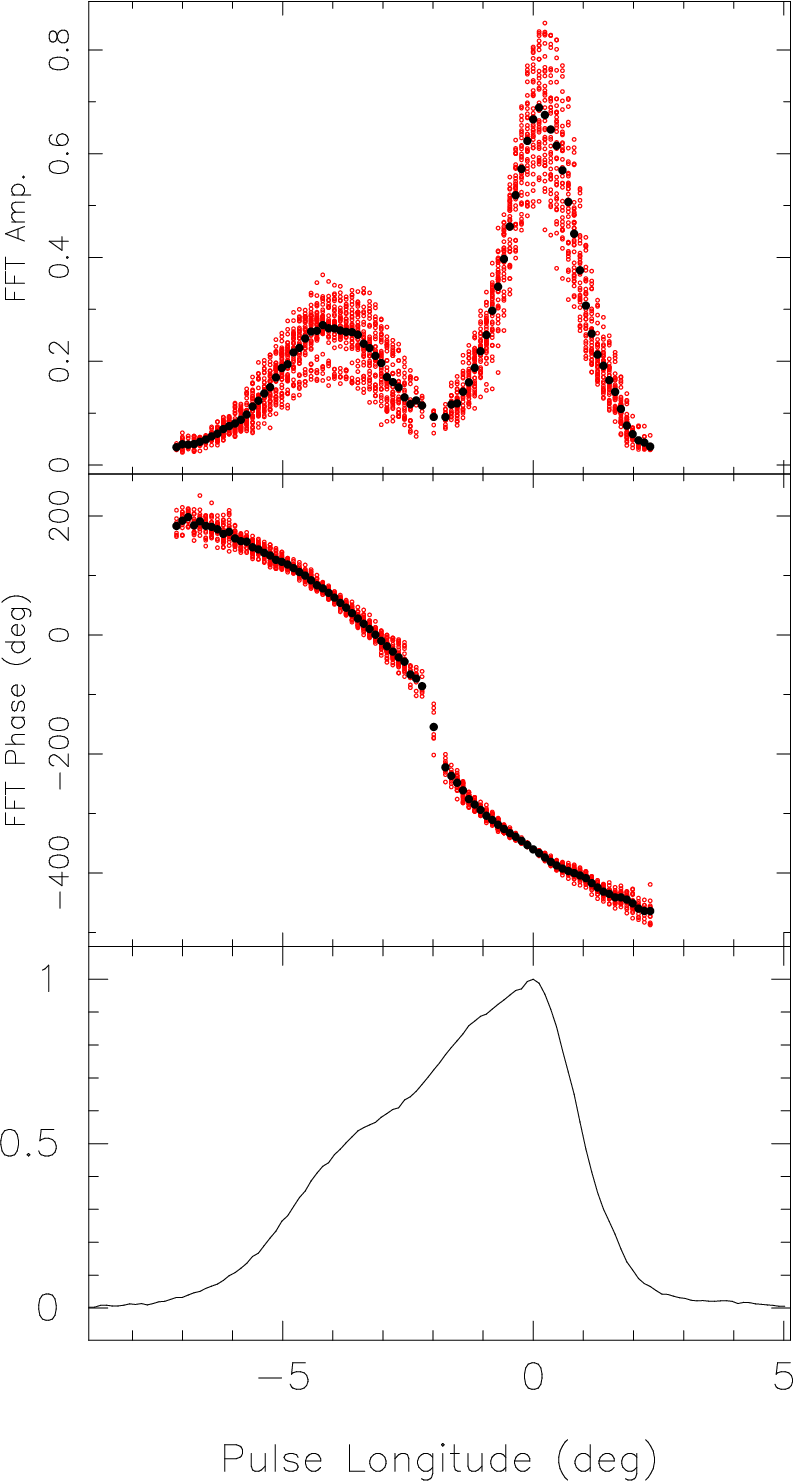}}} \\
\end{tabular}
\caption{The fluctuation spectral analysis for PSR J0323+3944 at 339 MHz. The 
left panel shows time variation of LRFS. The right panel shows the variation of
the peak frequency across the pulse window; peak amplitude (top window), phase 
variations (middle window), profile (bottom window).}
\label{fig_J0323}
\end{figure*}

% J1842-0359 - 618 MHz
\begin{figure*}
\begin{tabular}{@{}lr@{}}
{\mbox{\includegraphics[scale=0.4,angle=0.]{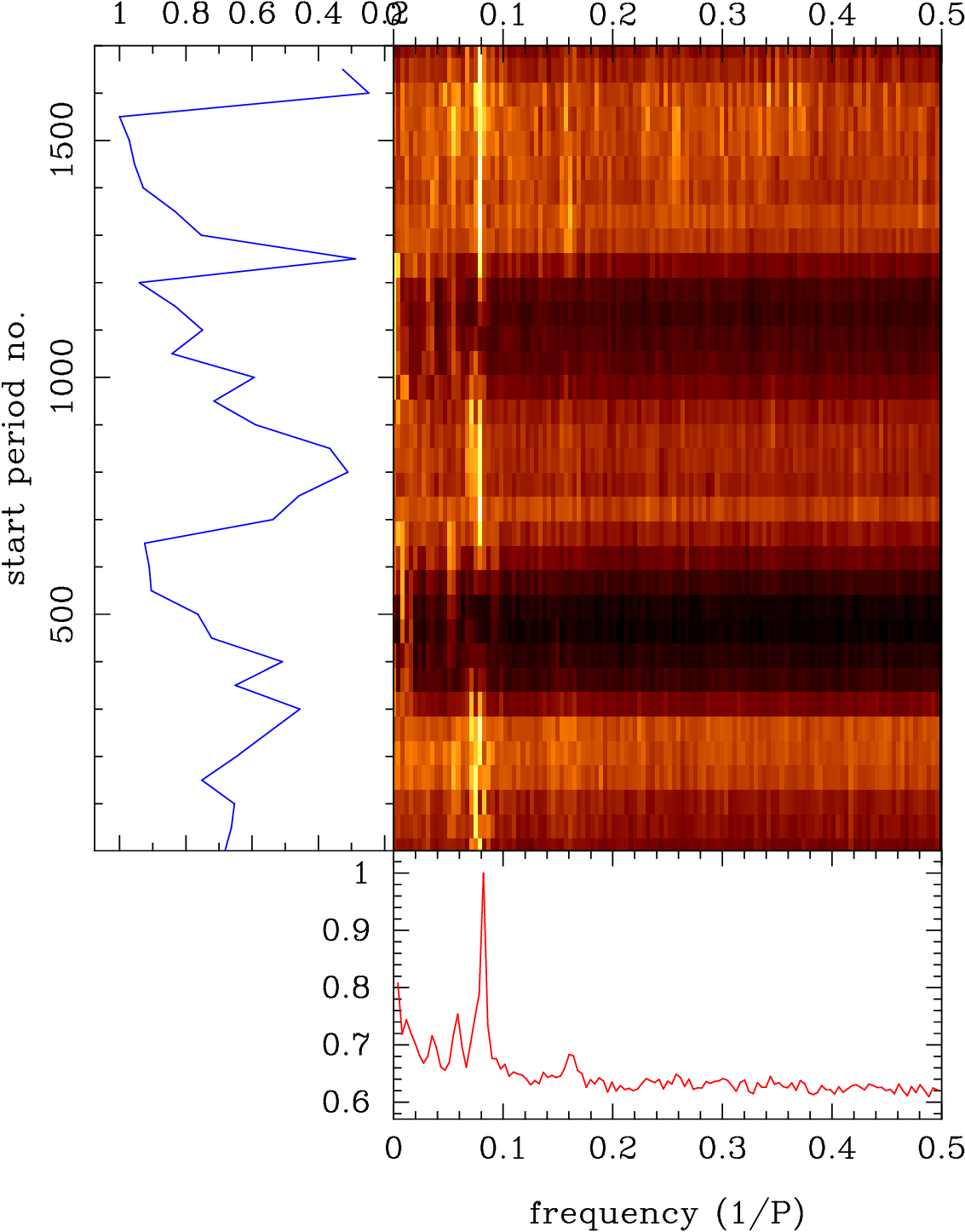}}} &
\hspace{50px}
{\mbox{\includegraphics[scale=0.4,angle=0.]{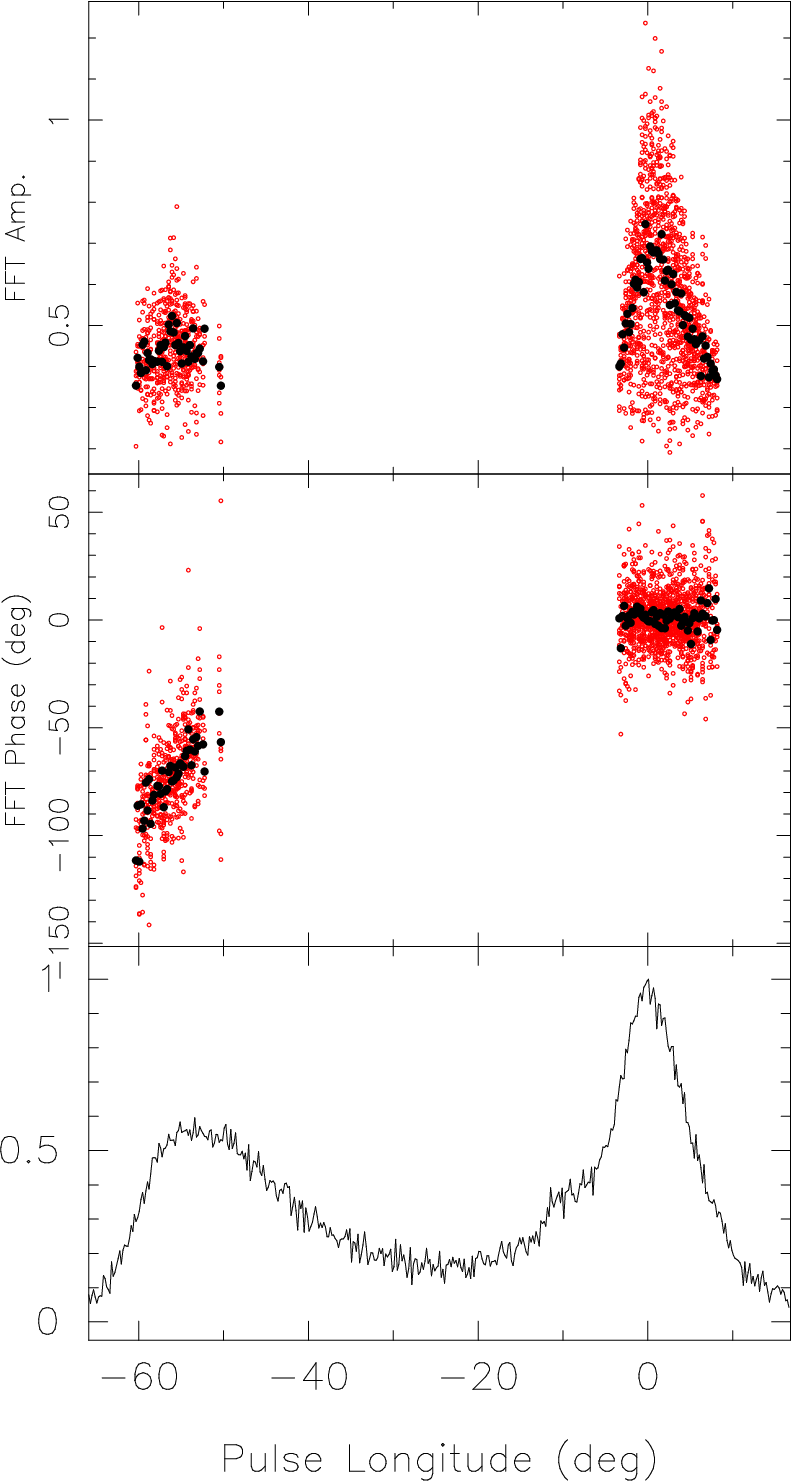}}} \\
\end{tabular}
\caption{The fluctuation spectral analysis for PSR J1842$-$0359 at 618 MHz. The
left panel shows time variation of LRFS. The right panel shows the variation of
the peak frequency across the pulse window; peak amplitude (top window), phase
variations (middle window), profile (bottom window).}
\label{fig_J1842}
\end{figure*}

% J1921+2153 - 618 MHz
\begin{figure*}
\begin{tabular}{@{}lr@{}}
{\mbox{\includegraphics[scale=0.4,angle=0.]{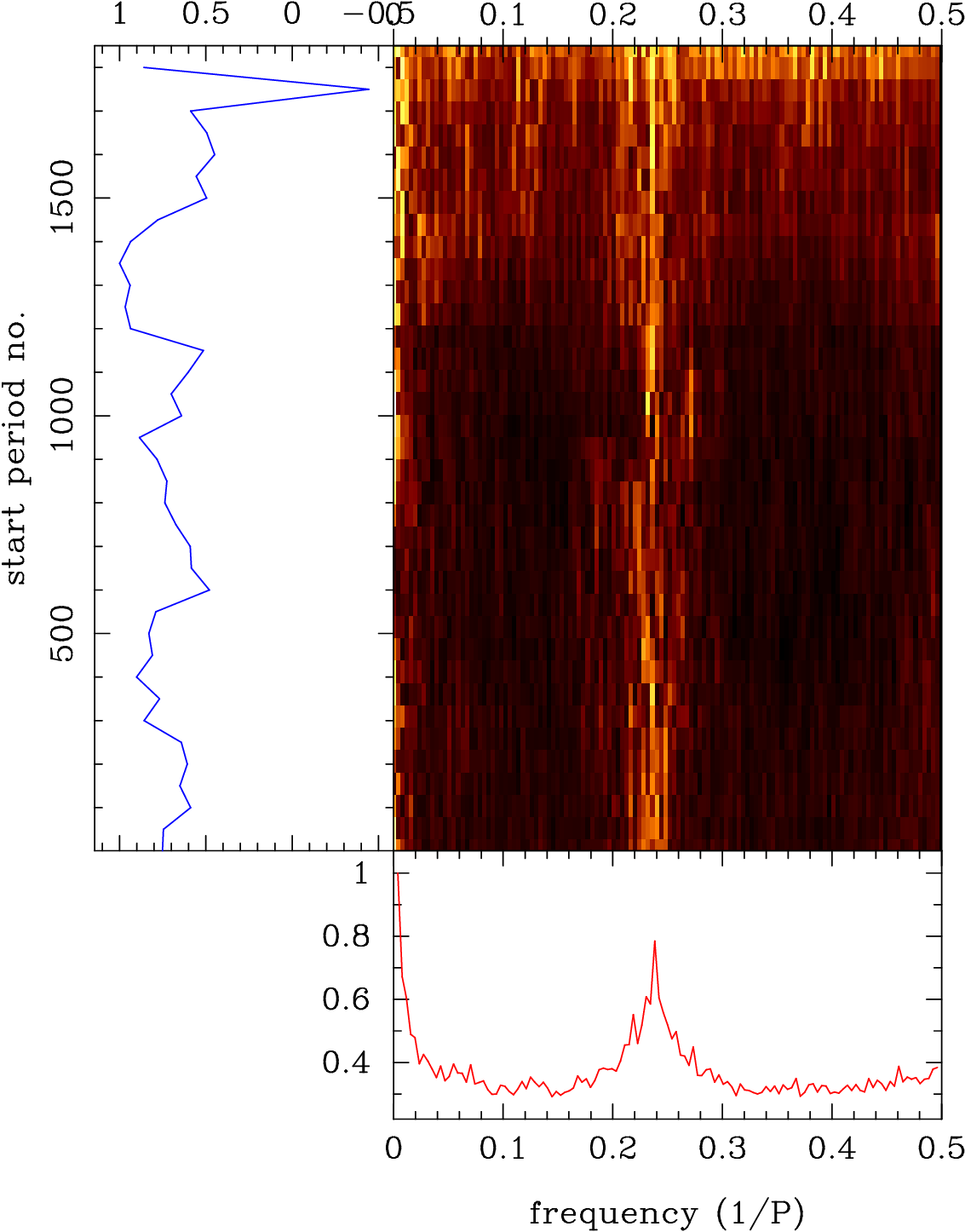}}} &
\hspace{50px}
{\mbox{\includegraphics[scale=0.4,angle=0.]{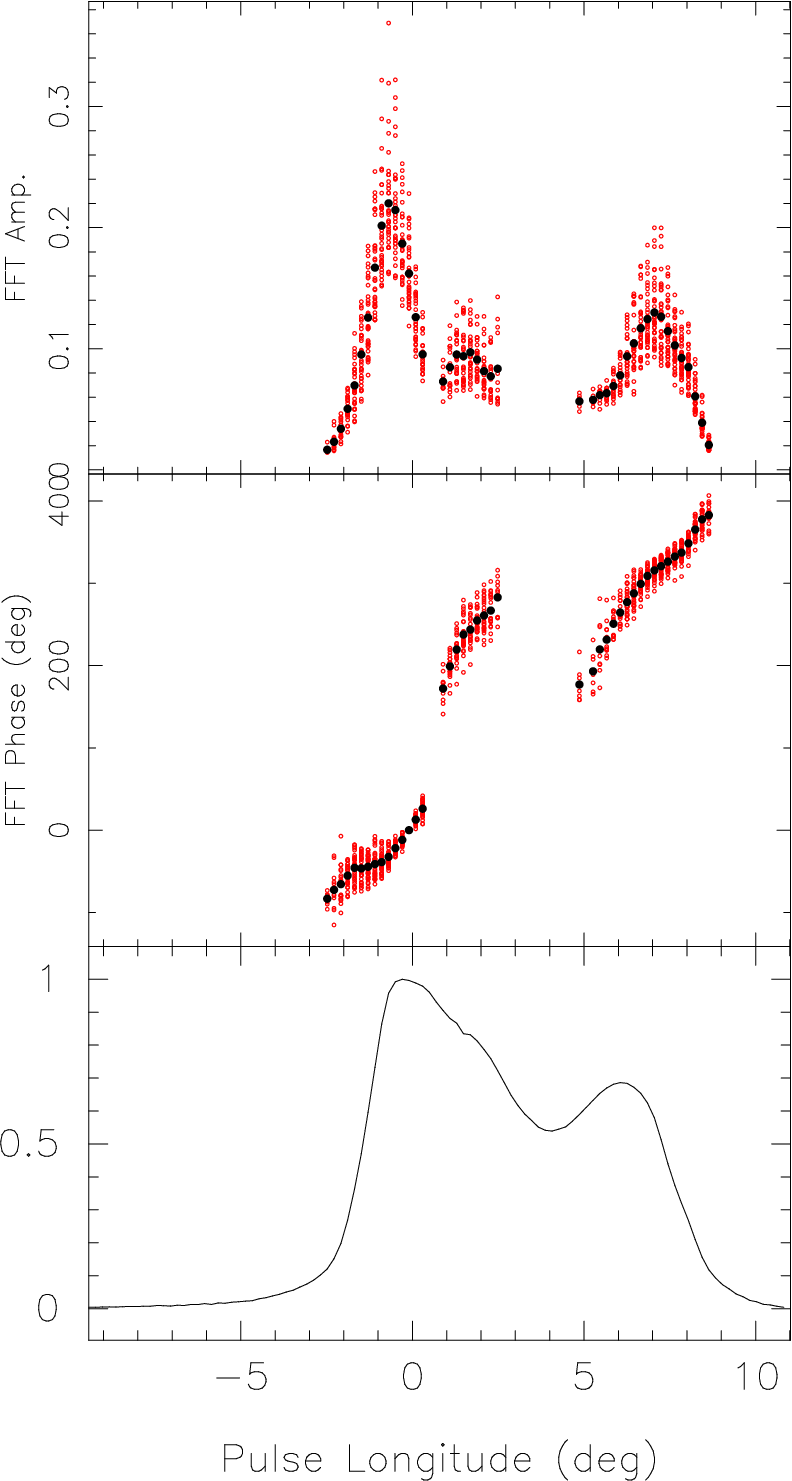}}} \\
\end{tabular}
\caption{The fluctuation spectral analysis for PSR J1921+2153 at 618 MHz. The
left panel shows time variation of LRFS. The right panel shows the variation of
the peak frequency across the pulse window; peak amplitude (top window), phase
variations (middle window), profile (bottom window).}
\label{fig_J1921_P2}
\end{figure*}

% J2321+6024 - 339 MHz
\begin{figure*}
\begin{tabular}{@{}lr@{}}
{\mbox{\includegraphics[scale=0.4,angle=0.]{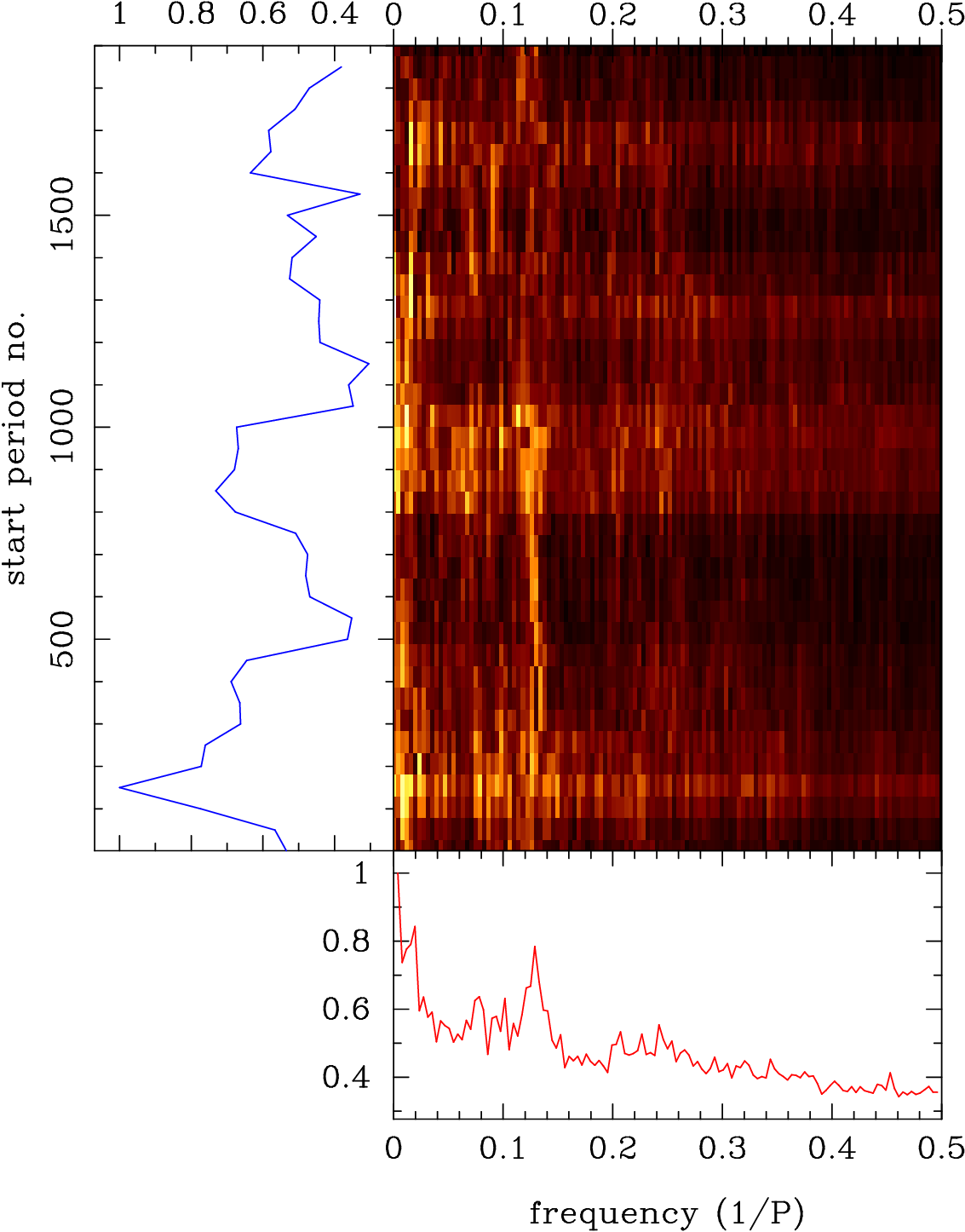}}} &
\hspace{50px}
{\mbox{\includegraphics[scale=0.4,angle=0.]{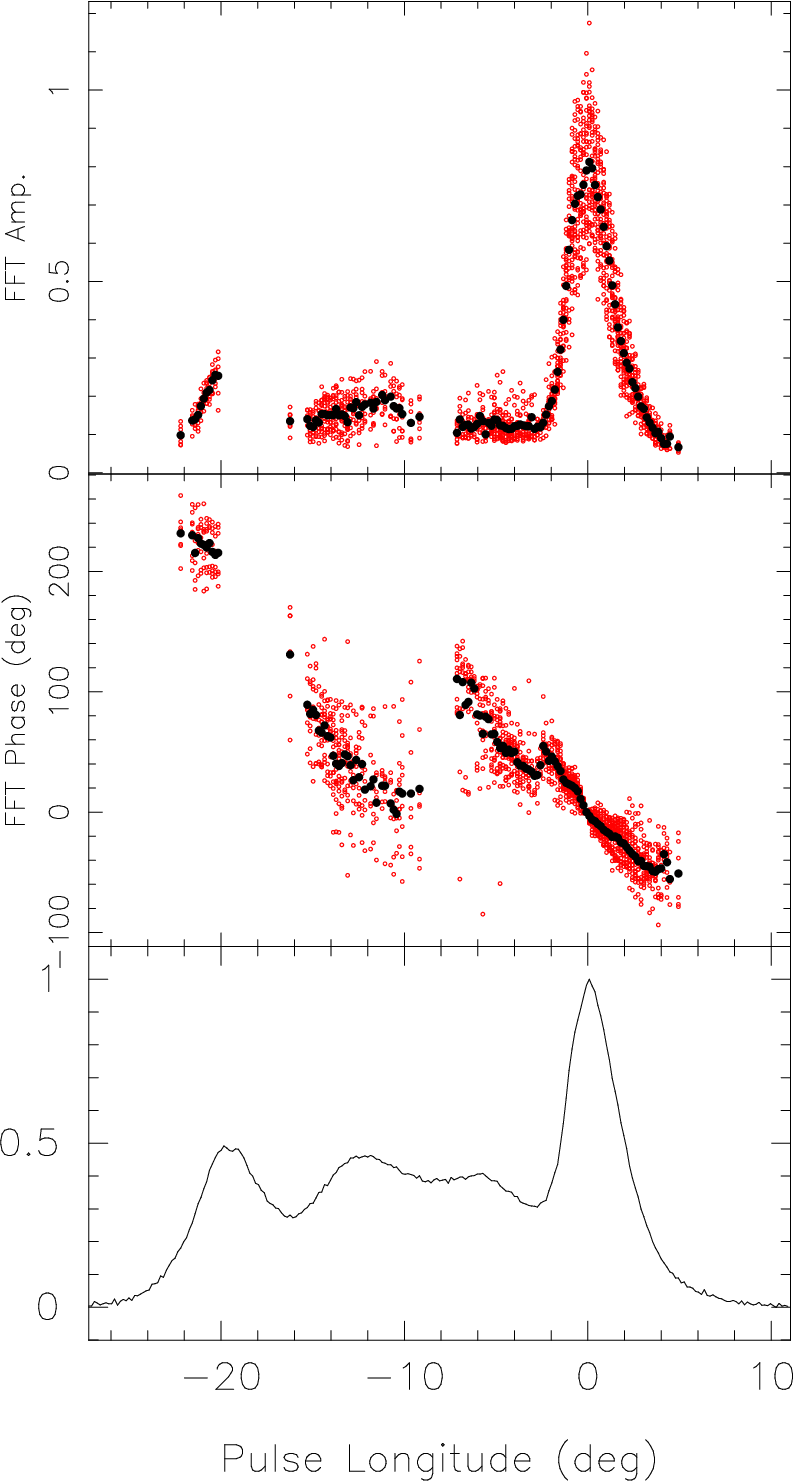}}} \\
\end{tabular}
\caption{The fluctuation spectral analysis for PSR J2321+6024 at 339 MHz. The 
left panel shows time variation of LRFS. The right panel shows the variation of
the peak frequency across the pulse window; peak amplitude (top window), phase
variations (middle window), profile (bottom window). The peak frequency used in
the right plot corresponds to $P_3$ = 8$P$.}
\label{fig_J2321}
\end{figure*}

\section{The diffuse phase modulated drifting : Fluctuation Spectral Plots}

% J0823+0159 - 339 MHz
\begin{figure*}
\begin{tabular}{@{}lr@{}}
{\mbox{\includegraphics[scale=0.4,angle=0.]{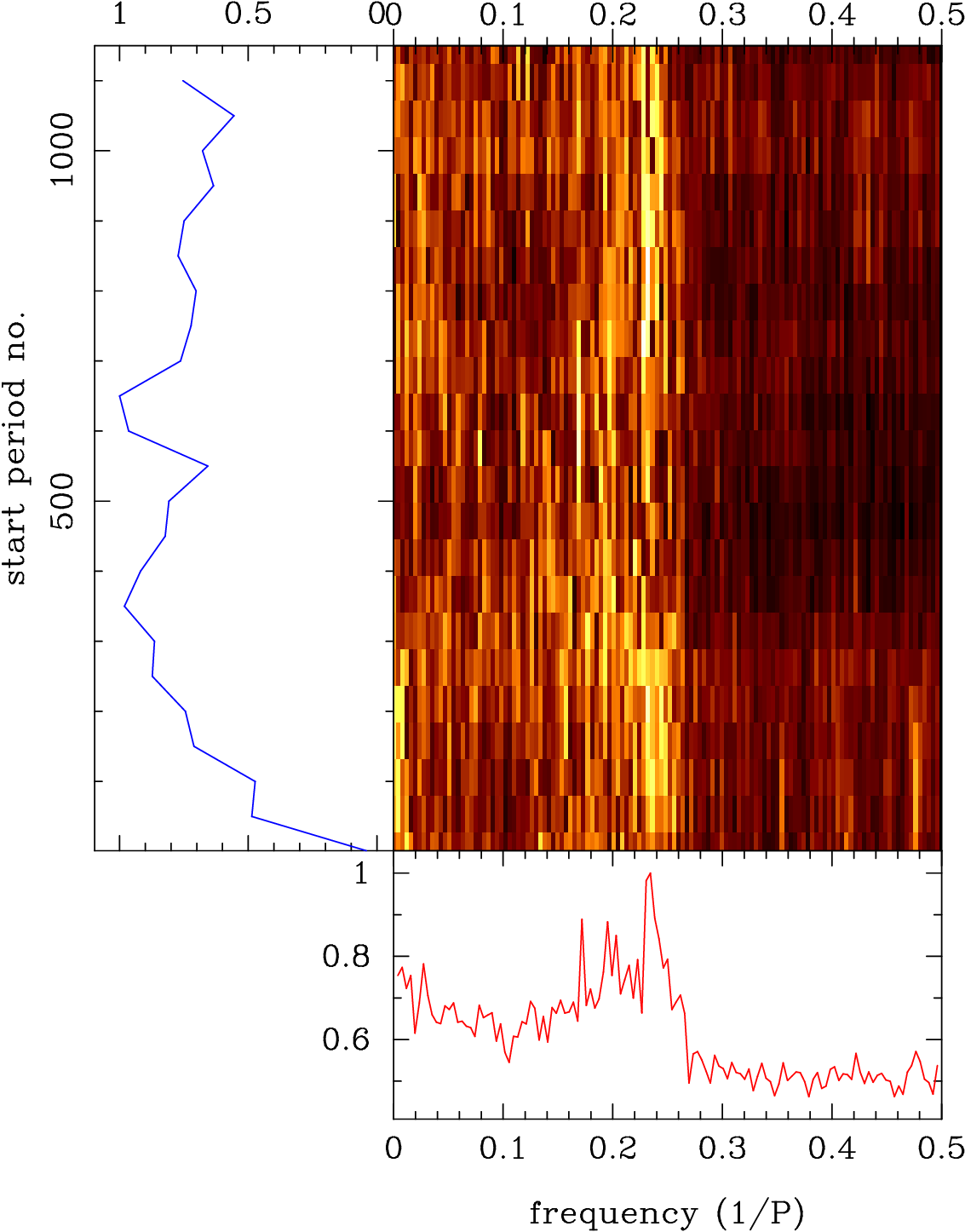}}} &
\hspace{50px}
{\mbox{\includegraphics[scale=0.4,angle=0.]{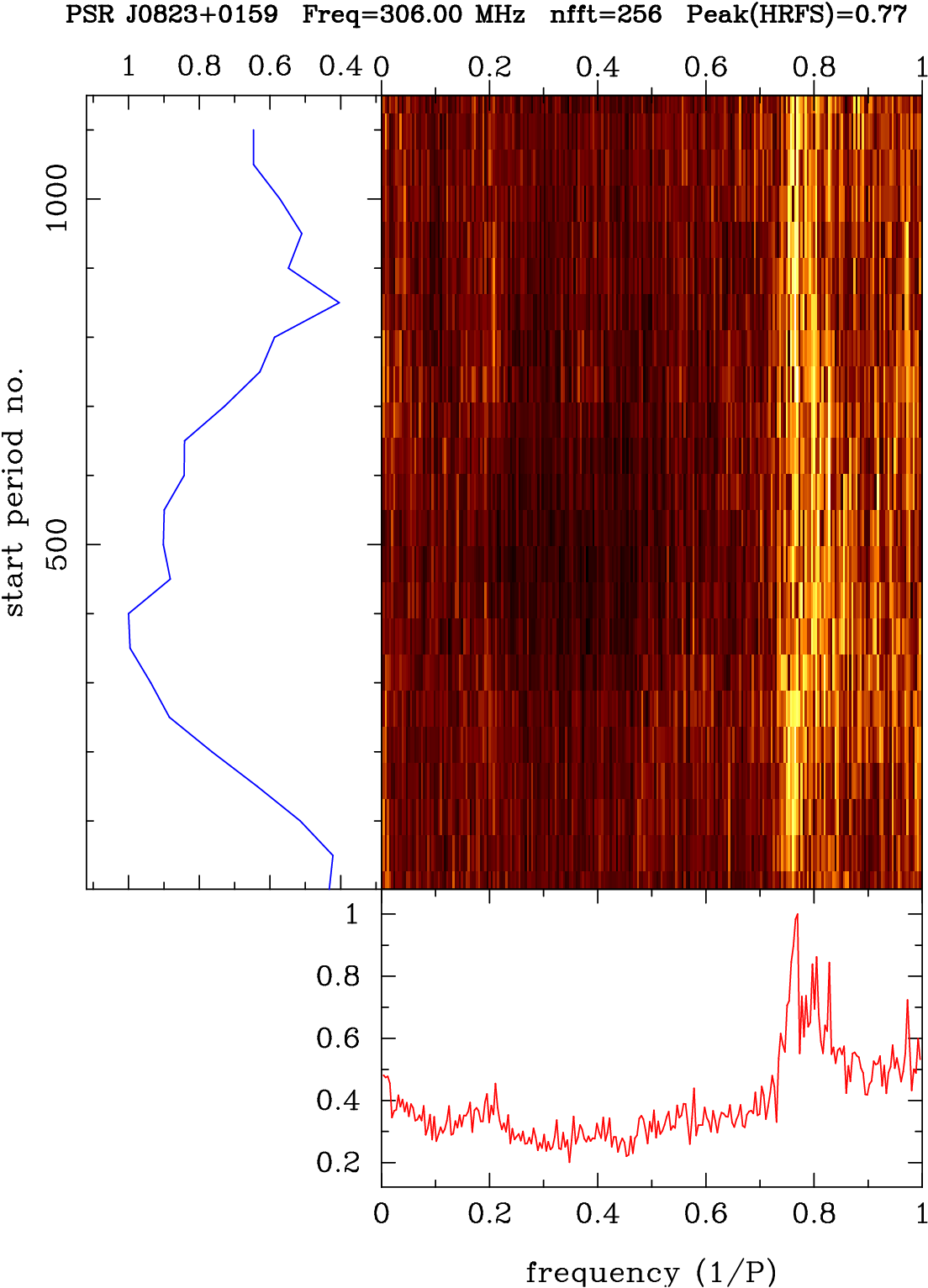}}} \\
\end{tabular}
\caption{The fluctuation spectral analysis for PSR J0823+0159 at 339 MHz. The
left panel shows time variation of LRFS. The right panel shows the time 
variation of HRFS.}
\label{fig_J0823}
\end{figure*}

% J2018+2839 - 339 MHz
\begin{figure*}
\begin{tabular}{@{}lr@{}}
{\mbox{\includegraphics[scale=0.4,angle=0.]{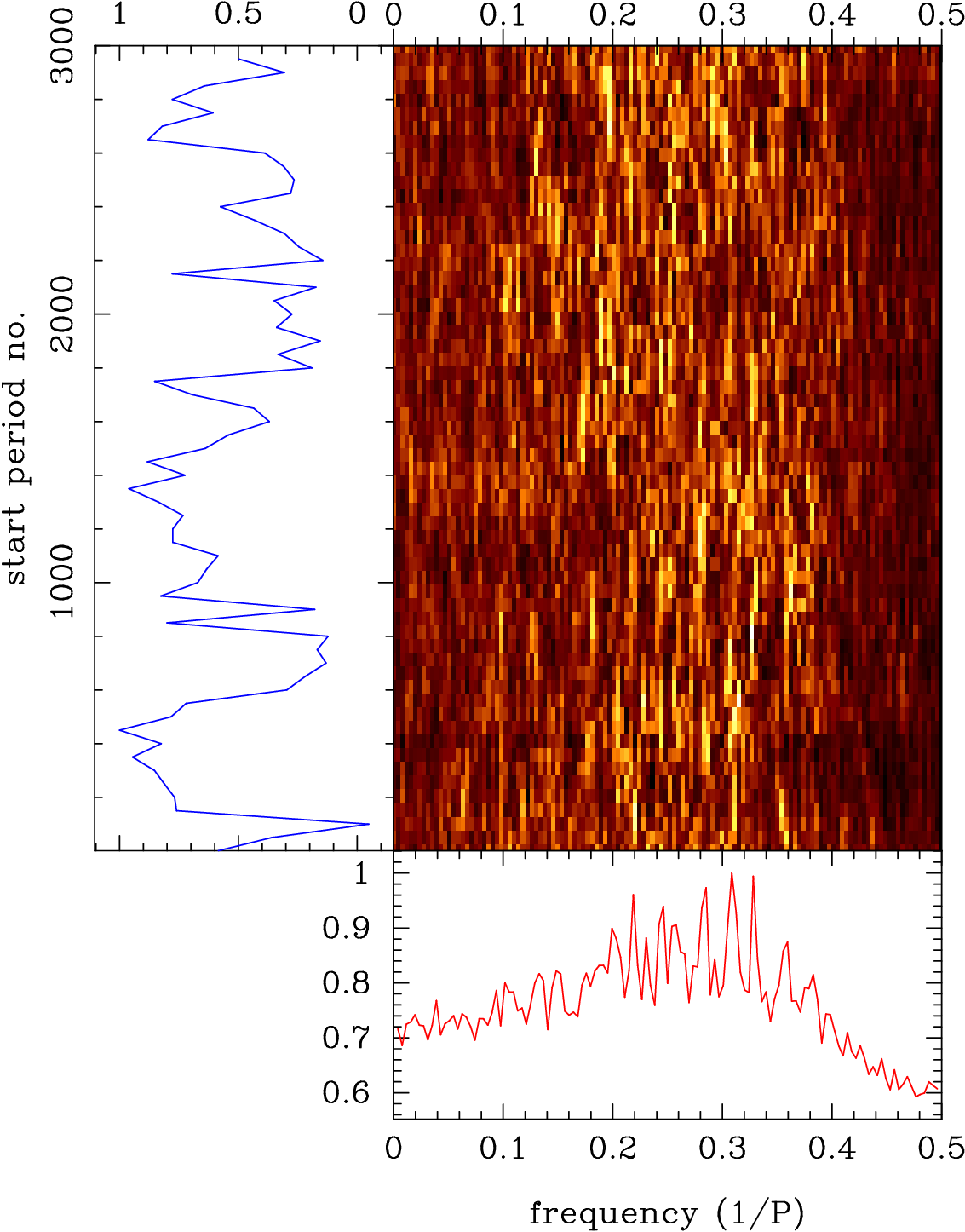}}} &
\hspace{50px}
{\mbox{\includegraphics[scale=0.4,angle=0.]{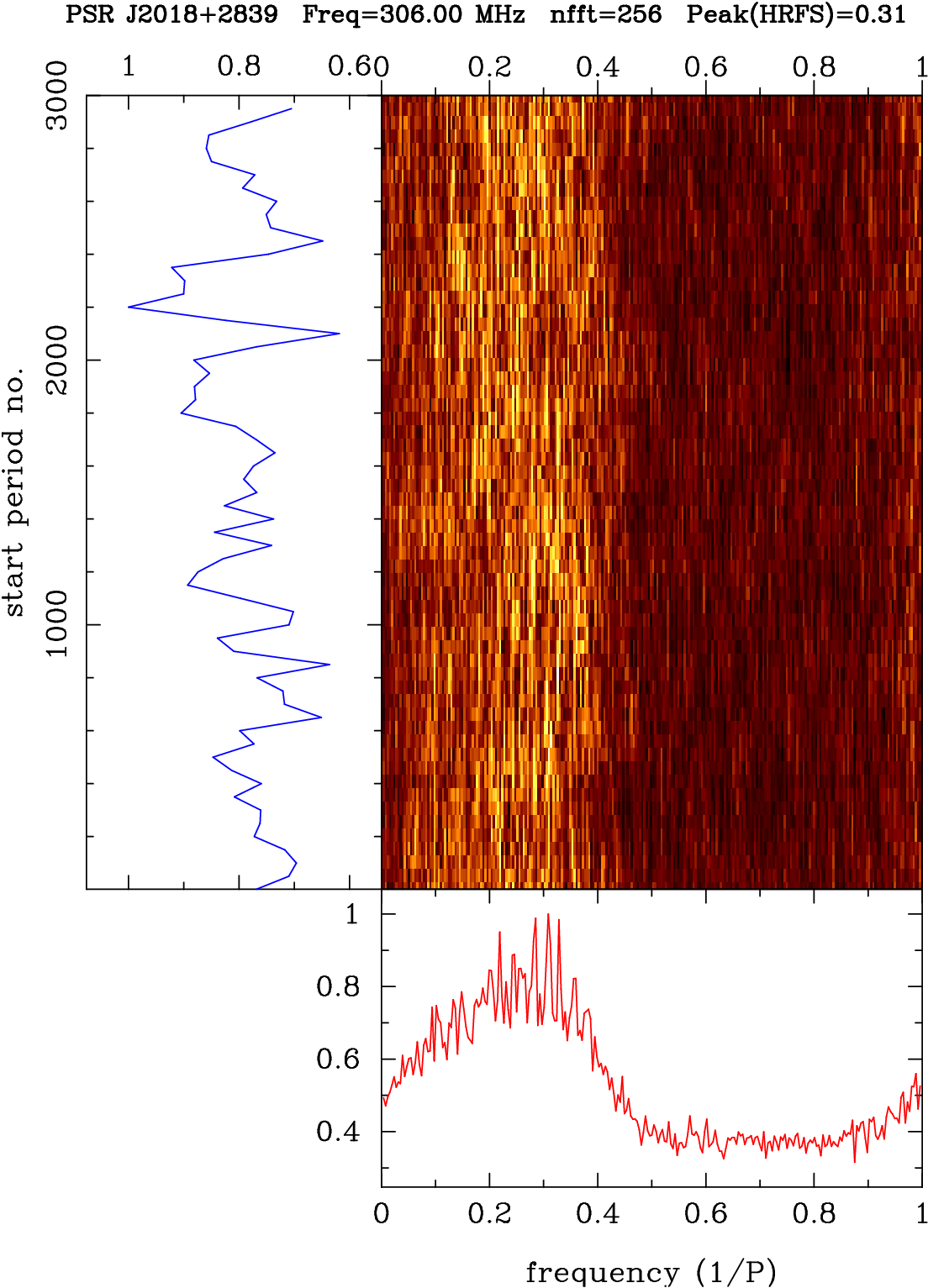}}} \\
\end{tabular}
\caption{The fluctuation spectral analysis for PSR J2018+2839 at 339 MHz. The
left panel shows time variation of LRFS. The right panel shows the time
variation of HRFS.}
\label{fig_J2018}
\end{figure*}

\section{The phase stationary drifting : Fluctuation Spectral Plots}

% J0624-0424 - 339 MHz
\begin{figure*}
\begin{tabular}{@{}lr@{}}
{\mbox{\includegraphics[scale=0.4,angle=0.]{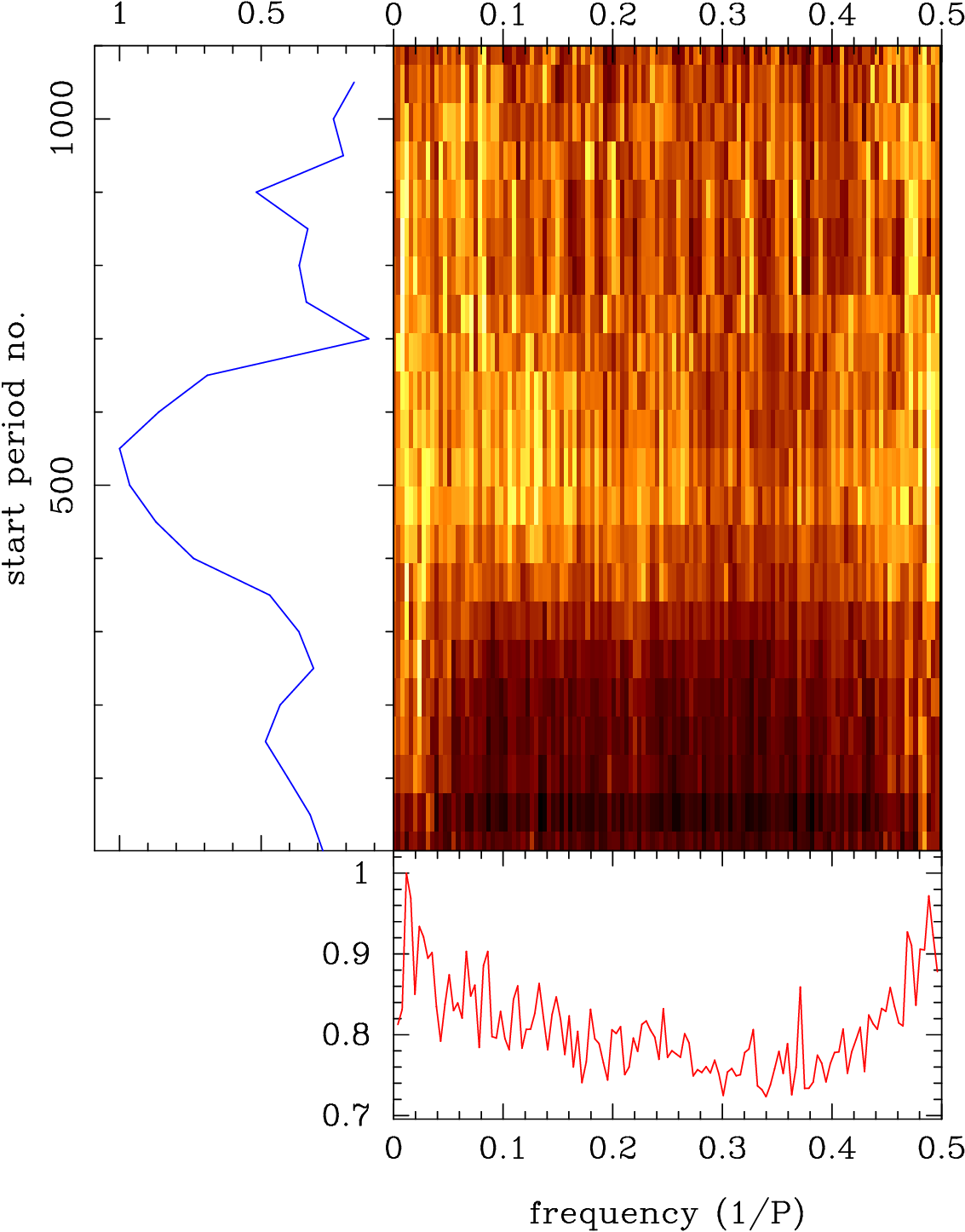}}} &
\hspace{50px}
{\mbox{\includegraphics[scale=0.4,angle=0.]{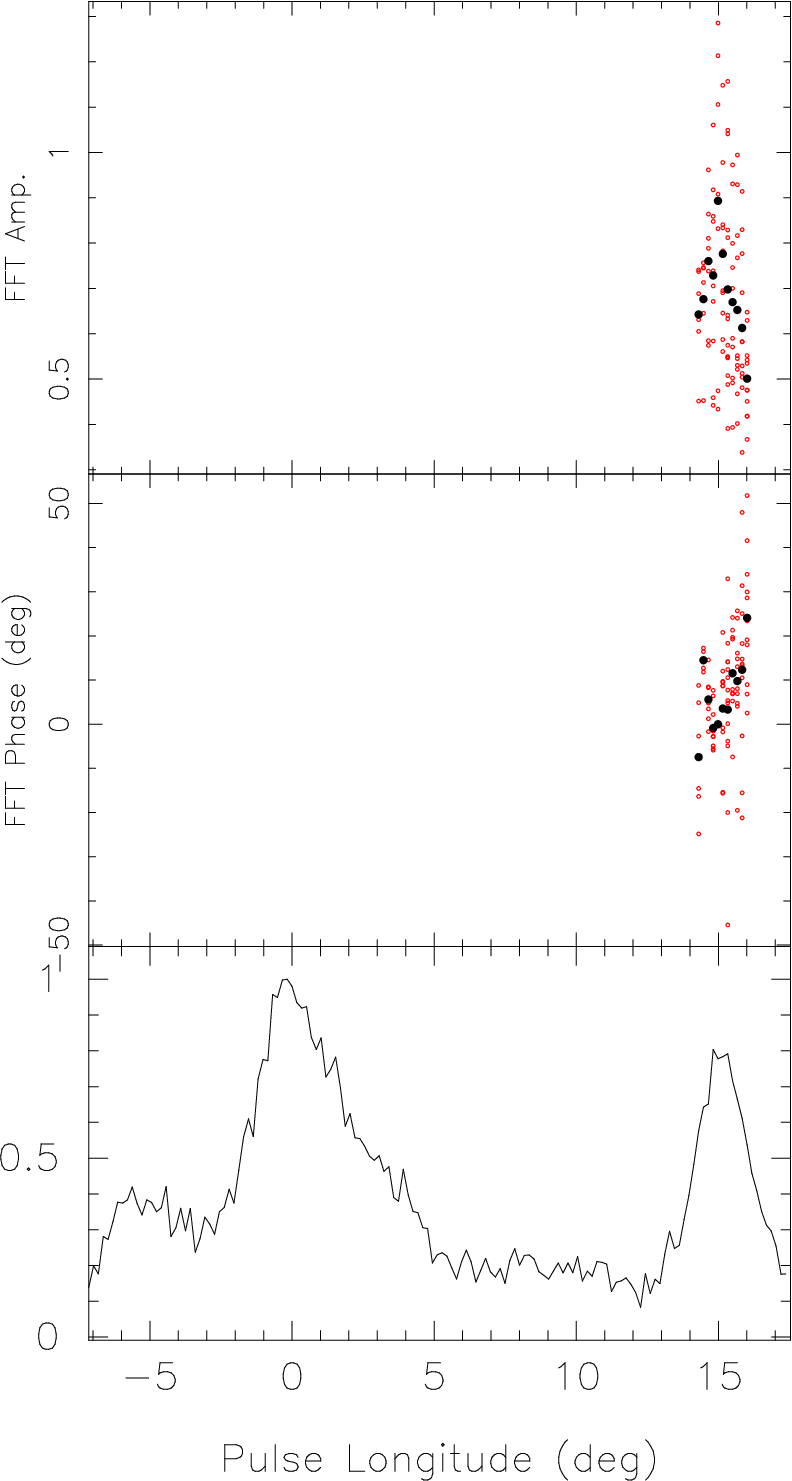}}} \\
\end{tabular}
\caption{The fluctuation spectral analysis for PSR J0624$-$0424 at 339 MHz. The
left panel shows time variation of LRFS. The right panel shows the variation of
the peak frequency across the pulse window; peak amplitude (top window), phase
variations (middle window), profile (bottom window).}
\label{fig_J0624}
\end{figure*}

% J0837+0610 - 333 MHz
\begin{figure*}
\begin{tabular}{@{}lr@{}}
{\mbox{\includegraphics[scale=0.4,angle=0.]{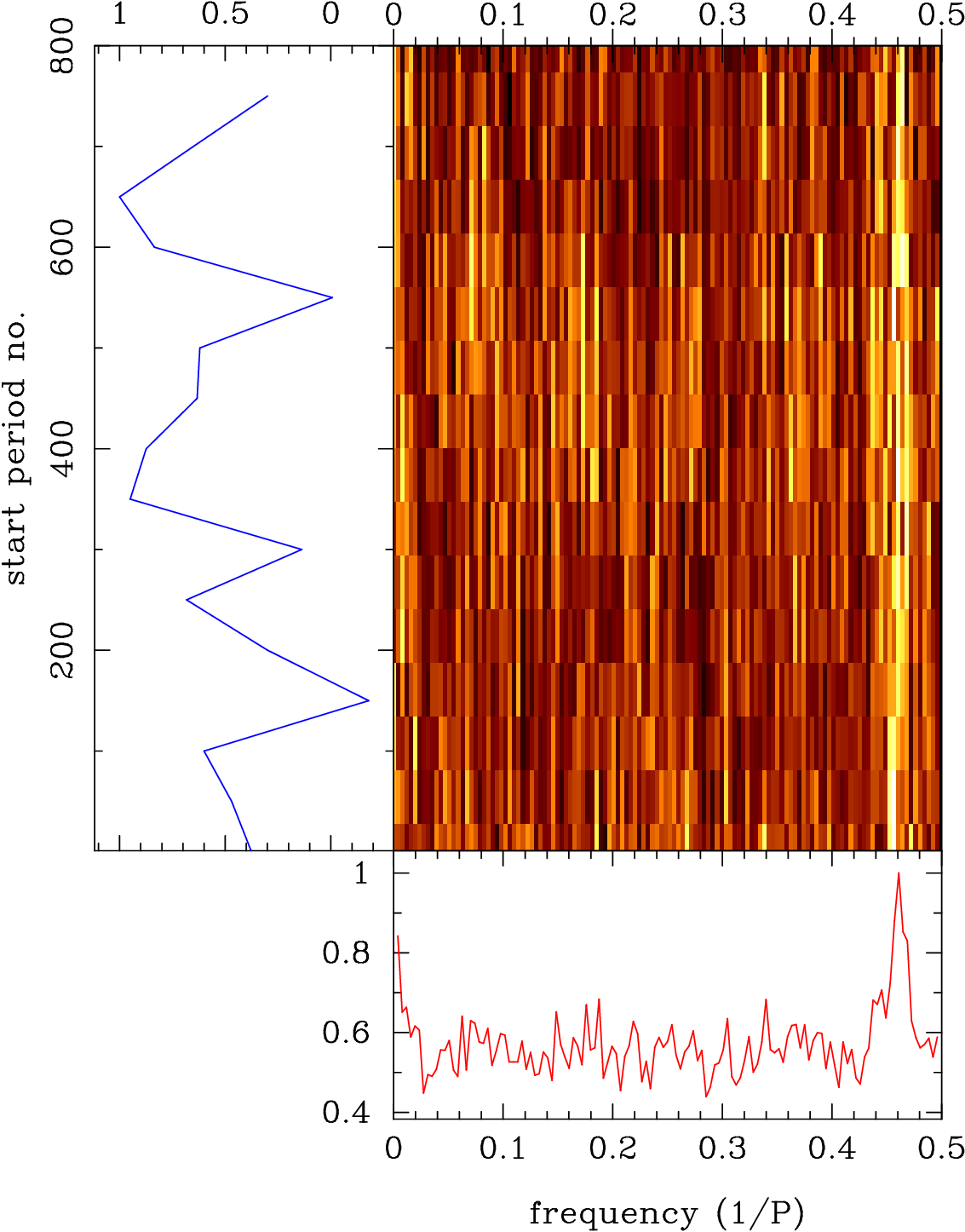}}} &
\hspace{50px}
{\mbox{\includegraphics[scale=0.4,angle=0.]{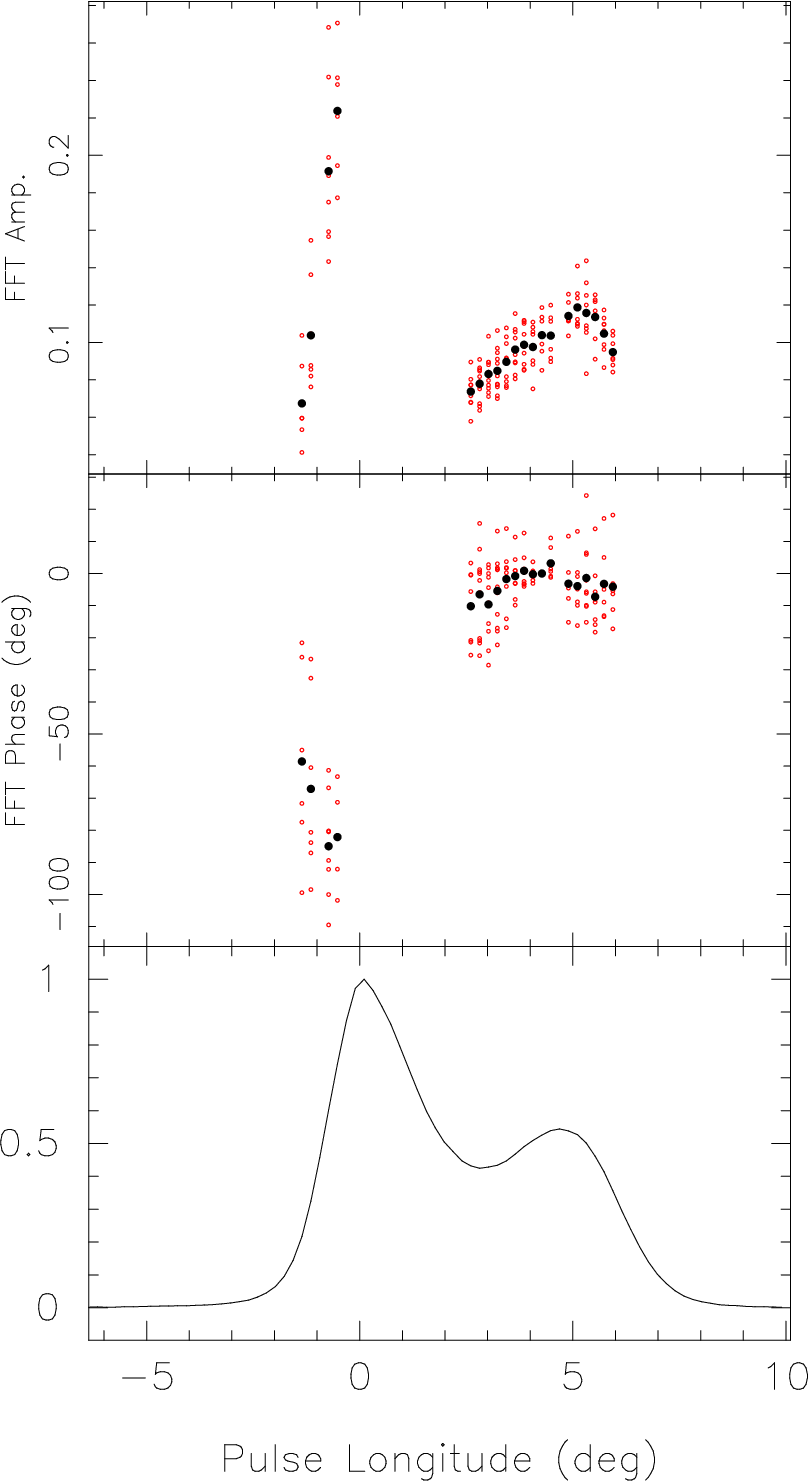}}} \\
\end{tabular}
\caption{The fluctuation spectral analysis for PSR J0837+0610 at 333 MHz. The
left panel shows time variation of LRFS. The right panel shows the variation of
the peak frequency across the pulse window; peak amplitude (top window), phase
variations (middle window), profile (bottom window).}
\label{fig_J0837_1}
\end{figure*}

% J0837+0610 - 618 MHz
\begin{figure*}
\begin{tabular}{@{}lr@{}}
{\mbox{\includegraphics[scale=0.4,angle=0.]{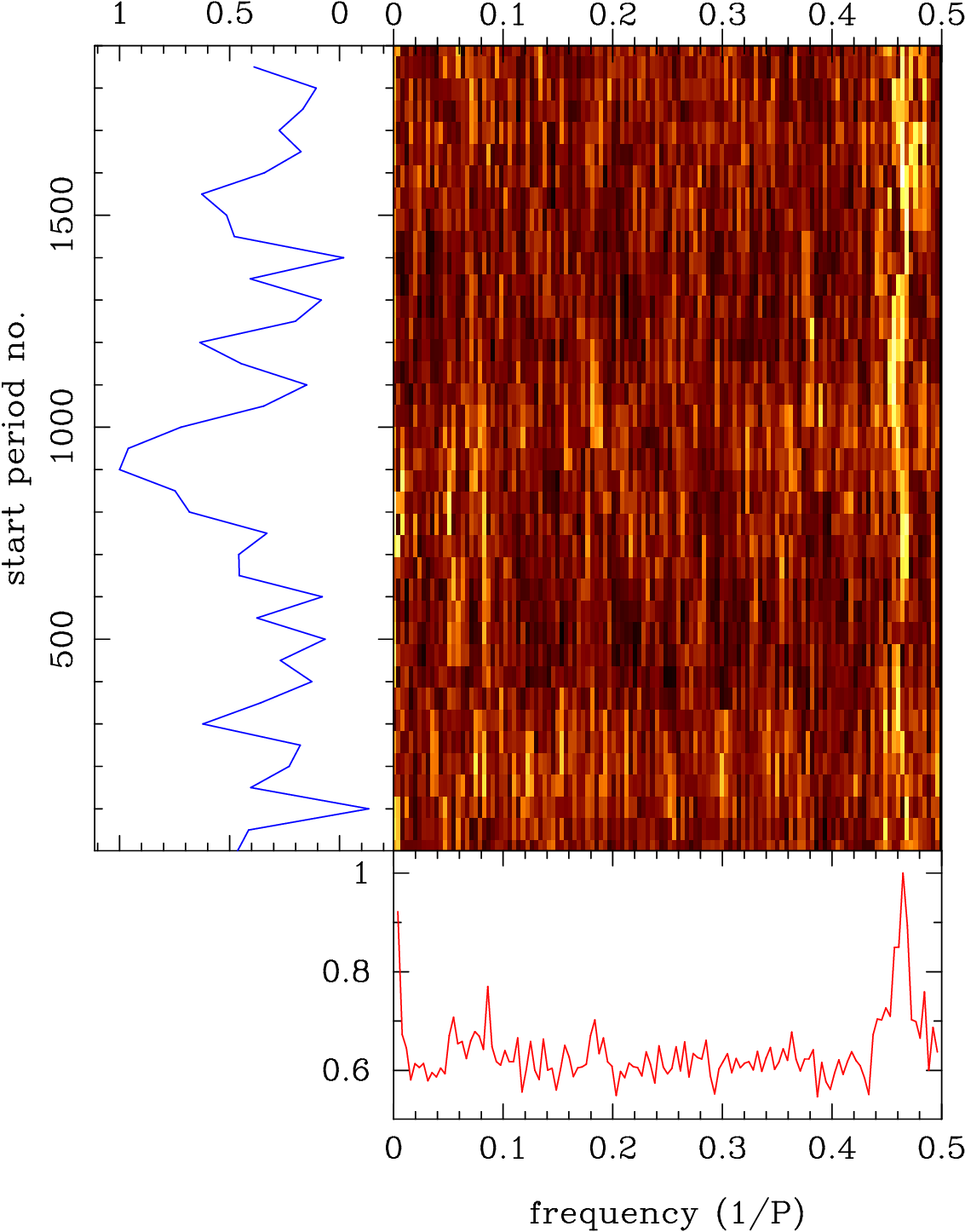}}} &
\hspace{50px}
{\mbox{\includegraphics[scale=0.4,angle=0.]{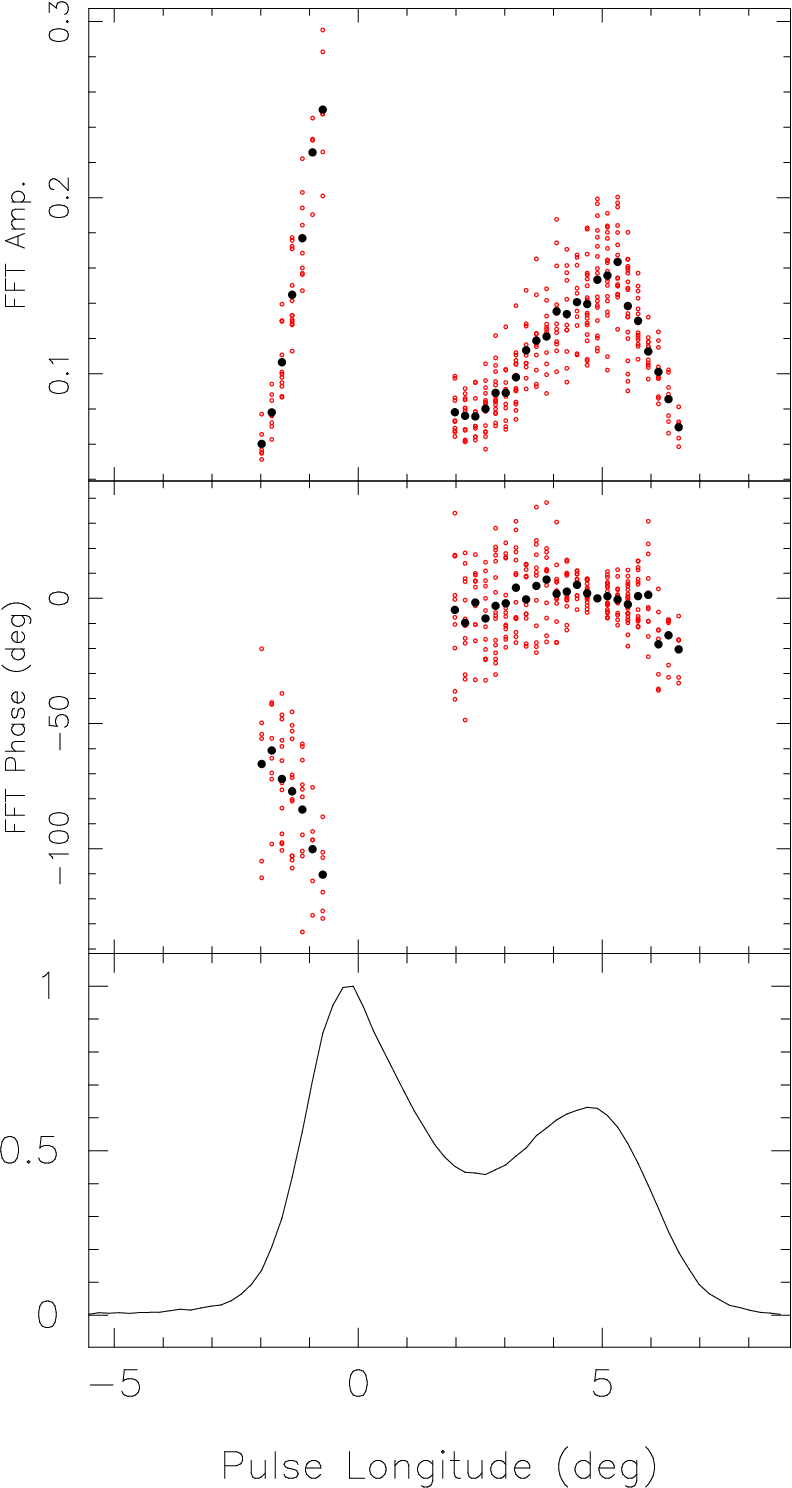}}} \\
\end{tabular}
\caption{The fluctuation spectral analysis for PSR J0837+0610 at 618 MHz. The
left panel shows time variation of LRFS. The right panel shows the variation of
the peak frequency across the pulse window; peak amplitude (top window), phase
variations (middle window), profile (bottom window).}
\label{fig_J0837_2}
\end{figure*}

% J0846-3533 - 333 MHz
\begin{figure*}
\begin{tabular}{@{}lr@{}}
{\mbox{\includegraphics[scale=0.4,angle=0.]{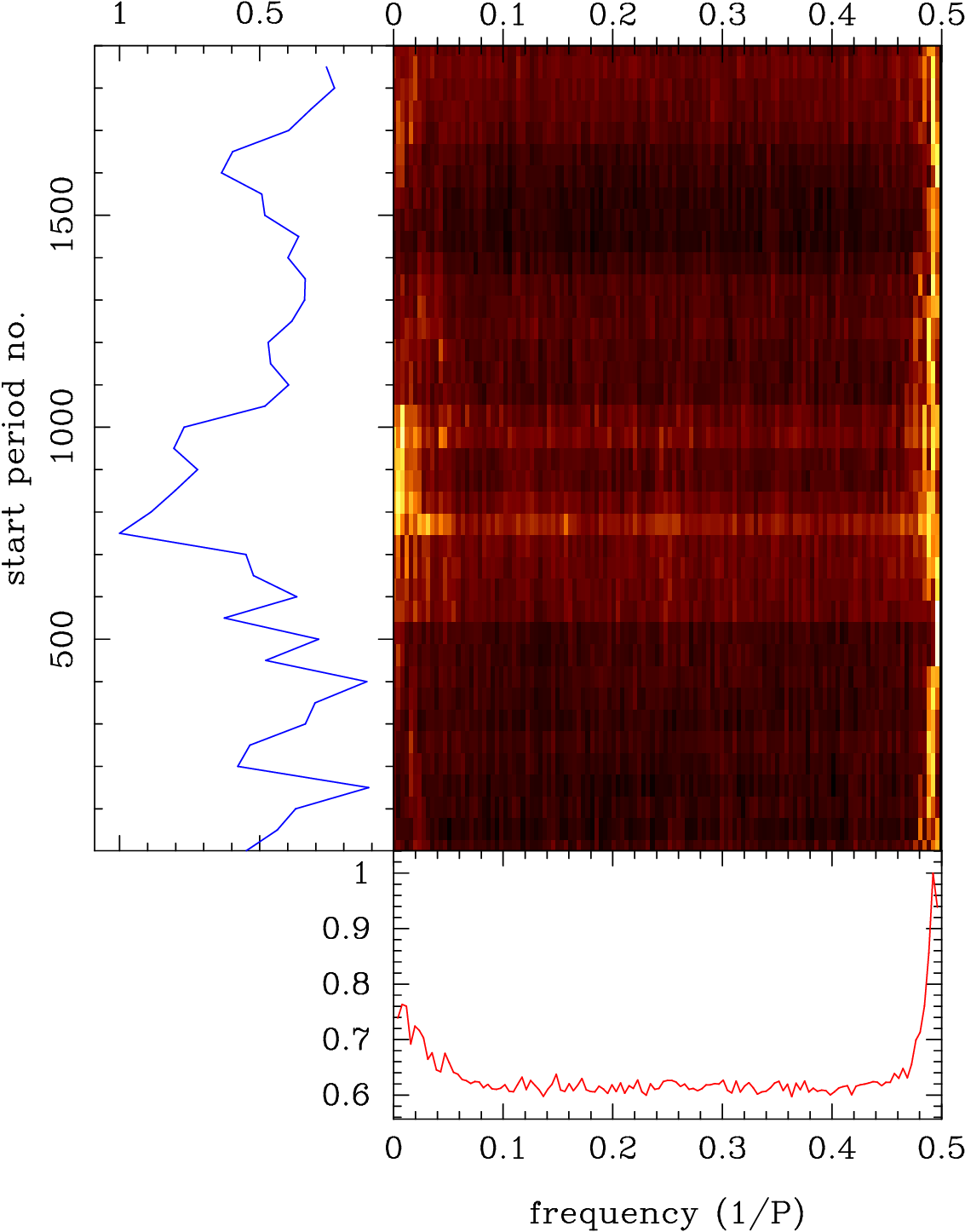}}} &
\hspace{50px}
{\mbox{\includegraphics[scale=0.4,angle=0.]{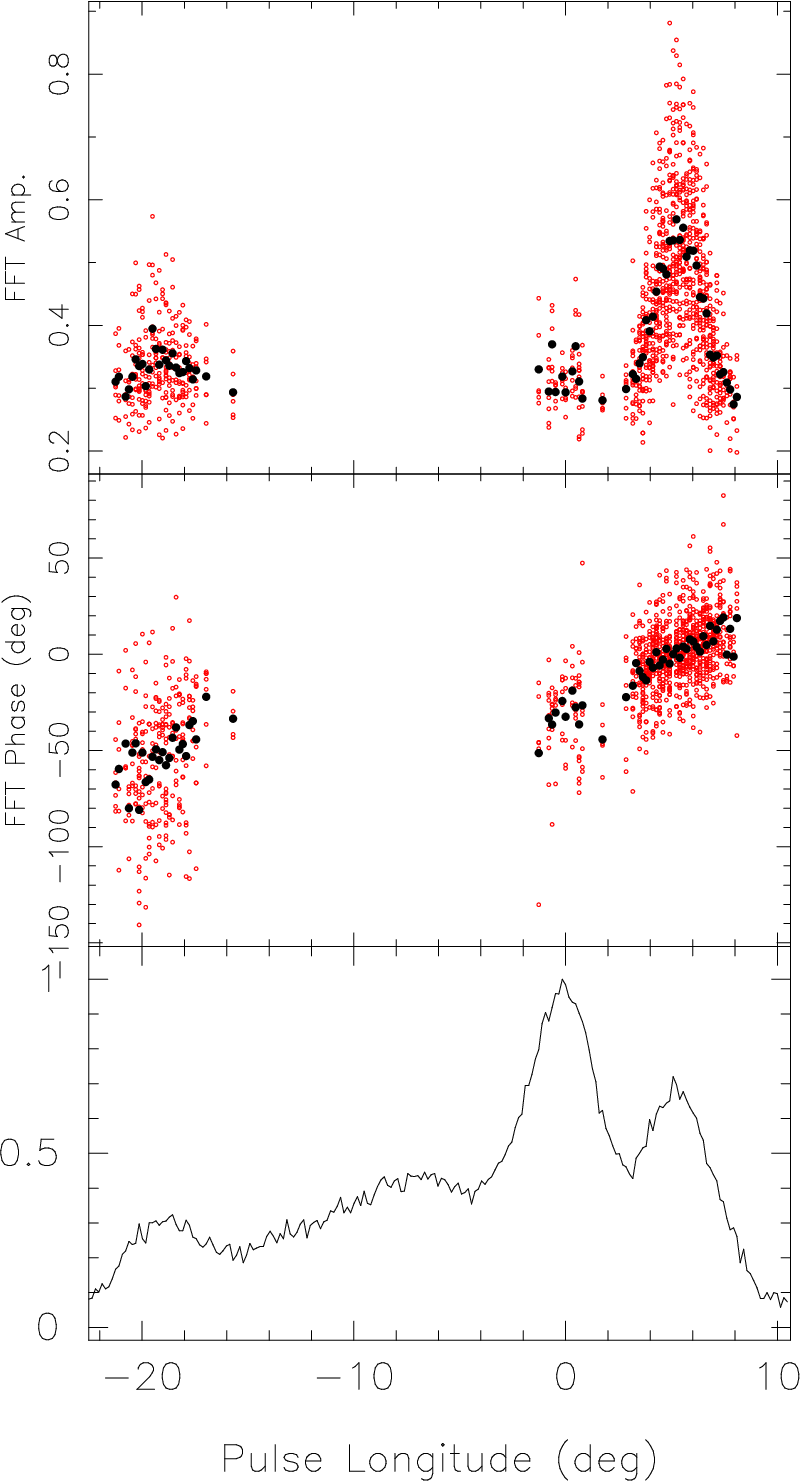}}} \\
\end{tabular}
\caption{The fluctuation spectral analysis for PSR J0846$-$3533 at 333 MHz. The
left panel shows time variation of LRFS. The right panel shows the variation of
the peak frequency across the pulse window; peak amplitude (top window), phase
variations (middle window), profile (bottom window).}
\label{fig_J0846_1}
\end{figure*}

% J0846-3533 - 618 MHz
\begin{figure*}
\begin{tabular}{@{}lr@{}}
{\mbox{\includegraphics[scale=0.4,angle=0.]{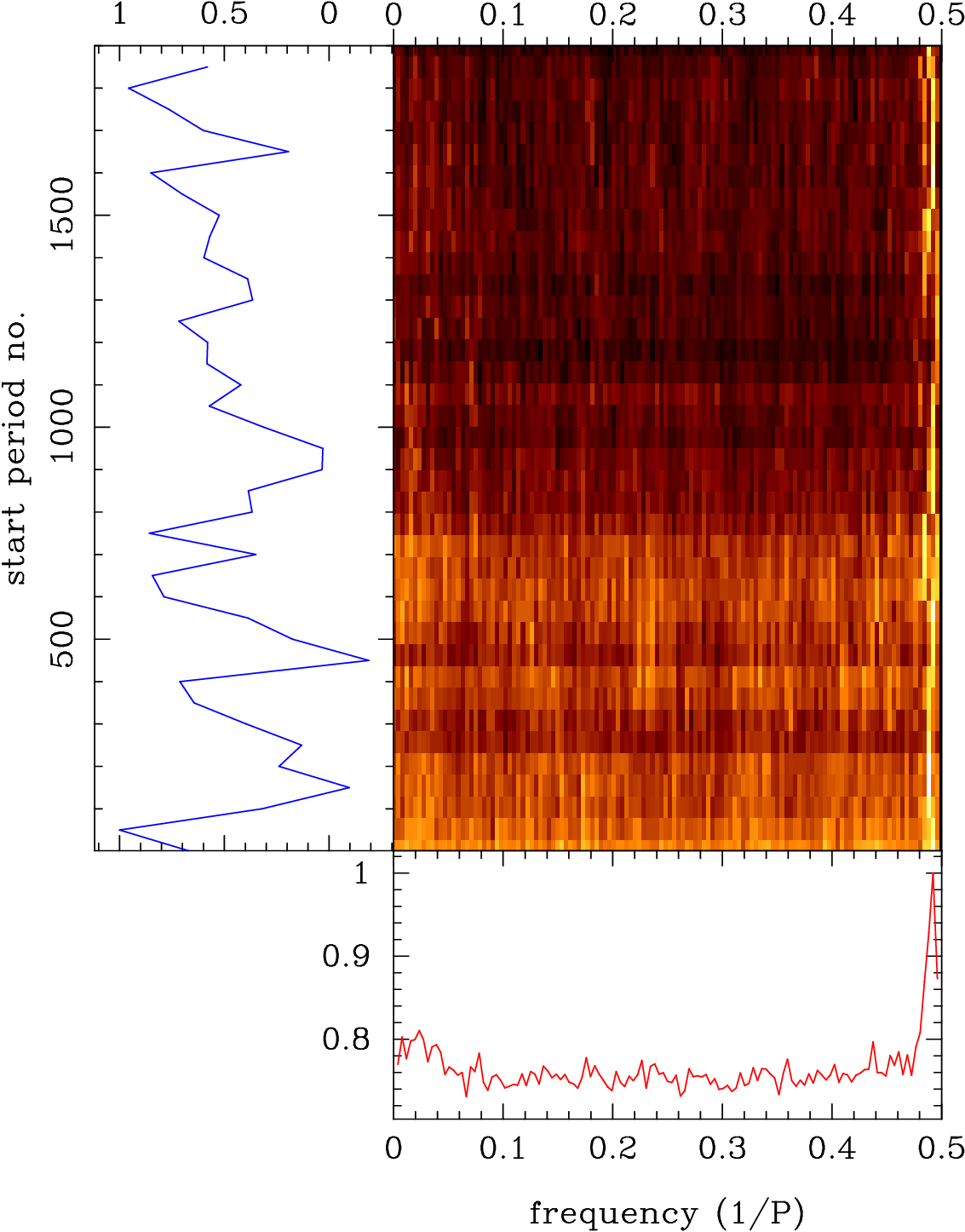}}} &
\hspace{50px}
{\mbox{\includegraphics[scale=0.4,angle=0.]{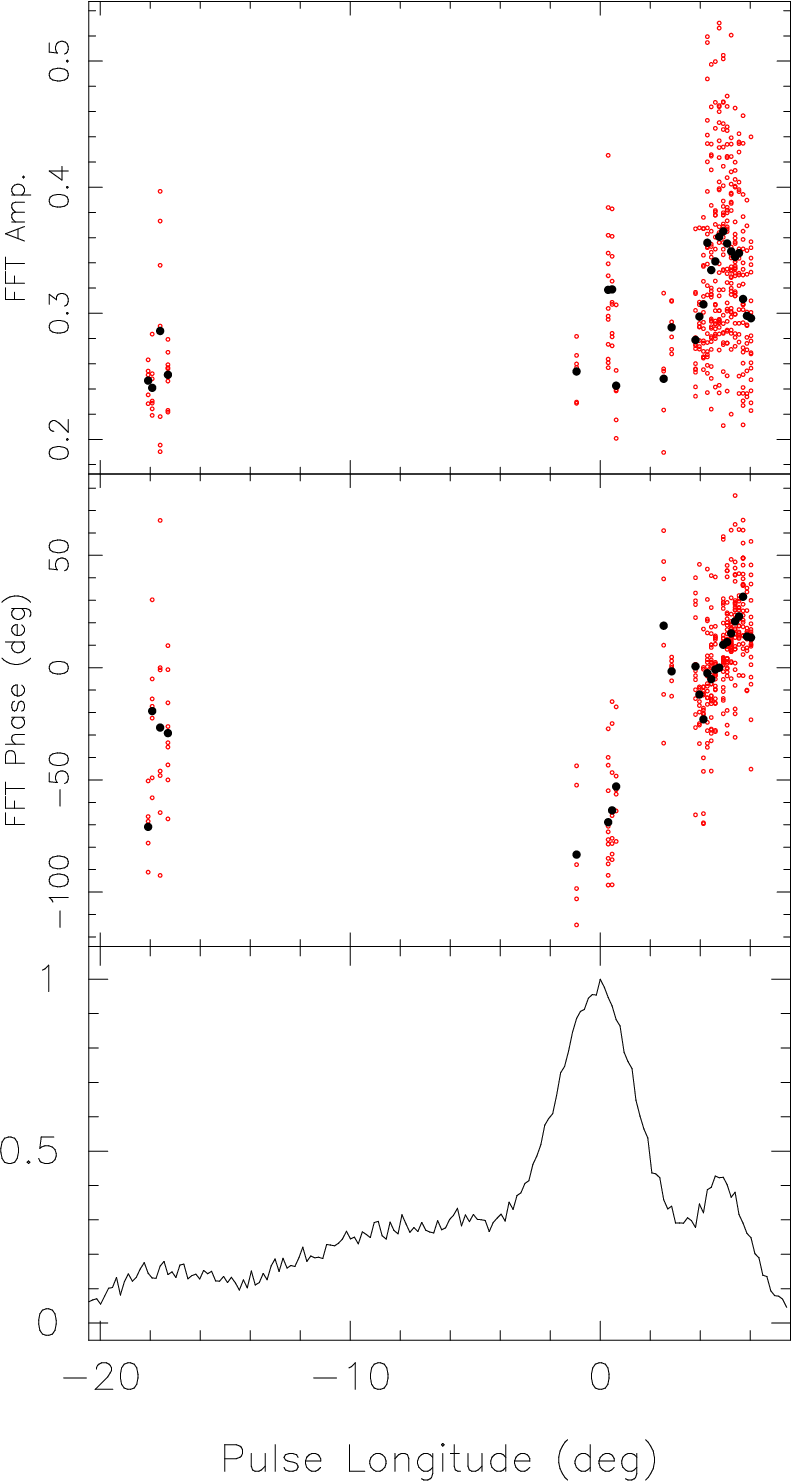}}} \\
\end{tabular}
\caption{The fluctuation spectral analysis for PSR J0846$-$3533 at 618 MHz. The
left panel shows time variation of LRFS. The right panel shows the variation of
the peak frequency across the pulse window; peak amplitude (top window), phase
variations (middle window), profile (bottom window).}
\label{fig_J0846_2}
\end{figure*}

% J1239+2453 - 333 MHz
\begin{figure*}
\begin{tabular}{@{}lr@{}}
{\mbox{\includegraphics[scale=0.4,angle=0.]{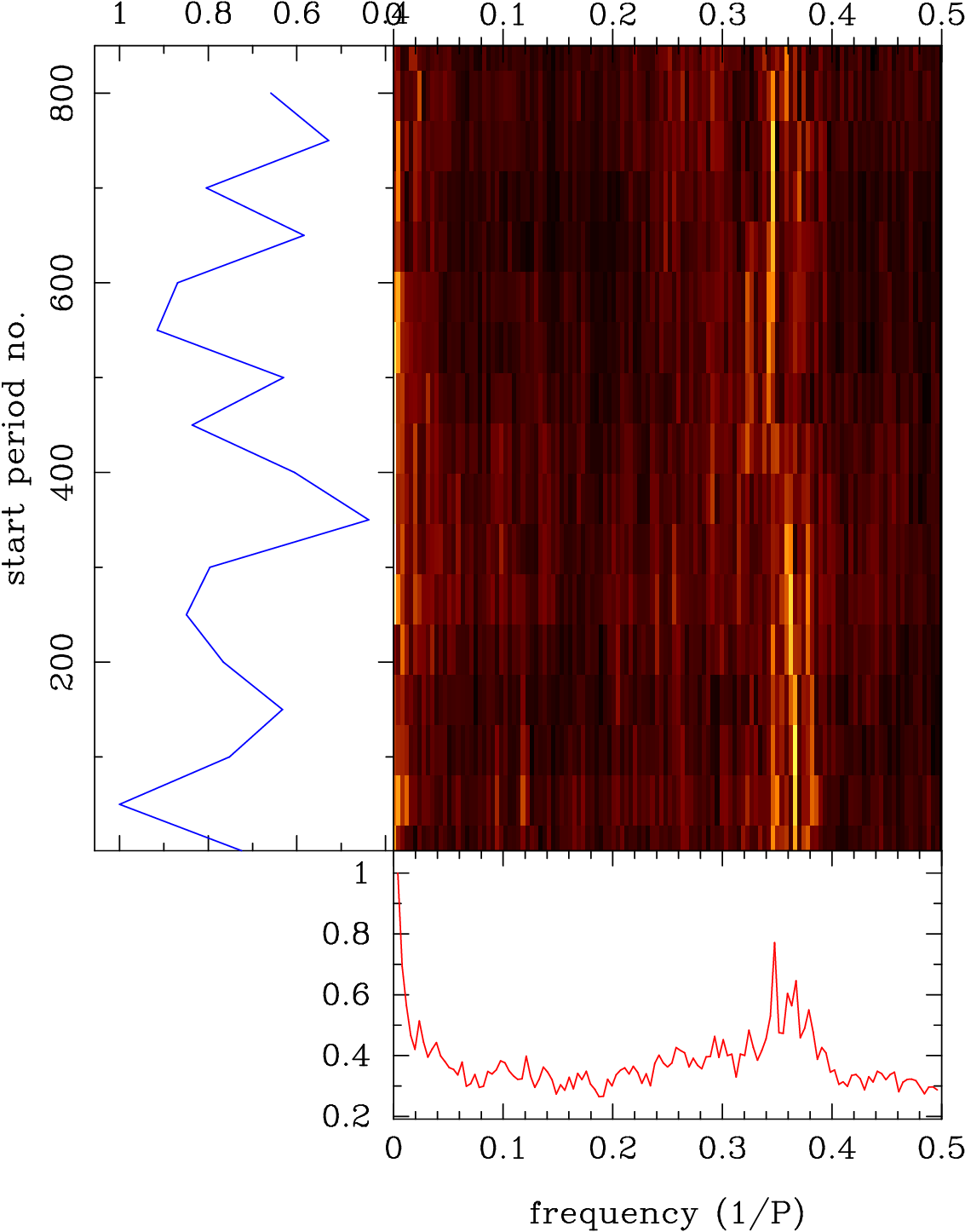}}} &
\hspace{50px}
{\mbox{\includegraphics[scale=0.4,angle=0.]{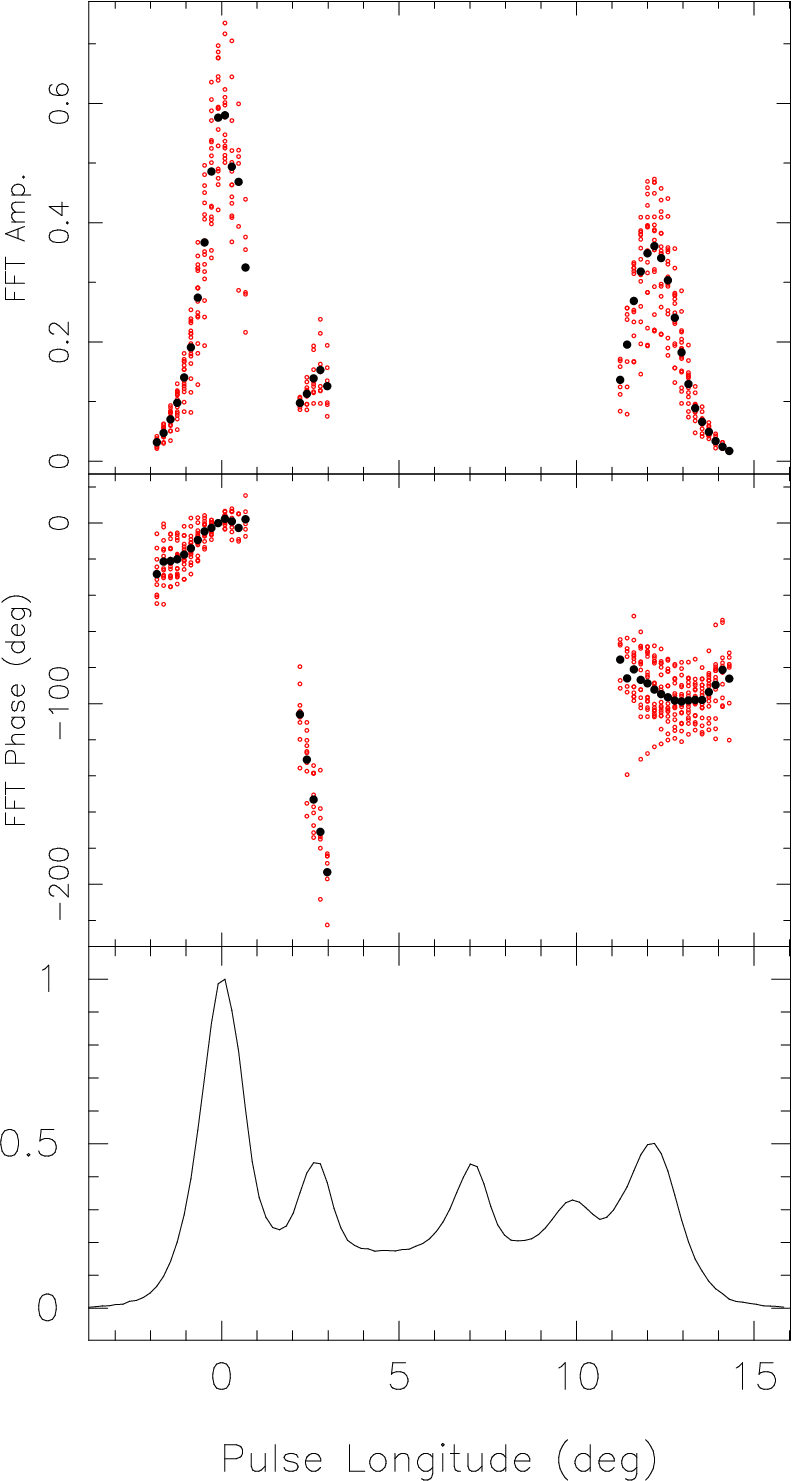}}} \\
\end{tabular}
\caption{The fluctuation spectral analysis for PSR J1239+2453 at 333 MHz. The
left panel shows time variation of LRFS. The right panel shows the variation of
the peak frequency across the pulse window; peak amplitude (top window), phase
variations (middle window), profile (bottom window).}
\label{fig_J1239_1}
\end{figure*}

% J1239+2453 - 618 MHz
\begin{figure*}
\begin{tabular}{@{}lr@{}}
{\mbox{\includegraphics[scale=0.4,angle=0.]{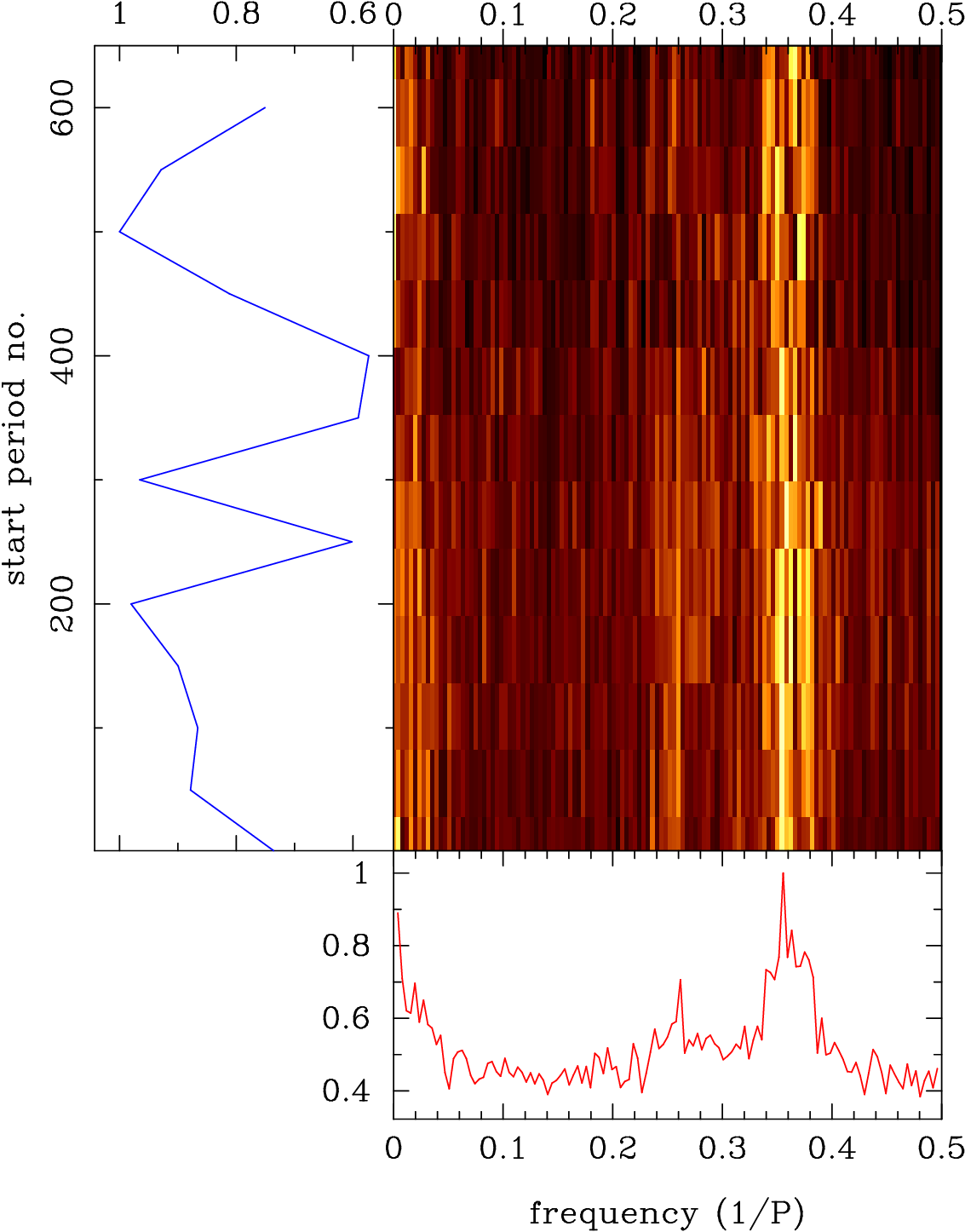}}} &
\hspace{50px}
{\mbox{\includegraphics[scale=0.4,angle=0.]{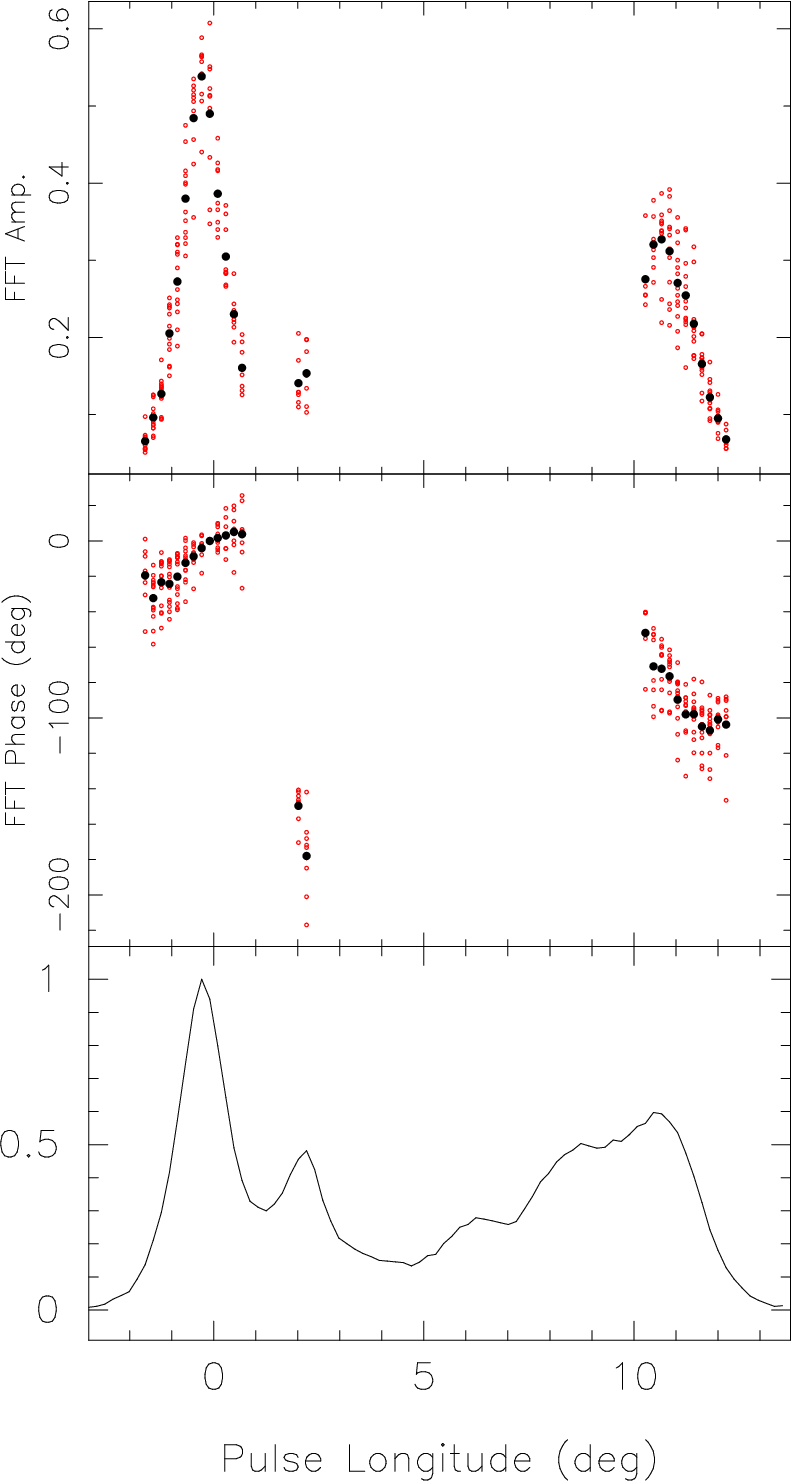}}} \\
\end{tabular}
\caption{The fluctuation spectral analysis for PSR J1239+2453 at 618 MHz. The
left panel shows time variation of LRFS. The right panel shows the variation of
the peak frequency across the pulse window; peak amplitude (top window), phase
variations (middle window), profile (bottom window).}
\label{fig_J1239_2}
\end{figure*}

% J1328-4921 - 333 MHz
\begin{figure*}
\begin{tabular}{@{}lr@{}}
{\mbox{\includegraphics[scale=0.4,angle=0.]{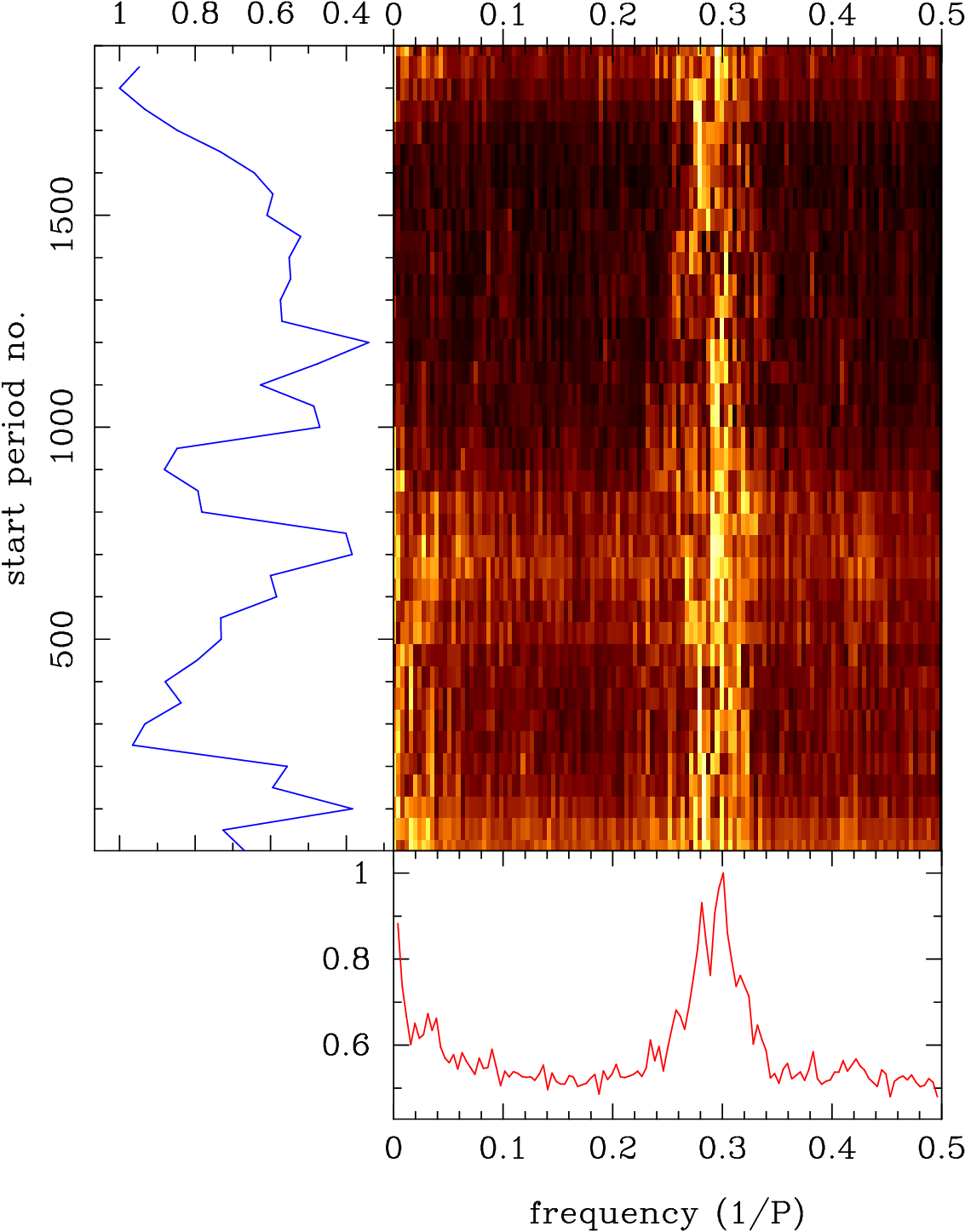}}} &
\hspace{50px}
{\mbox{\includegraphics[scale=0.4,angle=0.]{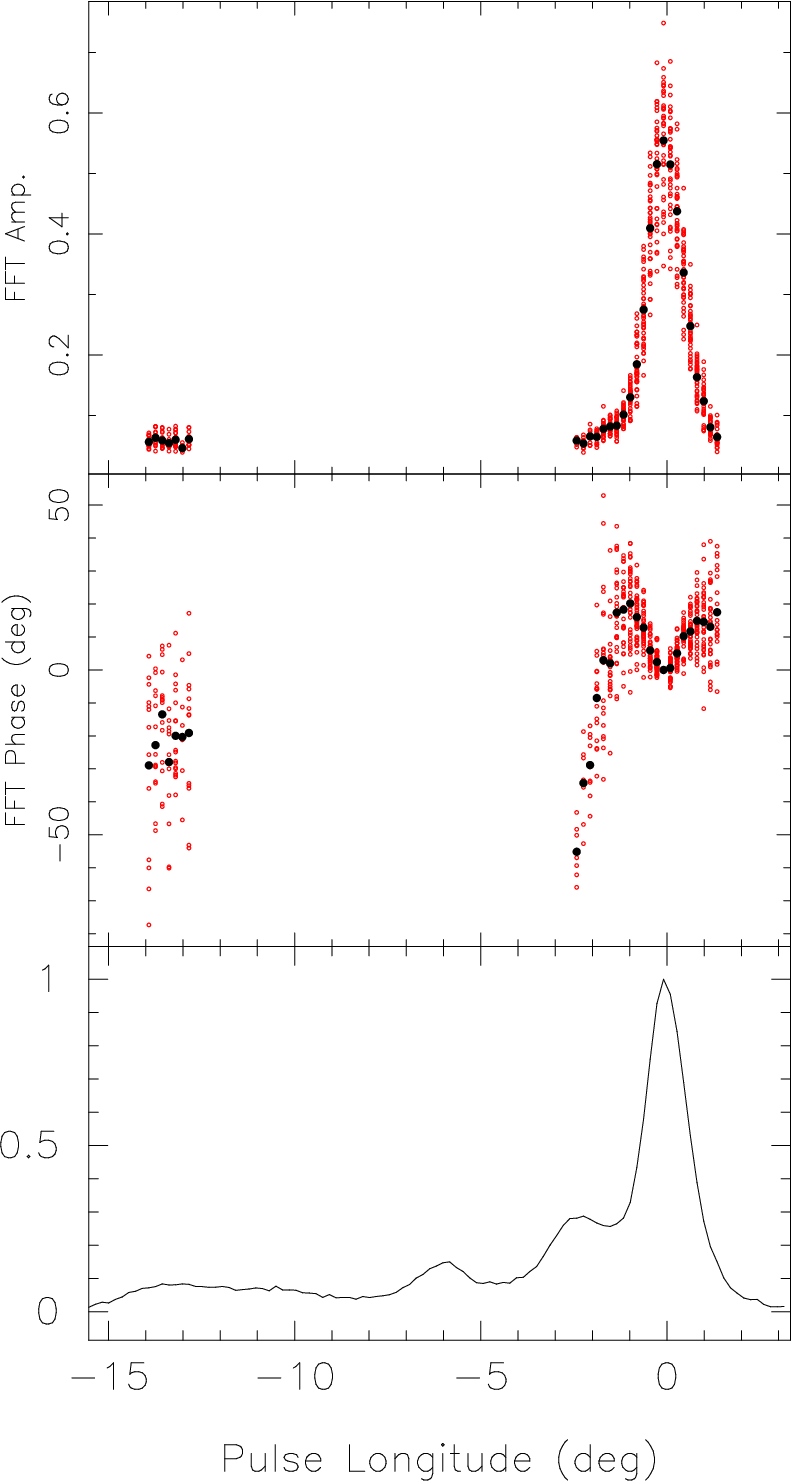}}} \\
\end{tabular}
\caption{The fluctuation spectral analysis for PSR J1328$-$4921 at 333 MHz. The
left panel shows time variation of LRFS. The right panel shows the variation of
the peak frequency across the pulse window; peak amplitude (top window), phase
variations (middle window), profile (bottom window).}
\label{fig_J1328_1}
\end{figure*}

% J1328-4921 - 618 MHz
\begin{figure*}
\begin{tabular}{@{}lr@{}}
{\mbox{\includegraphics[scale=0.4,angle=0.]{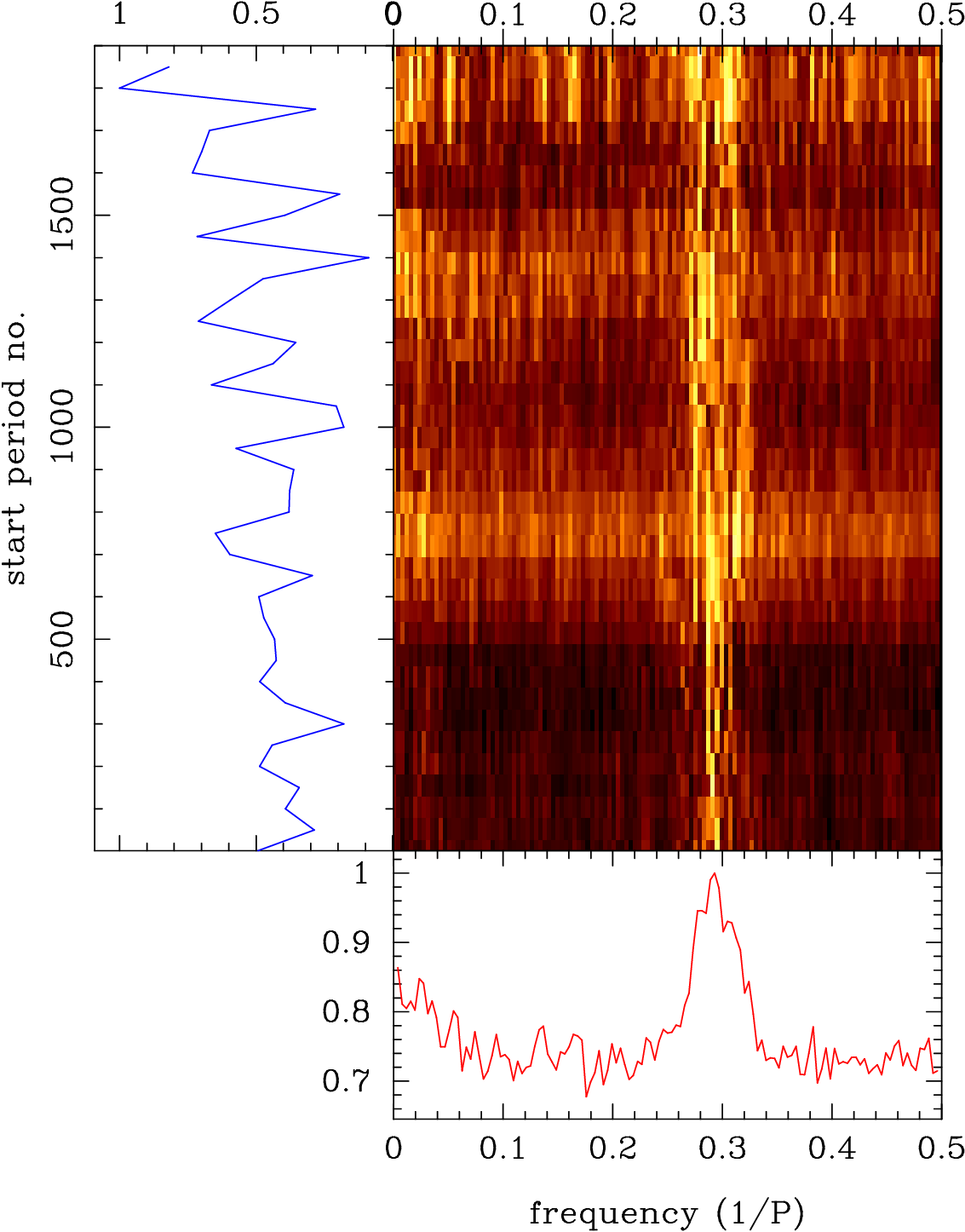}}} &
\hspace{50px}
{\mbox{\includegraphics[scale=0.4,angle=0.]{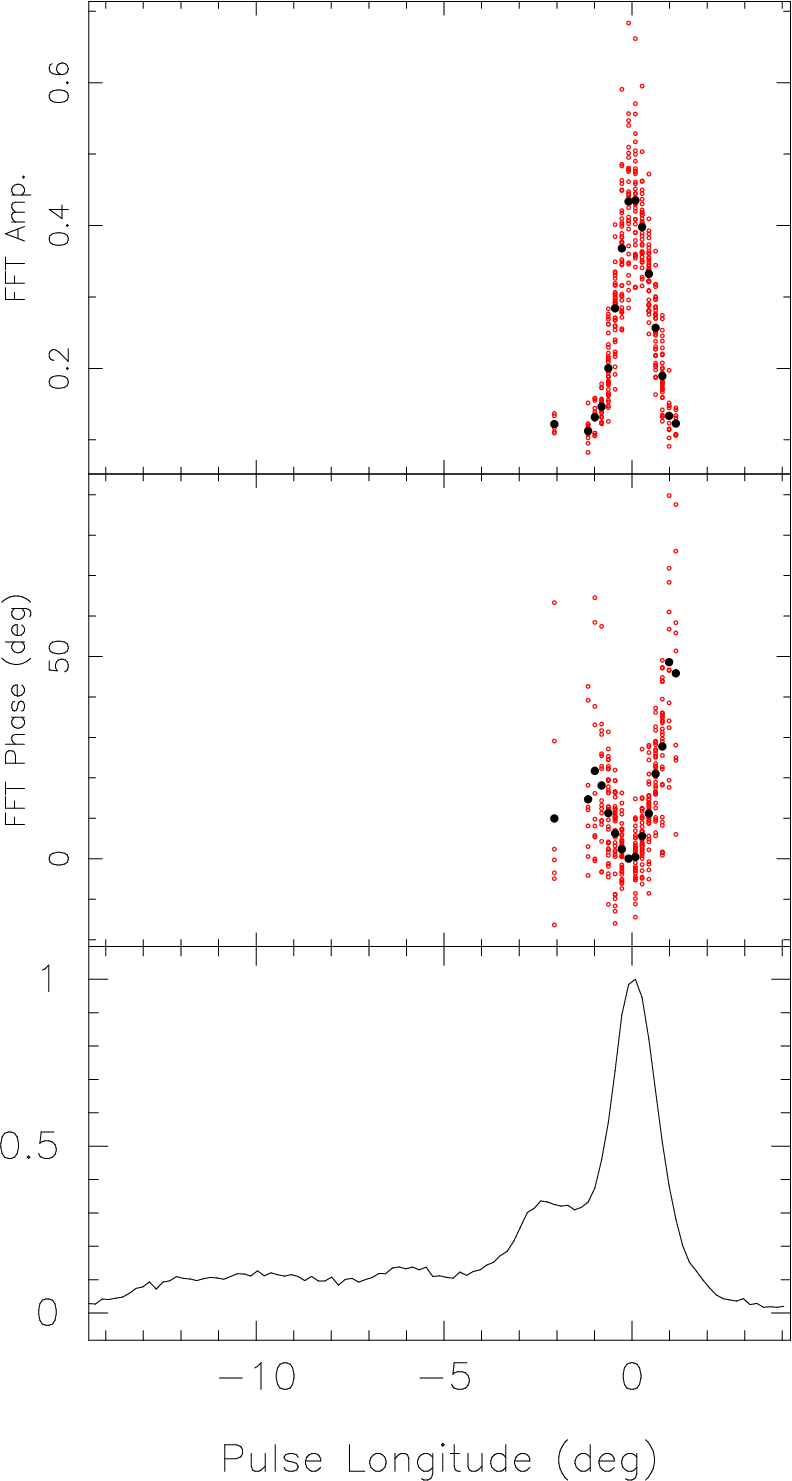}}} \\
\end{tabular}
\caption{The fluctuation spectral analysis for PSR J1328$-$4921 at 618 MHz. The
left panel shows time variation of LRFS. The right panel shows the variation of
the peak frequency across the pulse window; peak amplitude (top window), phase
variations (middle window), profile (bottom window).}
\label{fig_J1328_2}
\end{figure*}

% J1625-4048 - 618 MHz
\begin{figure*}
\begin{tabular}{@{}lr@{}}
{\mbox{\includegraphics[scale=0.4,angle=0.]{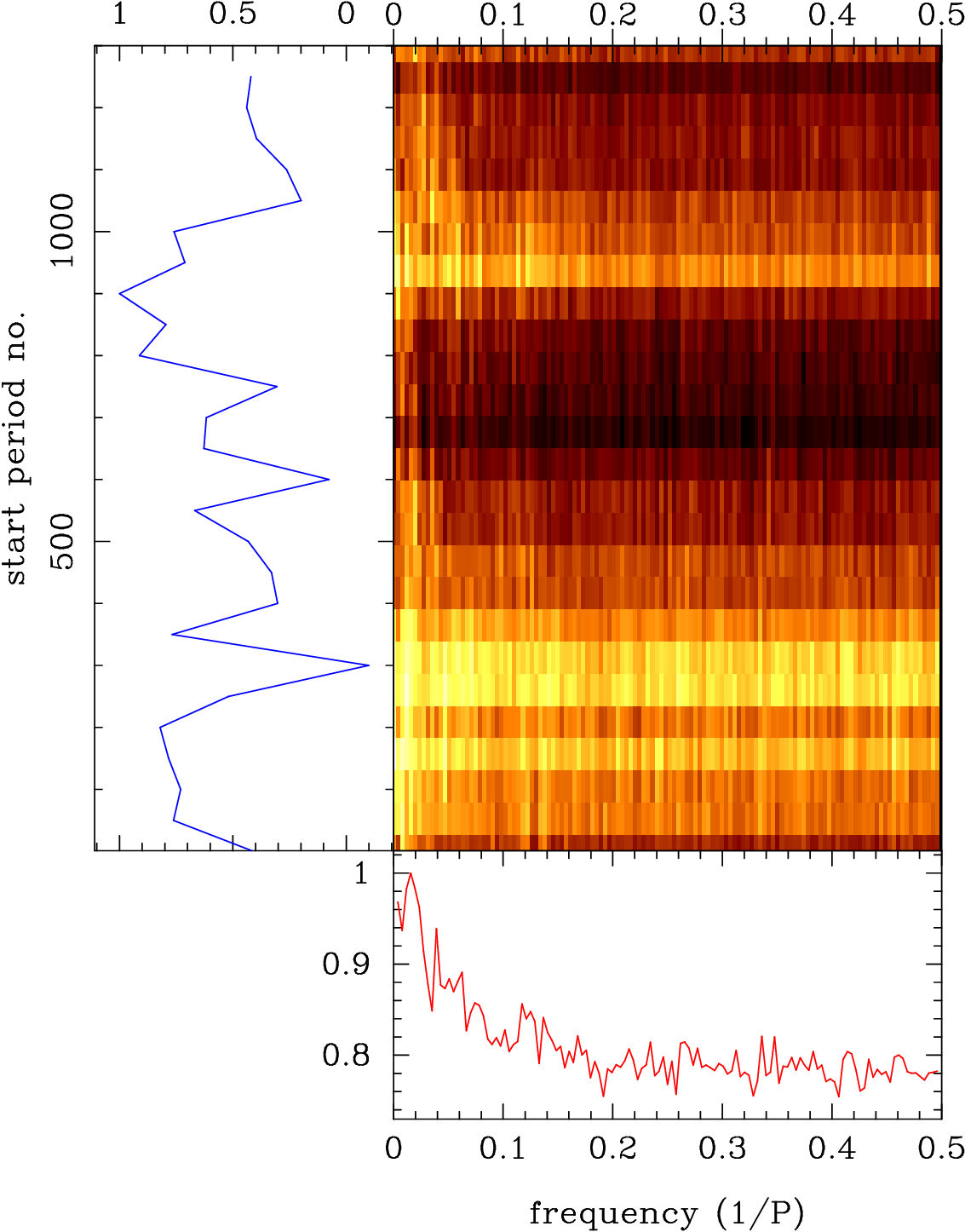}}} &
\hspace{50px}
{\mbox{\includegraphics[scale=0.4,angle=0.]{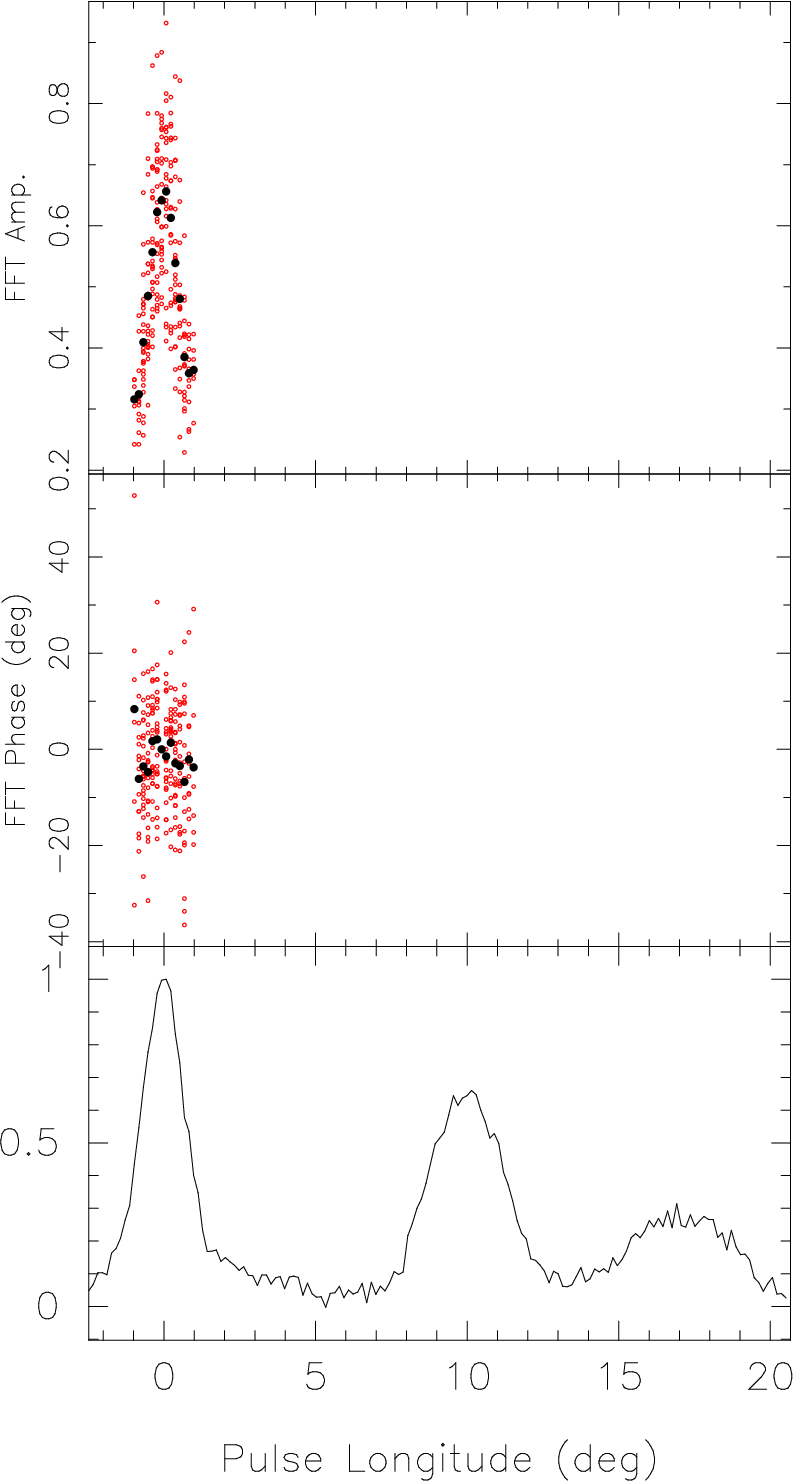}}} \\
\end{tabular}
\caption{The fluctuation spectral analysis for PSR J1625$-$4048 at 618 MHz. The
left panel shows time variation of LRFS. The right panel shows the variation of
the peak frequency across the pulse window; peak amplitude (top window), phase
variations (middle window), profile (bottom window).}
\label{fig_J1625}
\end{figure*}

% J1650-1654 - 339 MHz
\begin{figure*}
\begin{tabular}{@{}lr@{}}
{\mbox{\includegraphics[scale=0.4,angle=0.]{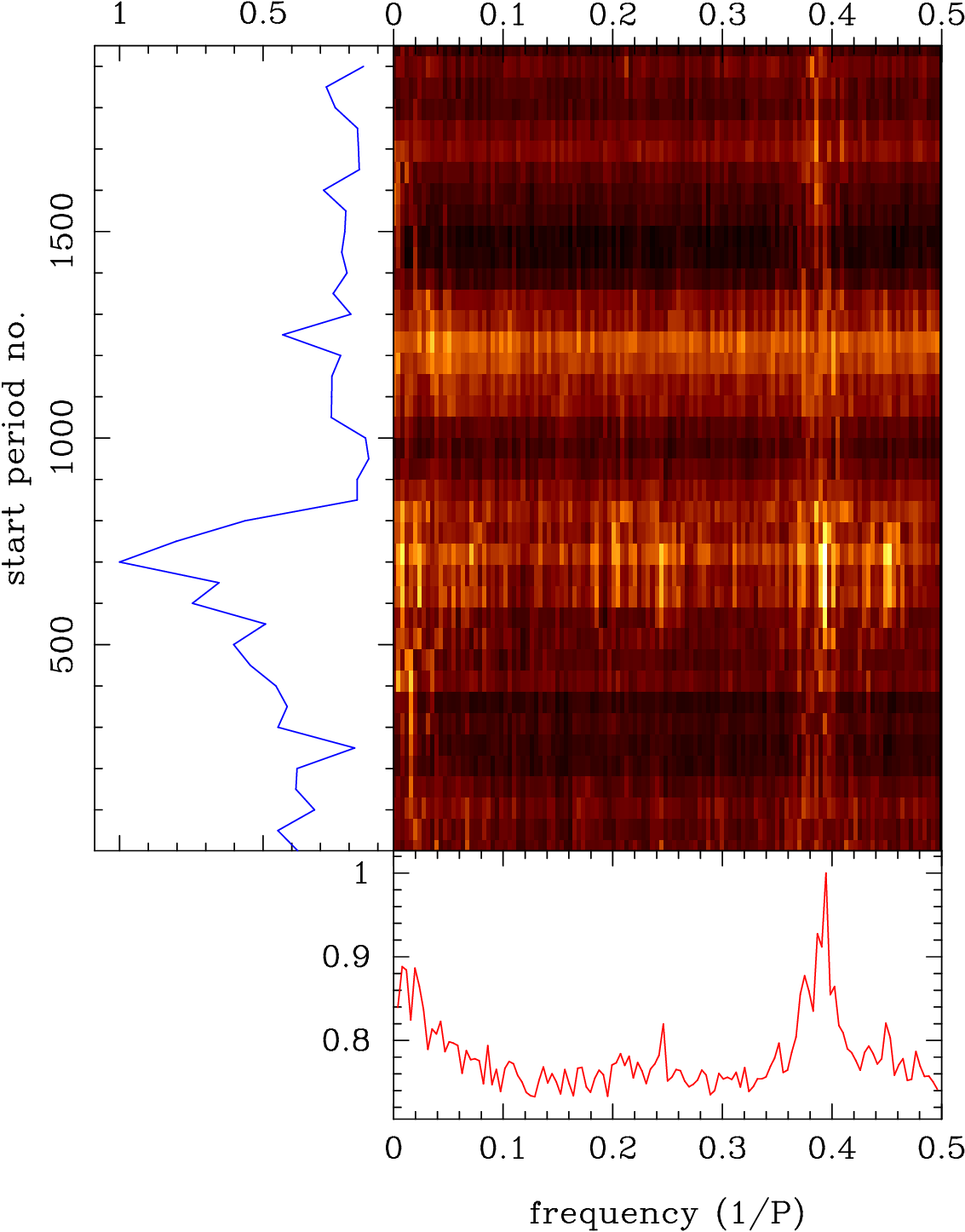}}} &
\hspace{50px}
{\mbox{\includegraphics[scale=0.4,angle=0.]{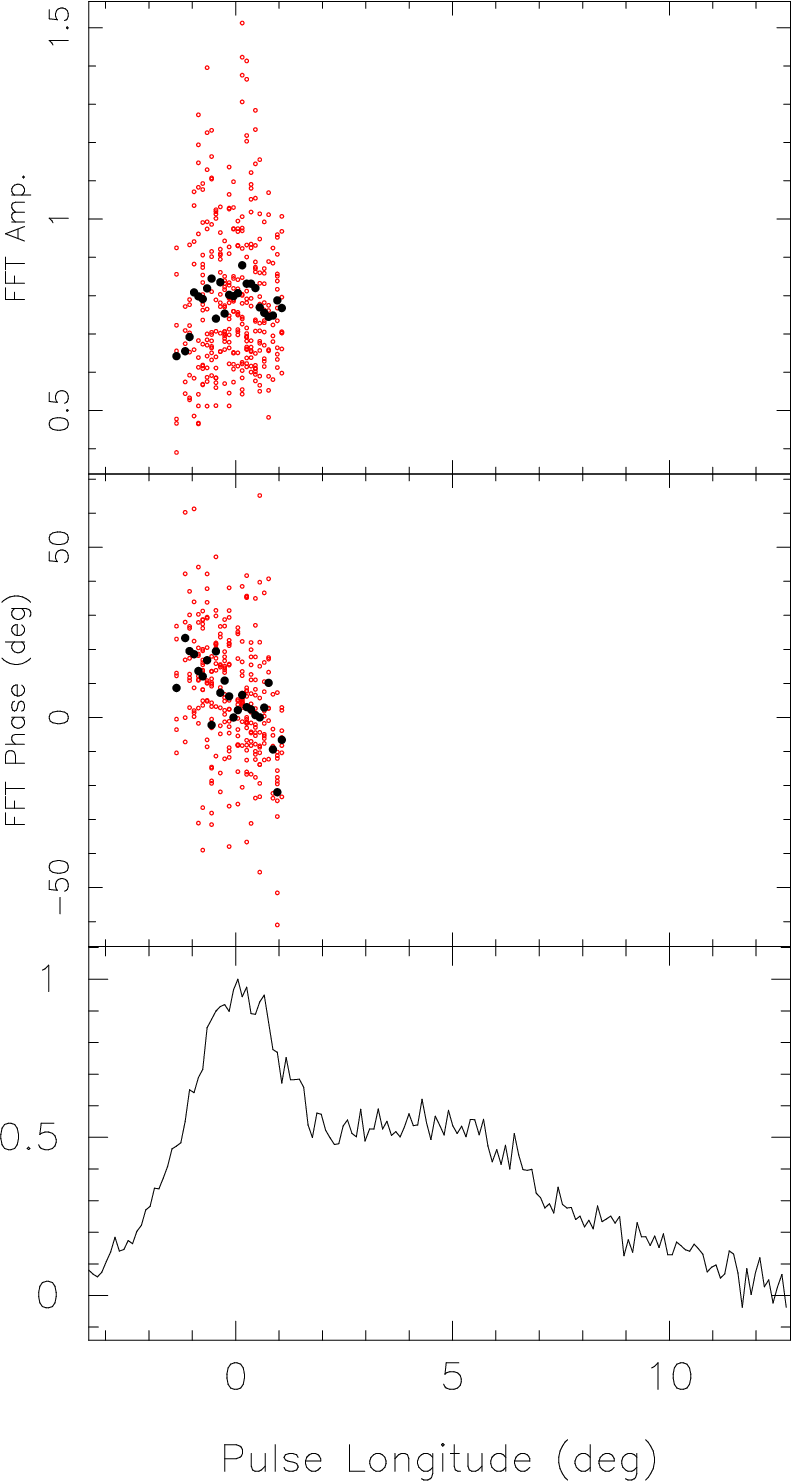}}} \\
\end{tabular}
\caption{The fluctuation spectral analysis for PSR J1650$-$1654 at 339 MHz. The
left panel shows time variation of LRFS. The right panel shows the variation of
the peak frequency across the pulse window; peak amplitude (top window), phase
variations (middle window), profile (bottom window).}
\label{fig_J1650}
\end{figure*}

% J1733-2228 - 333 MHz
\begin{figure*}
\begin{tabular}{@{}lr@{}}
{\mbox{\includegraphics[scale=0.4,angle=0.]{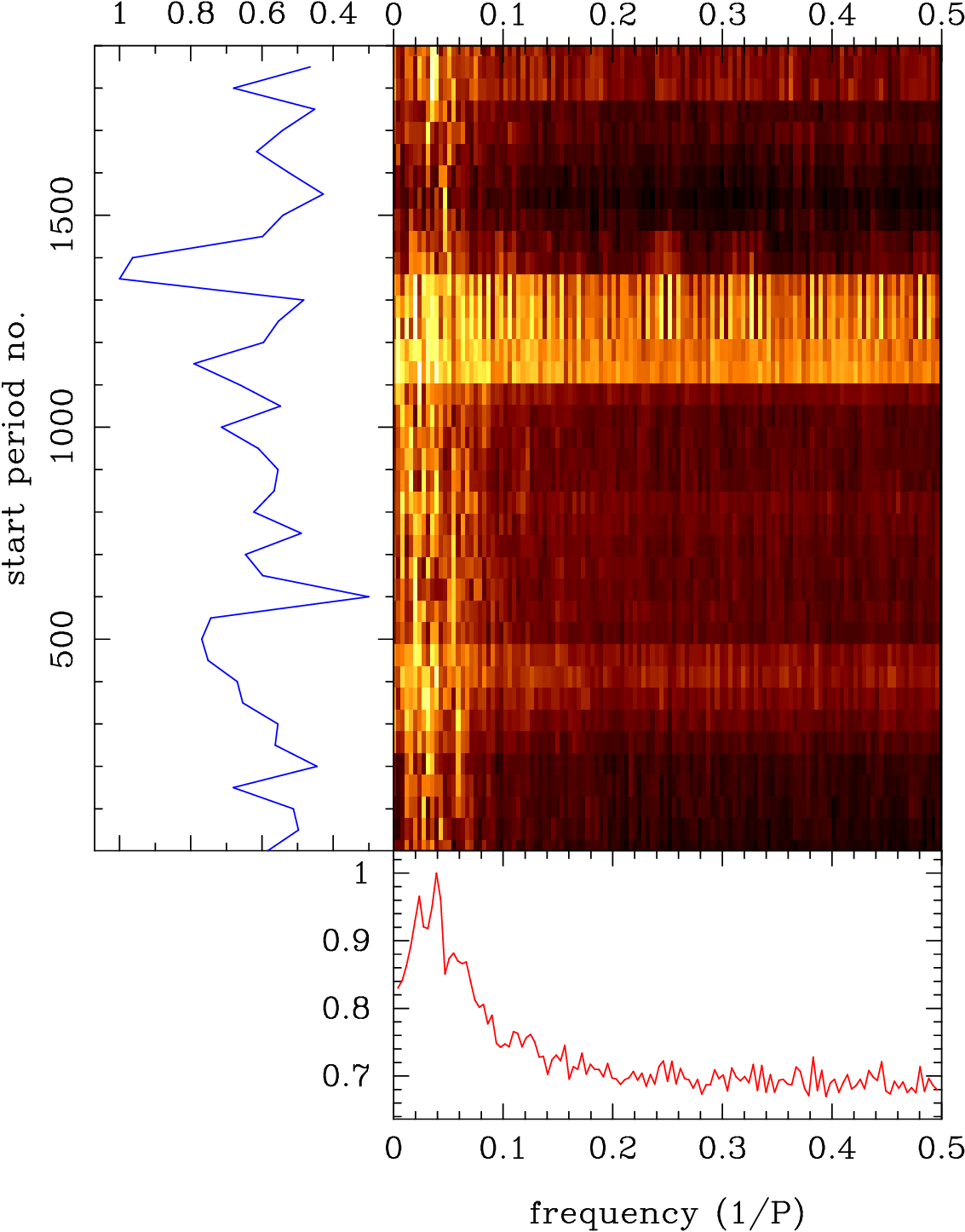}}} &
\hspace{50px}
{\mbox{\includegraphics[scale=0.4,angle=0.]{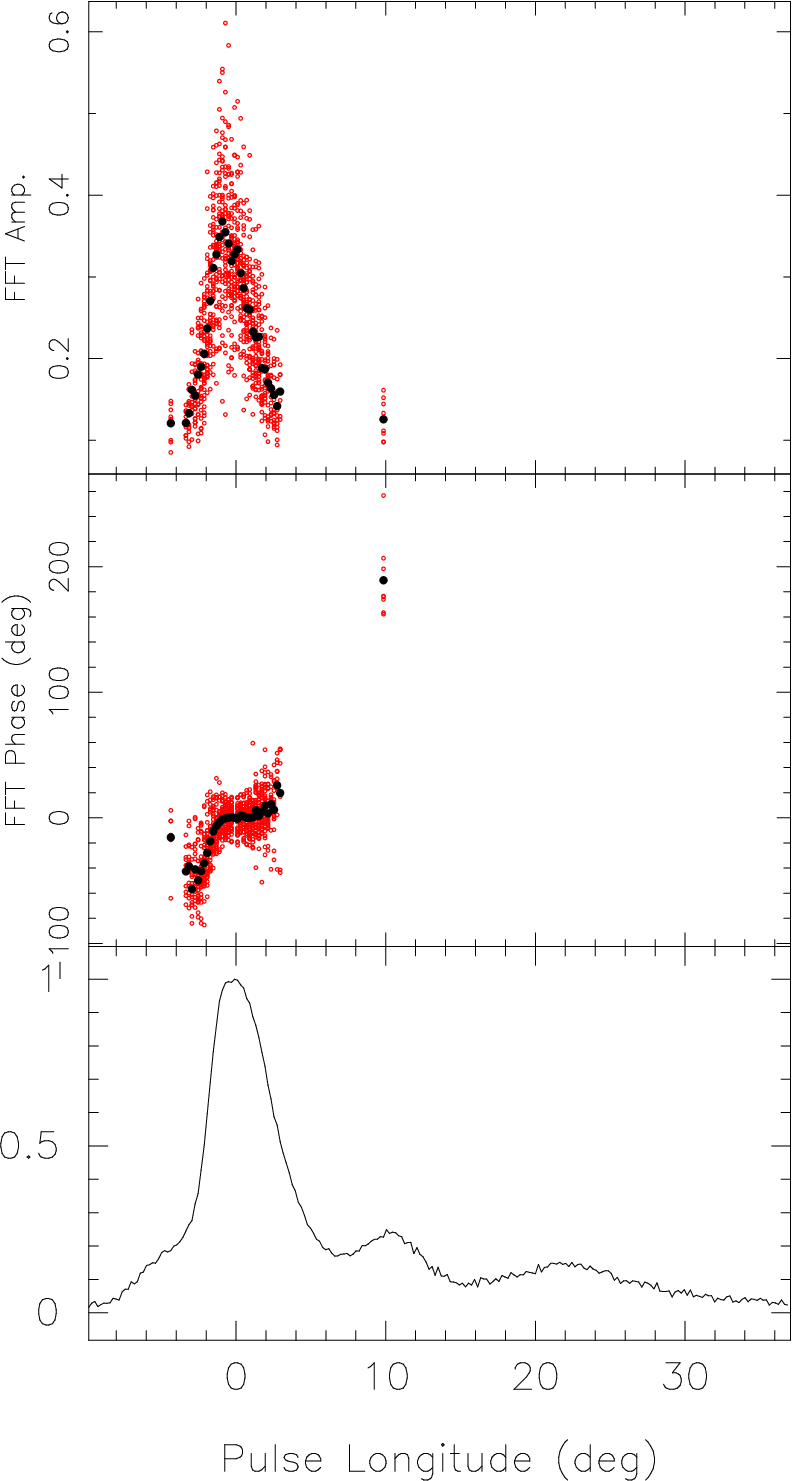}}} \\
\end{tabular}
\caption{The fluctuation spectral analysis for PSR J1733$-$2228 at 333 MHz. The
left panel shows time variation of LRFS. The right panel shows the variation of
the peak frequency across the pulse window; peak amplitude (top window), phase
variations (middle window), profile (bottom window).}
\label{fig_J1733}
\end{figure*}

% J1900-2600 - 325 MHz
\begin{figure*}
\begin{tabular}{@{}lr@{}}
{\mbox{\includegraphics[scale=0.4,angle=0.]{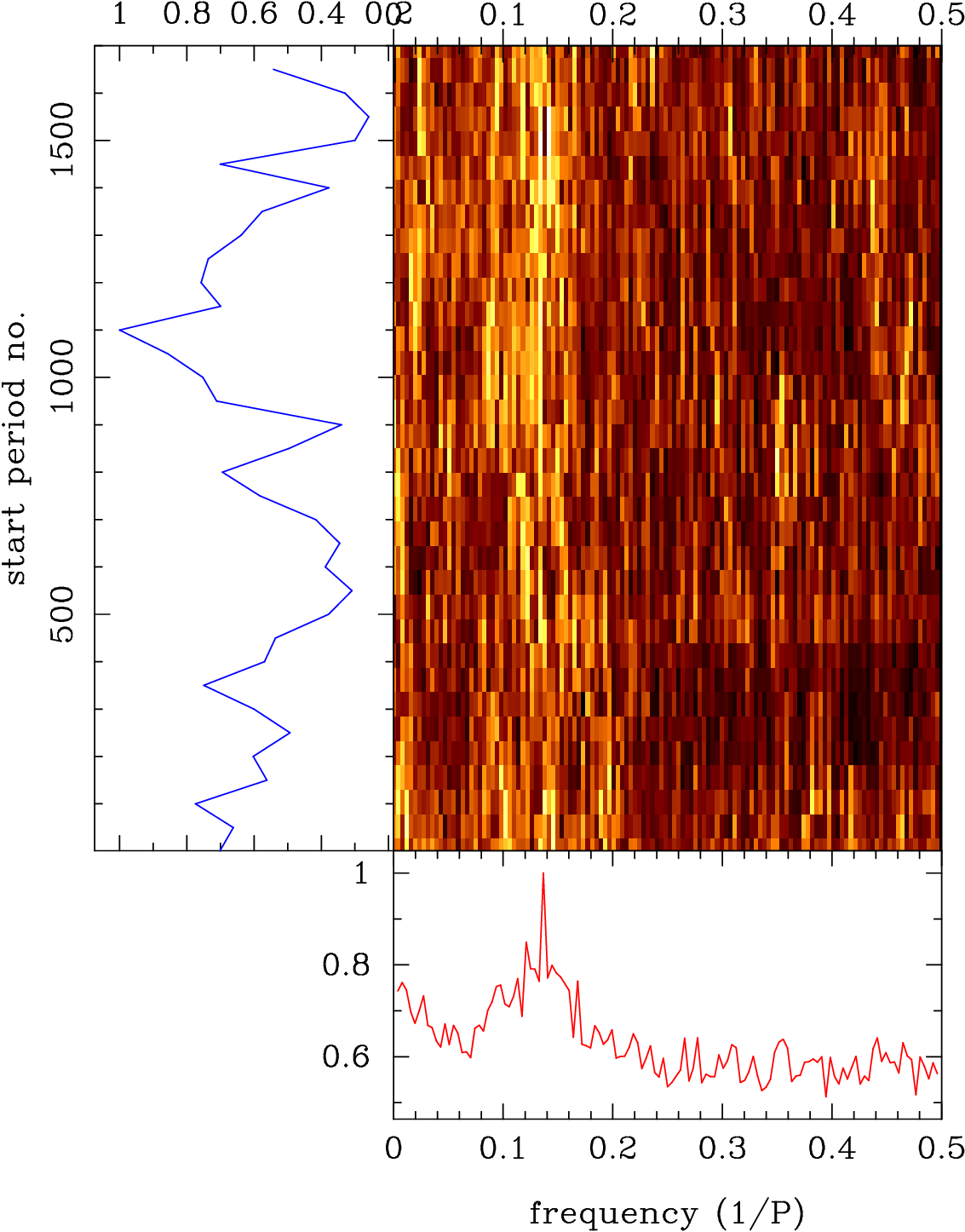}}} &
\hspace{50px}
{\mbox{\includegraphics[scale=0.4,angle=0.]{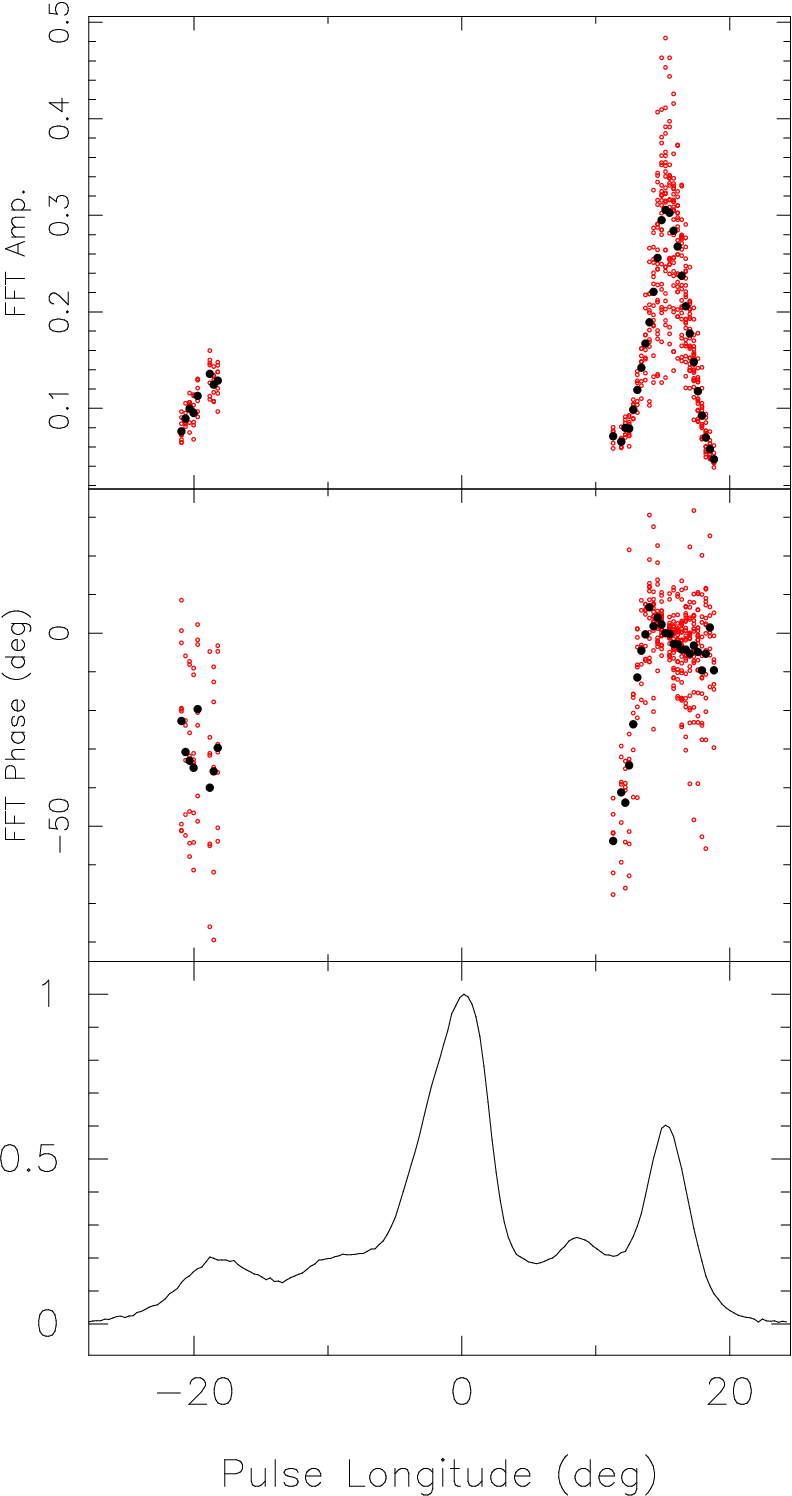}}} \\
\end{tabular}
\caption{The fluctuation spectral analysis for PSR J1900$-$2600 at 325 MHz. The
left panel shows time variation of LRFS. The right panel shows the variation of
the peak frequency across the pulse window; peak amplitude (top window), phase
variations (middle window), profile (bottom window).}
\label{fig_J1900_1}
\end{figure*}

% J1900-2600 - 618 MHz
\begin{figure*}
\begin{tabular}{@{}lr@{}}
{\mbox{\includegraphics[scale=0.4,angle=0.]{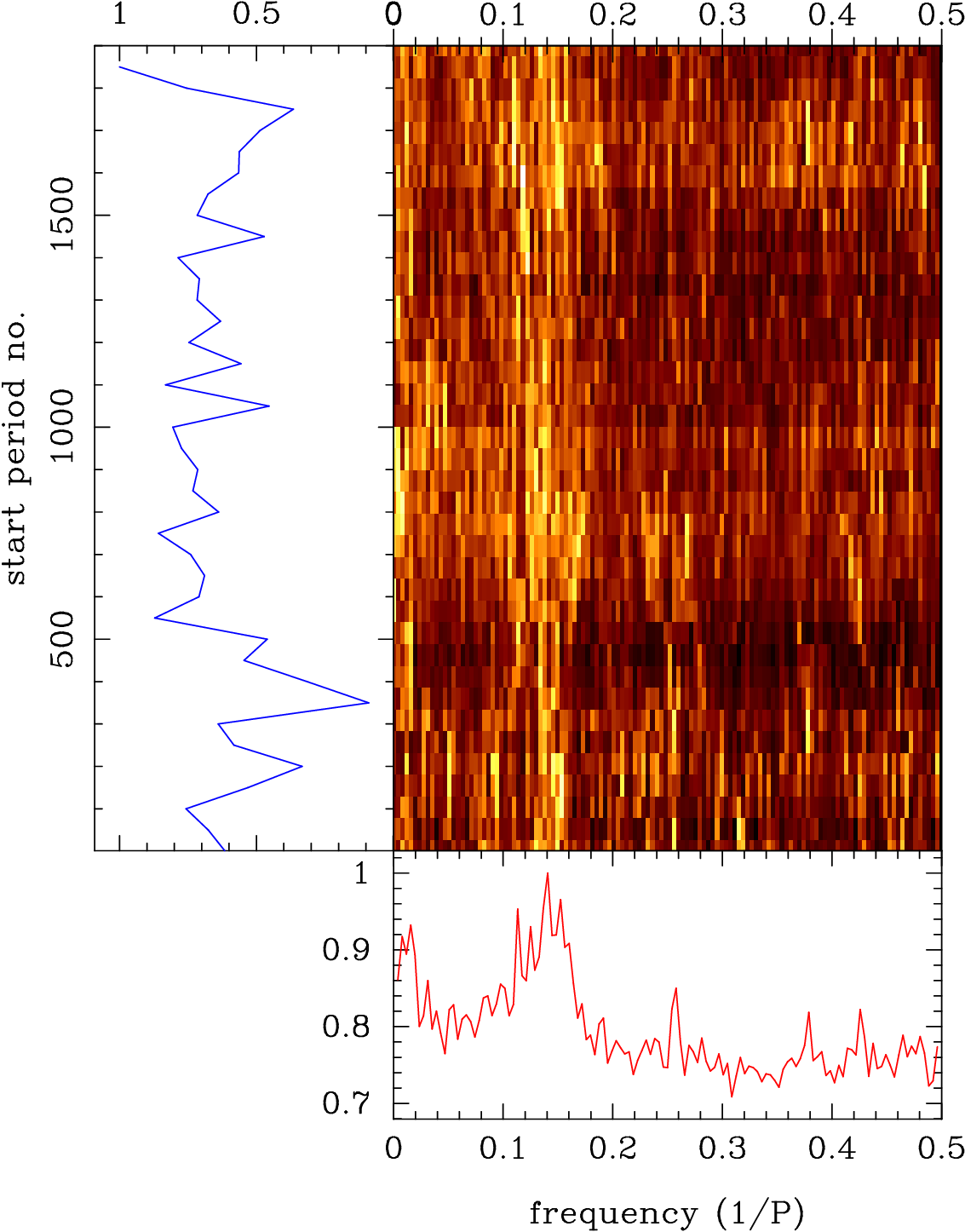}}} &
\hspace{50px}
{\mbox{\includegraphics[scale=0.4,angle=0.]{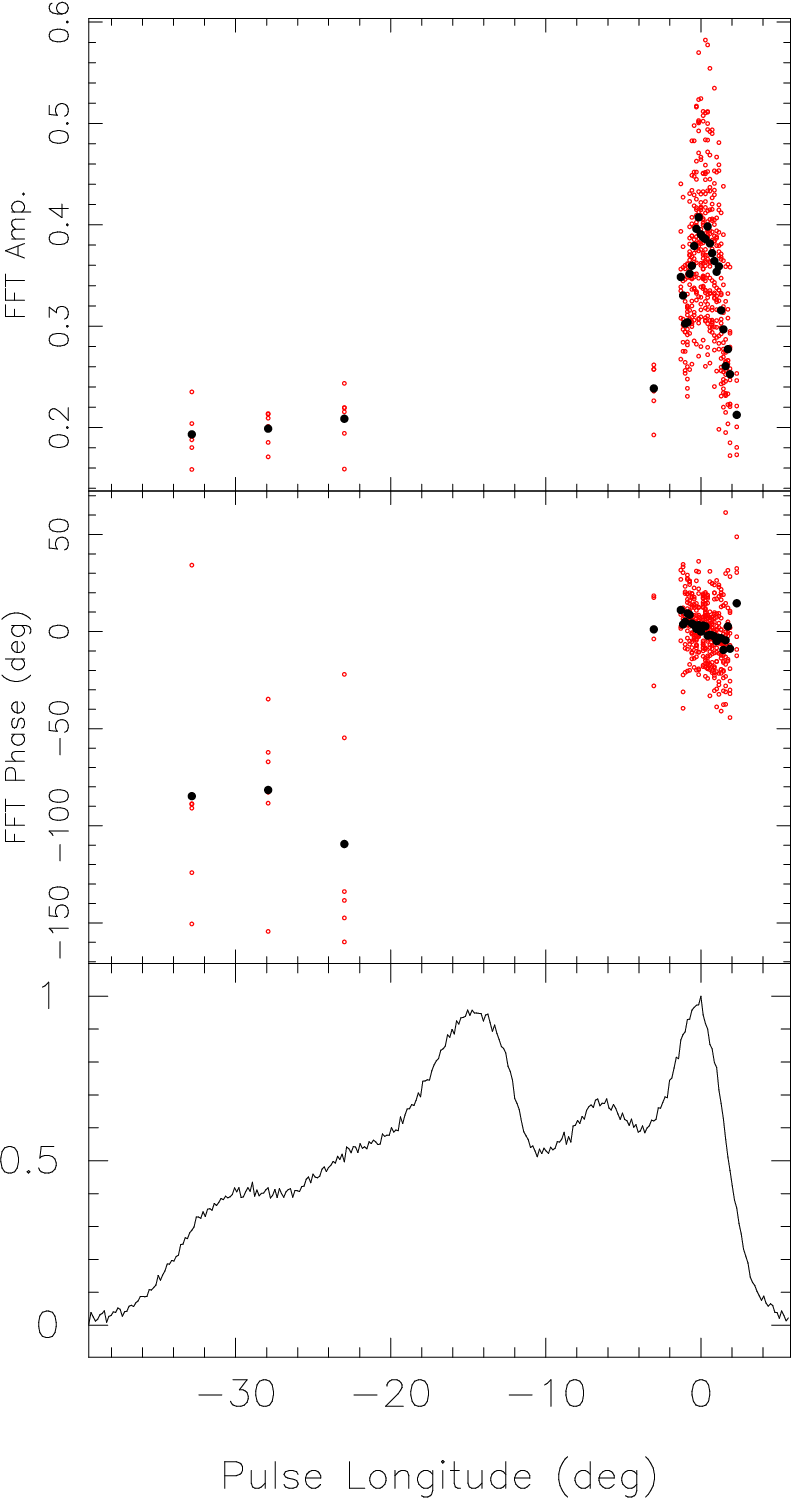}}} \\
\end{tabular}
\caption{The fluctuation spectral analysis for PSR J1900$-$2600 at 618 MHz. The
left panel shows time variation of LRFS. The right panel shows the variation of
the peak frequency across the pulse window; peak amplitude (top window), phase
variations (middle window), profile (bottom window).}
\label{fig_J1900_2}
\end{figure*}

\end{document}